\documentclass[structabstract]{aa}
\usepackage{graphicx}
\usepackage{rotating}
\usepackage{txfonts}
\usepackage{natbib}
\usepackage{lscape}
\usepackage{booktabs}
\usepackage{dcolumn}
\usepackage{float}

\bibpunct{(}{)}{;}{a}{}{,} 

\begin{document}
\renewcommand{\topfraction}{0.99}
\renewcommand{\bottomfraction}{0.99}
\renewcommand{\floatpagefraction}{0.99}
\renewcommand{\dbltopfraction}{0.99}
\renewcommand{\dblfloatpagefraction}{0.99}

\title{Catching the Radio Flare in CTA\,102}
\subtitle{III. Core-Shift and Spectral Analysis}

\author{C. M. Fromm
\inst{1}, E. Ros\inst{2,3,1}, M. Perucho\inst{3}, T. Savolainen\inst{1}, P. Mimica\inst{3}, M. Kadler\inst{4}, A. P. Lobanov\inst{1} \and J. A. Zensus\inst{1}}
\institute{Max-Planck-Institut f\"ur Radioastronomie, Auf dem H\"ugel 69, D-53121 Bonn, Germany\
\email{cfromm@mpifr.de}
\and Observatori Astron\`omic, Universitat de Val\`encia, Apartat de Correus 22085, E-46071 Val\`encia, Spain
\and Departament d'Astronomia i Astrof\'\i sica, Universitat de Val\`encia, Dr. Moliner 50, E-46100 Burjassot, Val\`encia, Spain\
\and Institut f\"ur Theoretische Physik und Astrophysik, Universit\"at W\"urzburg, Am Hubland, 97074 W\"urzburg, Germany}

\abstract
   {The temporal and spatial spectral evolution of the jets of Active Galactic Nuclei (AGN) can be studied with multi-frequency, multi-epoch very-long-baseline-interferometry (VLBI) observations. The combination of both, morphological (kinematical) and spectral parameters can be used to derive source intrinsic physical properties such as the magnetic field and the non-thermal particle density. Furthermore, we can trace the temporal variations of the source intrinsic parameters during the flare, which may reflect the interaction between the underlying plasma and a traveling shock wave. The source CTA\,102 exhibited such a radio flare around 2006}
   {In the first two papers of this series (Papers I \& II) we analyzed the single-dish light curves and the VLBI kinematics of the blazar CTA\,102 and suggested a shock-shock interaction between a traveling and a standing shock wave as a possible scenario to explain the observed evolution of the component associated to the 2006 flare. In this paper we investigate the core-shift and spectral evolution to test our hypothesis of a shock-shock interaction.}
   {We used eight multi-frequency Very Long Baseline Array (VLBA) observations to analyze the temporal and spatial evolution of the spectral parameters during the flare. We observed CTA\,102 between May 2005 and April 2007 using the VLBA at six different frequencies spanning from $2\,\mathrm{GHz}$ up to $86\,\mathrm{GHz}$. After the calibrated VLBA images were corrected for opacity, we performed a detailed spectral analysis. We developed methods for the alignment of the images and extraction of the uncertainties in the spectral parameters. From the derived values we estimated the magnetic field and the density of the relativistic particles and combined those values with the kinematical changes provided from the long-term VLBA monitoring (Paper II) and single-dish measurements (Paper I).}
   {The detailed analysis of the opacity shift reveals that the position of the jet core is proportional to $\nu^{-1}$ with some temporal variations. The value suggests possible equipartition between magnetic field energy and particle kinetic energy densities at the most compact regions. From the variation of the physical parameters we deduced that the 2006 flare in CTA\,102 is connected to the ejection of a new traveling feature $(t_\mathrm{ej}=2005.9)$ and the interaction between this shock wave and a stationary structure (interpreted as a recollimation shock) around 0.1\,$\mathrm{mas}$ from the core (de-projected 18\,$\mathrm{pc}$ at a viewing angle of $\vartheta=2.6^\circ$). The source kinematics together with the spectral and structural variations can be described by helical motions in an over-pressured jet.}
   {}

\titlerunning{CTA\,102  Core-Shift and Spectral Analysis}
\authorrunning{C. M. Fromm et al.}

\maketitle
\section{Introduction}
The spectral evolution of radio jets in active galactic nuclei (AGN) is probed by multi-frequency single-dish observations and their modeling can provide estimates on the evolution of physical conditions in the source \citep[see e.g.,][]{Marscher:1985p50,Turler:2000p1}. However, these observations cannot resolve the structure of the jet, and therefore only the overall spectral evolution can be obtained. One approach to extract the spatial and temporal spectral evolution of AGN, used in the case of blazars, includes multi-epoch and multi-frequency VLBI observations. These high-resolution VLBI observations offer the unique possibility to resolve the structure of AGN jets and allow us to study the morphological as well as the spectral evolution at parsec-scale resolutions. The analysis of these observations provides a tool to test several jet models and to extract the source intrinsic parameters.

Simultaneous VLBI observations at several frequencies can be used to analyze the variations in the observed base of the jet, i.e., the core, with frequency (core-shift analysis) \citep{Lobanov:1998p2152, Hirotani:2005p23}. These studies allow the observed core position to be related to source intrinsic parameters such as the magnetic field and the particle density \citep{Lobanov:1998p2152,Hirotani:2005p23,Kovalev:2008p1999,OSullivan:2009p1877,Pushkarev:2009p5426}. Based on the core-shift corrected maps, spectral index maps can be produced which show the spectral slope between two adjacent frequencies. The complete spectral information i.e., turnover frequency, $\nu_m$, turnover flux density, $S_m$ and spectral index, $\alpha_0$, can be extracted from multi-frequency VLBI maps by performing a spectral analysis. Such studies have been performed by e.g., \citet{Lobanov:1998p2310}, \citet{Savolainen:2008p2958}, and yield the spatial distribution of the spectral values and source intrinsic parameters such as the particle density, $N$, and the magnetic field, $B$.

Our multi-frequency VLBI observations of CTA\,102 between May 2005 and April 2007 cover the period of a large flux density outburst in April 2006. This data set allows us to examine the changes in the physical properties of the source under these extreme circumstances. In \citet[][hereafter Paper I ]{Fromm:2011p4088},we analysed single-dish data of the 2006 radio flare in CTA\,102 and suggested that the interaction between a propagating and a standing shock could be the driving mechanism behind the strong flare. We derived the evolution of physical parameters according to a slightly modified shock-in-jet model. The kinematic analysis of the source during the flare is presented in \citet[][hereafter Paper II ]{2012arXiv1211.3606F}. This study revealed one apparently stationary component at $r\sim0.1\,\mathrm{mas}$ from the core and the connection of the 2006 radio flare to the ejection of a new component at epoch $t_\mathrm{ej}=2005.9\pm0.2$. Based on the gradients in the evolution of the brightness temperature we tested several jet configurations and derived estimates for the variation of the Doppler factor along the jet. From the results, we found evidence for bends in the jet as well as standing components that could be associated to recollimation shocks.

In this third paper of the series we perform a full core-shift and spectral analysis to gain additional insights to clarify the jet nature. For this study we use eight multi-frequency VLBI observations with a frequency range between $2\,\mathrm{GHz}$ and $86\,\mathrm{GHz}$ during the 2006 radio flare.

{The organization of the paper is the following: The data analysis tools developed for the spectral analysis of VLBI observations are described in Sect.~\ref{data}. The observational results are divided into three sections: the core-shift analysis is presented in Sect.~\ref{coreshiftres}, the measure of the jet ridge-line and width is included in Sect.~\ref{jetridge}, and the results of the spectral analysis are presented in Sect.~\ref{resspecana}. We present the derivation of the physical parameters in the jet from the observational results in Sect.~\ref{physpara}. Finally, we discuss our results in Sect.~\ref{disc} and present our conclusions in Sect.~\ref{sum}. A brief summary of the basic synchrotron self-absorption mechanism and the required equations for the calculation of the source intrinsic parameters are given in Appendix A.}

Throughout the paper we define the spectral index, $\alpha$, using the relation $S_\nu \propto \nu^{\alpha}$. The optically thin spectral index, $\alpha_0$, can be derived from the spectral slope, $s$ of the relativistic electron distribution ($N\propto E^{-s}$), via the relation $\alpha_{0}=-(s-1)/2$. We define the optically thin spectral index as $\alpha_0<0$. We adopt the following cosmological parameters: $\Omega_m=0.27$, $\Omega_\Lambda=0.73$ and $H_0=71\,\mathrm{km\,s^{-1}\, Mpc^{-1}}$. This results in a linear scale of $8.11\,\mathrm{pc\,mas^{-1}}$ or $26.45\,\mathrm{ly\,mas^{-1}}$ for CTA\,102 ($z$=1.037). With these conventions, $1\,\mathrm{mas}\,\mathrm{yr}^{-1}$ corresponds to $52.9\,\mathrm{c}$. We use $R$ for the radius of the jet and $r$ for the distance along the jet. If not explicit mentioned we use \textit{cgs} units throughout this paper.

\section{Data analysis}
\label{data}
\begin{figure}[h!]
\resizebox{\hsize}{!}{\includegraphics{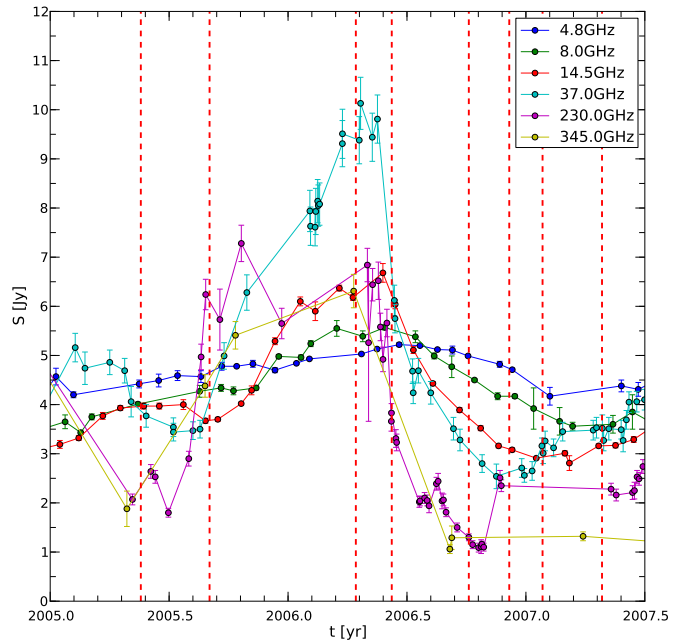}} 
\caption{$\mathrm{cm}$-$\mathrm{mm}$ single dish light curves for CTA\,102, covering the 2006 radio flare (see Paper I). The red dashed lines correspond to the epochs of multi-frequency VLBI observations presented in this paper. (Taken from Paper II)} 
\label{lc} 
\end{figure}

For our analysis we used eight VLBI observations spanning from May 2005 until April 2007. The coverage of the 2006 radio flare in CTA\,102 by our VLBI observations is presented in Fig. \ref{lc}, where we show the temporal correspondence between the $\mathrm{cm}$-$\mathrm{mm}$ light curves and the VLBA observations of  CTA\,102. The red dashed lines correspond to the epochs of multi-frequency VLBA observations. More details about calibration, imaging and model fitting of these VLBI images are presented in section~2 of Paper II. 

Following the convention used in Paper II, the structure of the jet in CTA\,102 can be studied by dividing it into four regions labeled as A ($r>8\,\mathrm{mas}$), B ($4\,\mathrm{mas}<r<8\,\mathrm{mas}$), D ($2\,\mathrm{mas}<r<4\,\mathrm{mas}$) and C ($r<2\,\mathrm{mas}$, the core region). Throughout our data set we could clearly cross-identify seven features within regions A, B, and D. The analysis of the long-term monitoring of the source at $43\,\mathrm{GHz}$ revealed four additional features in region C (Paper II).

Figure \ref{allcont} shows the VLBI observation of CTA\,102 in June 2006 together with the cross-identified circular Gaussian components.  A more detailed view into region C is provided in Fig. \ref{allcontcore}. All images of the source at different frequencies together with the fitted components are presented in the Appendix of Paper II. Table \ref{averpara} presents a list of typical (average) image parameters for the CTA\,102 observations at different frequencies.

\begin{table}[h!]
\caption{Typical image parameters for the CTA\,102 observations}  
\label{averpara}
\centering  
\begin{tabular}{c c c c c} 
\hline\hline
$\nu$ &$\Theta_\mathrm{min}$ & $\Theta_\mathrm{maj}$ & P.A. & pixel size \\
{$[\mathrm{GHz}]$}& $[\mathrm{mas}]$ & $[\mathrm{mas}]$ & $[\deg]$ &$[\mathrm{mas}]$ \\
\hline
2	&	3.73	&	8.17	&   $-$5	&	0.70	\\
5	&	1.52	&   	3.65	&   $-$8	&	0.30	\\
8	&	0.97	&	2.32	&   $-$7	&	0.20	\\
15	&	0.52	&	1.33	&   $-$9	&	0.10	\\
22	&	0.33	&	0.95	&   $-$13	&	0.07	\\
43	&	0.18	&	0.45	&   $-$11	&	0.04	\\
86	&	0.11	&	0.25	&   $-$19	&	0.03	\\
\hline
\end{tabular} 
\end{table}

\subsection{Image Alignment}
\label{align}
The absolute position of the source is lost during the calibration process of VLBI observations due to the use of closure phases. However, a proper relative alignment of the different frequency images is a basic requirement for a reliable core-shift and spectral analysis. The ideal technique to restore the initial position of the source would be based on phase-referencing observations.

\clearpage

\begin{figure*}[t!]
\centering 
\includegraphics[width=17cm]{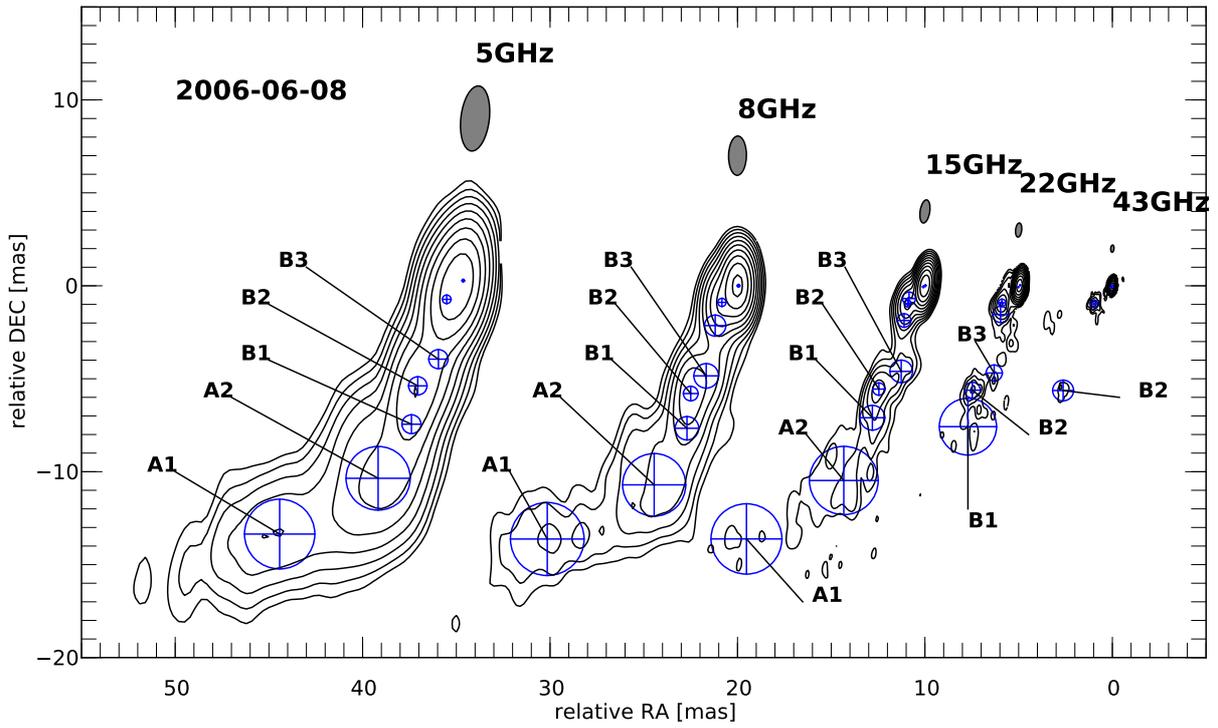} 
\caption{Uniform weighted VLBA CLEAN images with fitted circular Gaussian components at different frequencies for the July 2006 observation of CTA102. The lowest contour is plotted at $10\times$ the off-source $\mathrm{rms}$ at $43\,\mathrm{GHz}$ and increases in steps of 2. The observing frequency and the restoring beam size are plotted above each map. For the labeling we use capital letters for the same physical region in the jet and the numbers increase with inverse distance from the core. For a more detailed picture of the core region see Fig. \ref{allcontcore}.} 
\label{allcont} 
\end{figure*}

\begin{figure*}[b!]
\centering 
\includegraphics[width=17cm]{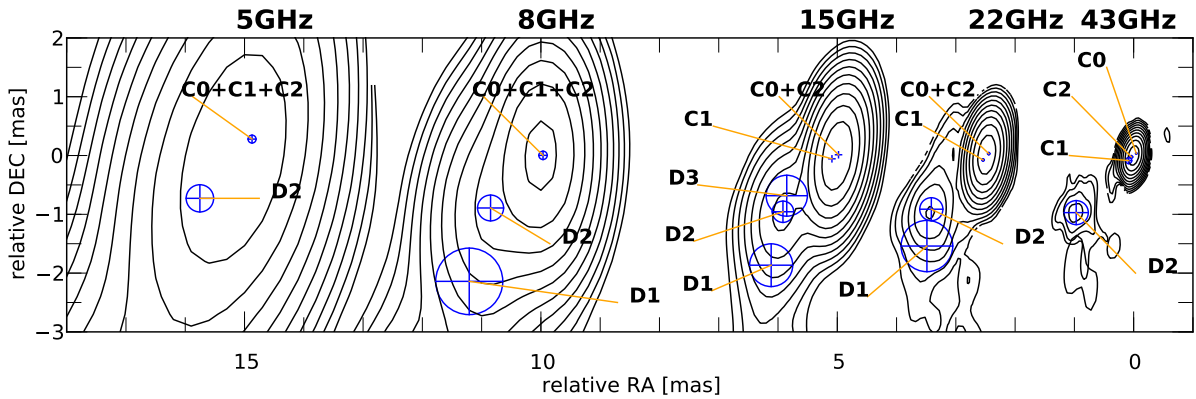} 
\caption{Zoom into the core region for the images presented in Fig.\ref{allcont} corresponding to epoch 2006-06-08. Notice the splitting of the C-components with increasing frequency. For map details see the caption of Fig.\ref{allcont}. } 
\label{allcontcore} 
\end{figure*}
\clearpage
Such experiments require a compact object in the neighborhood of the source. In most of the cases this object would be not within the primary antenna beam of the interferometer, and this requires a nodding between the main source and the calibrator \citep[e.g.,][]{Ros:2005p2974}. Our observations were not designed in phase-referencing mode making another approach is necessary.

A common procedure to correct for the frequency dependent position of the source and especially the core, is based on the assumption that the position of an optically thin region ($S\propto v^{\alpha_0}$) does not depend on the observing frequency.
There are two different approaches for the registration of multi-frequency VLBI observations; one based on cross-identification of fitted Gaussian functions to the visibilities (denoted as components) and another based on 2D cross-correlation of the optically thin emission regions in the image plane. The advantage of the 2D approach is that the whole optically thin emission region is taken into account for the alignment, whereas the one based on fitted components takes only a single region into account. Nevertheless, if the source does not exhibit an extended jet region the approach based on fitted features would be a better choice \citep[see,e.g.,][]{2004A&A...426..481K,Kovalev:2008p1999, Croke:2008p1889}.
\newline In this paper we used a hybrid approach (2D cross-correlation and fitted components) depending on the amount of extended structure and on the resulting spectral index maps. We followed the work of \citet{Croke:2008p1889} and adjusted their technique to our needs. For the alignment based on cross-identified components we used the feature labeled as B2 for frequencies $\nu<22\,\mathrm{GHz}$ and component D2 for higher frequencies. Both features could be cross-identified throughout our entire data-set (see Fig. \ref{allcont} and Paper II).

\subsubsection{2D cross-correlation of VLBI maps}
The image parameters of a VLBI map, e.g., map size ($ms$), pixel size ($ps$) and convolving beam size ($bs$), increase with decreasing frequency (to optimize them with the resolution in each case). Before we perform the cross-correlation of the VLBI images we select a set of common image parameters. The best results for the cross correlation of two VLBI images with frequencies, $\nu_1$ and $\nu_2$, where $\nu_1>\nu_2$, are achieved by using the map size and beam size of the low frequency image, $ms(\nu_{2})$ and $bs(\nu_{2})$, and $\frac{1}{20}$ of the high frequency beam size as pixel size $ps$ \citep[see also][]{Pushkarev:2012p6264}. These settings guarantee that no structure information is lost during the alignment process. After both images are convolved with the common beam, the optically thick core region is masked in both maps. During the cross-correlation process one of the images is shifted in $\Delta x$ and $\Delta y$ direction and the cross-correlation coefficient (see Eq. \ref{cc}) at each shift position $(\Delta x,\Delta y)$ is calculated. The cross-correlation coefficient $cc_{\Delta x,\Delta y}$, between the two maps is defined as (assuming that only the second map is shifted):
 \begin{equation}
 cc=\frac{\sum_{i=1}^n\sum_{j=1}^m\left(S_{\nu_{1}}^{i,j}-\bar{S}_{\nu_1}\right)\left(S_{\nu_2}^{\Delta x\cdot i,\Delta y\cdot j}-\bar{S}_{\nu_2}\right)}{\left[\sum_{i=1}^n\sum_{j=1}^m\left(S_{\nu_1}^{i,j}-\bar{S}_{\nu_1}\right)^2\sum_{i=1}^n\sum_{j=1}^m\left(S_{\nu_2}^{\Delta x\cdot i,\Delta y\cdot j}-\bar{S}_{\nu_2}\right)^2\right]^{\frac{1}{2}}},
 \label{cc}
 \end{equation}  
 where {$i$ and $j$ are the indices of pixels in the different images}, $S_{\nu_{1}}^{i,j}$ is the flux density of the non-shifted image at the position $(i,j)$ and $S_{\nu_{2}}^{\Delta x\cdot i,\Delta y \cdot j}$ is the flux density of the shifted map by $\Delta x$ and $\Delta y$.  Finally, $\bar{S}_{\nu_{1,2}}$ is the average flux density in the images. The result of this approach is the distribution of cross-correlation coefficients within the shift area. The best shift position corresponds to the maximum cross-correlation coefficient. The finite width of the pixel size leads to discrete values for the best shift position in steps of the used pixel size. This dependence on the selected pixel size can be overcome by fitting a 2D elliptical Gaussian to the cross-correlation distribution achieved. A 2D elliptical Gaussian distribution can be written in the following form
\begin{equation}
g(x,y)=Ae^{-\left(a\left(x-x_0\right)^2+2b\left(x-x_0\right)\left(y-y_0\right)+c\left(y-y_0\right)^2\right)},
\label{egauss}
\end{equation}
\noindent where the coefficients $a,\,b,\,c$ are defined by:
\begin{eqnarray}
&a&=\frac{\cos{^2\Theta}}{2\sigma^2_x}+\frac{\sin{^2\Theta}}{2\sigma^2_y},\\
&b&=-\frac{\sin{2\Theta}}{4\sigma^2_x}+\frac{\sin{2\Theta}}{4\sigma^2_y},\\
&c&=\frac{\sin{^2\Theta}}{2\sigma^2_x}+\frac{\cos{^2\Theta}}{2\sigma^2_y},
\end{eqnarray}
with $\Theta$ the rotation angle, $x_0$ and $y_0$ the mean values, and $\sigma_x$ and $\sigma_y$ the standard deviations. The fitted parameters $x_0$ and $y_0$ can be regarded as the best shift position.
\newline If the distribution of the cross-correlation coefficient is smooth and symmetric around the mean values, the distribution of the cross correlation coefficient could be described as a Gaussian distribution and the mean values are the best choice for the correction of the core shift. Otherwise, the position of the highest cross correlation coefficient is a suitable choice.\\
The distribution of the spectral index, $\alpha$, computed from two adjacent frequencies is highly sensitive to positional changes:
\begin{equation}
\alpha=\frac{\log\left(S(\nu_1)/S(\nu_2)\right)}{\log\left(\nu_1/\nu_2\right)}
\label{specindex}
\end{equation}
In this way, a visual check on $\alpha$ for both alignment methods allows us to diagnose their performance. 
In principle, the distribution of the spectral index, $\alpha$, should be smooth across the jet with a decrease towards the extended structure and a slight transversal gradient.    
We computed several spectral index maps using the 2D cross-correlation approach and component approach in order to investigate the difference between both methods. Examples are presented in Figs. \ref{dist1} -\ref{spix1} for the alignment of two low frequency and two high frequency VLBI images.

\subsubsection{Low frequency alignment}
\begin{figure}[h!]
\resizebox{\hsize}{!}{\includegraphics{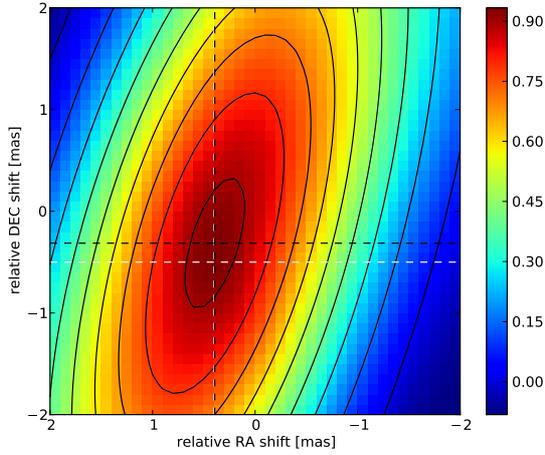}} 
\caption{Distribution of the cross-correlation coefficient for the alignment of the $8\,\mathrm{GHz}$ and $5\,\mathrm{GHz}$ VLBA maps. The {intersection of the} white dashed lines correspond to the maximum cross-correlation coefficient in the color-map and the {intersection of the} dashed black line for the peak position of the fitted 2D Gaussian.} 
\label{dist1} 
\end{figure}

For the alignment of the $8\,\mathrm{GHz}$ and $5\,\mathrm{GHz}$ maps of the May 2005 observations, we used a pixel size of $0.04\,\mathrm{mas}$ and beam size of $3.65\times1.52\,\mathrm{mas}$ and a P.A. of $-8^\circ$ for the 2D cross-correlation and for the component based approach we used the feature labeled as B2 (see contour plots in Fig. \ref{allcont}). Figure \ref{dist1} presents the distribution of the cross correlation coefficient between the $8\,\mathrm{GHz}$ and the $5\,\mathrm{GHz}$ image. The best shift position was found at $\Delta x=0.38\,\mathrm{mas}$ and $\Delta y=-0.28\,\mathrm{mas}$ {(black dashed line in Fig. \ref{dist1})}. From the component-based approach we obtained an image shift of $\Delta x=0.47\,\mathrm{mas}$ and $\Delta y=-0.34\,\mathrm{mas}$.

\begin{figure*}
\centering 
\includegraphics[width=17cm]{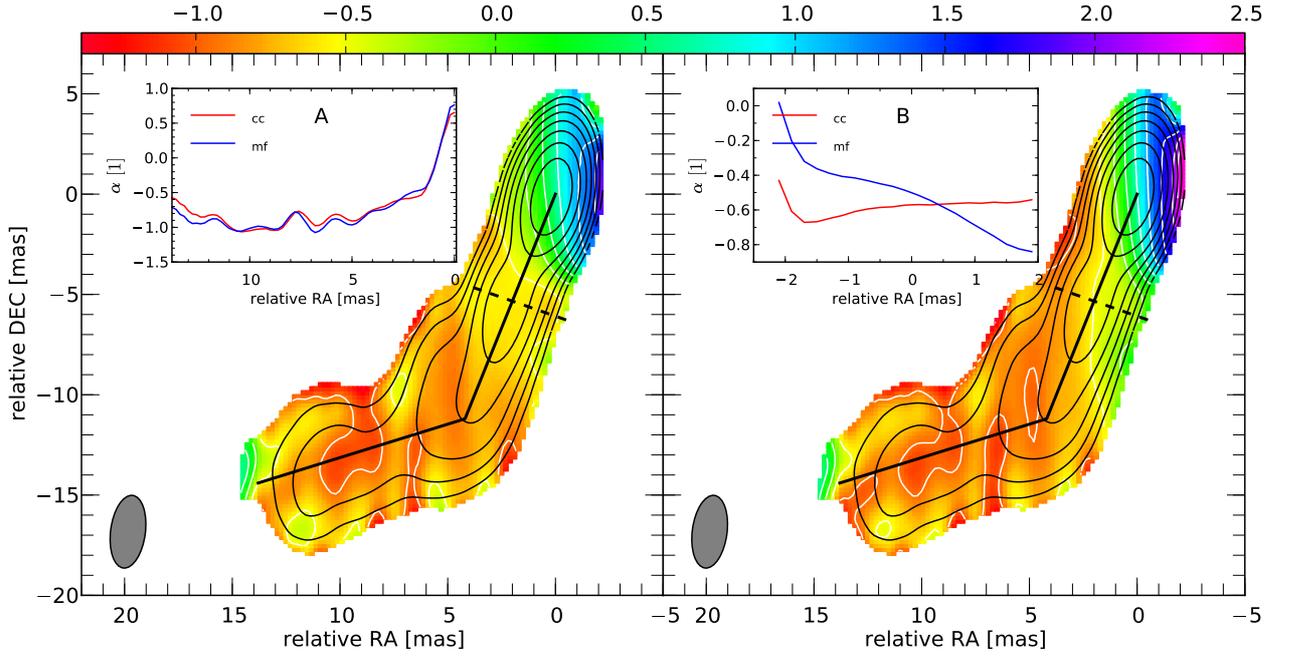} 
\caption{$5\,\mathrm{GHz}$-$8\,\mathrm{GHz}$ spectral index maps for the May 2005 observations of CTA\,102. The left panel shows the spectral index map produced by the 2D cross-correlation and the right panel the one by using the component based approach. The contours are drawn for the $8\,\mathrm{GHz}$ observations. The inlets show the distribution of the spectral index along the jet axis, black solid line, (inlet A) and transversal to the jet axis, dashed black line, (inlet B).} 
\label{spix} 
\end{figure*}

The distribution of the spectral index shows in both cases an optically thick core region $\left(\alpha>0\right)$ and an optically thin jet region $\left(\alpha<0\right)$, with the transition at a distance of $r\approx2.7\,\mathrm{mas}$ away from the core. The evolution of the spectral index along the jet axis is similar for both approaches with minor variations (see inlet A in Fig. \ref{spix}). However, there are major differences in the transversal distribution of the spectral index (inlet B in Fig. \ref{spix}). The modelfit based approach has a strong transversal gradient in the spectral index (blue line in inlet B), which is smoothed out by the 2D cross-correlation technique (red line in inlet B).

\subsubsection{High frequency alignment}
\begin{figure}[h!]
\resizebox{\hsize}{!}{\includegraphics{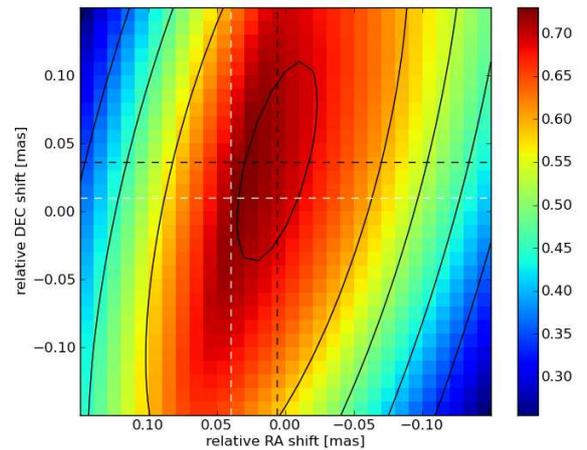}} 
\caption{Distribution of the cross-correlation coefficient for the alignment of the $43\,\mathrm{GHz}$ and $22\,\mathrm{GHz}$ VLBI map. The {intersection of the} white dashed lines correspond to the maximum cross-correlation coefficient in the color-map and the {intersection of the} dashed black line for the peak position of the fitted 2D Gaussian.} 
\label{dist2} 
\end{figure}

We used the $22\,\mathrm{GHz}$ and $43\,\mathrm{GHz}$ VLBA observations in May 2005 epoch to study the alignment of high frequency maps. The 2D cross-correlation technique resulted in a correction of $\Delta x=0.04\,\mathrm{mas}$ and $\Delta y=0.01\,\mathrm{mas}$, using a pixel size of $0.01\,\mathrm{mas}$, beam size of $0.95\times0.33\,\mathrm{mas}$, and a P.A. of $-13^\circ$ (see Fig. \ref{dist2} for the distribution of the cross-correlation coefficient). Fixing component D2 (see Fig. \ref{allcontcore}), we calculated a value of  $\Delta x=0.03\,\mathrm{mas}$ and $\Delta y=0.01\,\mathrm{mas}$ for the correction of the opacity shift. Since the distribution of the cross-correlation coefficient is not well described by a 2D-Gaussian, we selected the position of the highest cross-correlation coefficient as the best shift position {(see white  dashed line in Fig. \ref{dist2})}.

\begin{figure*}
\centering 
\includegraphics[width=17cm]{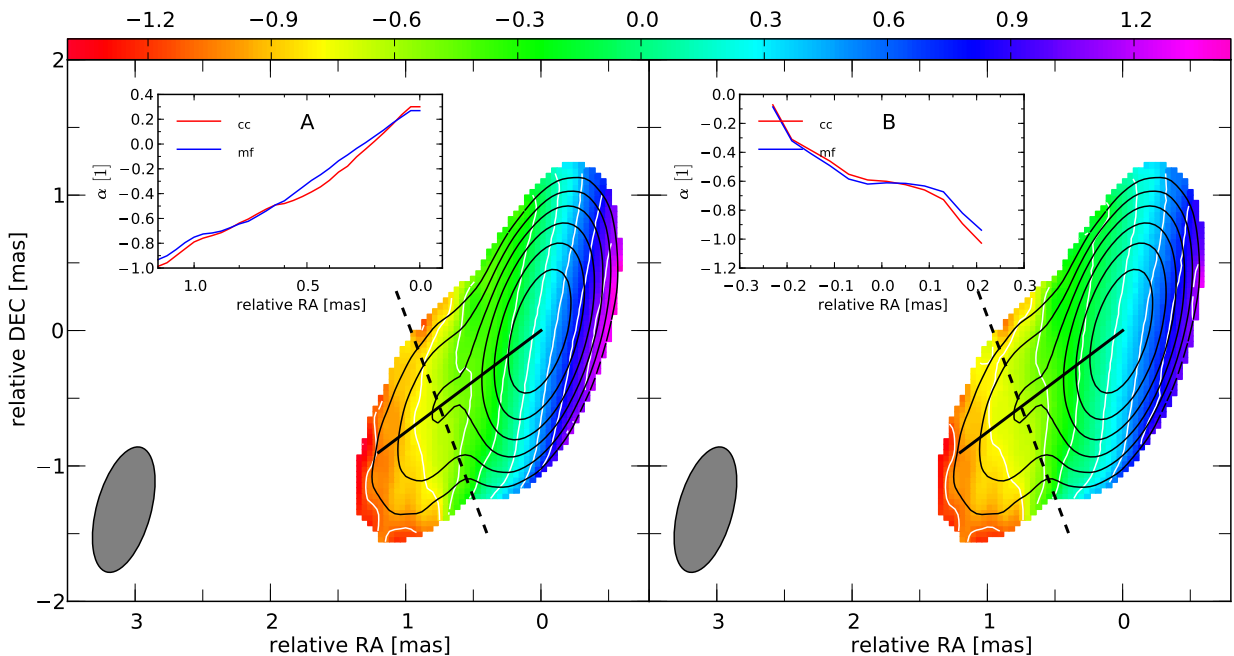} 
\caption{$22\,\mathrm{GHz}$-$43\,\mathrm{GHz}$ spectral index maps for the May 2005 observations of CTA\,102. The left panel shows the spectral index map produced by the 2D cross-correlation and the right panel the one by using the component based approach. The contours are drawn for the $43\,\mathrm{GHz}$ observations. The inlets show the distribution of the spectral index along the jet axis, black solid line, (inlet A) and transversal to the jet axis, dashed black line, (inlet B).} 
\label{spix1} 
\end{figure*}

The computed spectral index maps between $22\,\mathrm{GHz}$ and $43\,\mathrm{GHz}$ are presented in Fig. \ref{spix1}. The distribution of the spectral indices in the optically thick core region $\left(r<0.3\,\mathrm{mas}\right)$ follows the same decreasing trend as in the low-frequency case (see inlet A in Fig. \ref{spix1}). The distributions are slightly rotated by roughly $15^\circ$. The two different techniques, 2D cross-correlation and component-based alignment lead to nearly identical spectral index maps.

\subsection{Spectral Analysis}
\label{spectralana}
For the extraction of the spectral parameters, turnover frequency, $\nu_m$, turnover flux density, $S_m$, and optically thin spectral index, $\alpha_0$ we need at least three VLBI observations at different frequencies. In general, there are two approaches for the extraction of the spectral parameters, one based on the flux densities in each pixel \citep{Lobanov:1998p2310} and the other one on the fitted components \citep{Lobanov:1999p2299,Savolainen:2008p2958}. The advantage of the pixel-based approach is that it provides a continuous evolution of the spectral values along the jet. If the spectrum is not homogenous because several components are located within a region of the size of the convolving beam, or the turnover frequency, $\nu_m$, is out of our frequency range, we can no longer obtain the turnover values. Nevertheless, the spectral index, $\alpha$ ($S_\nu\propto \nu^{\alpha}$) can 
still de derived.\\

As mentioned in section~\ref{align}, the convolving beam and pixel size of the maps are decreasing with increasing frequency (see Table \ref{averpara}). An appropriate selection of the common beam and pixel size for all the maps involved in the analysis is needed to avoid the generation of image artifacts (small beam size) or the loss of structural information (too large pixel size). There are two possible techniques that can be used to avoid these difficulties. The first one uses the beam size of the lowest frequency map and pixel size of the highest frequency map. The second one is an adaptive method, where the beam and pixel size are selected depending on the distance to the core, i.e., the closer to the core, the smaller the beam and the pixel size. Using the latter implies that the lower frequency maps have to be excluded from the spectral analysis closer to the core region to avoid image artifacts.

After one of the aforementioned methods is selected and the opacity shift is corrected in the maps relative to a reference one (Sect. \ref{align}), the pixel-to-pixel spectral analysis can be performed. The procedure is the following: i) The aligned images are all convolved with a common beam and pixel size and are stacked on top of each other into a datacube. ii) The dimensions of this cube are the map dimensions in the $x-$ and $y-$direction and the number of involved frequencies in the $z-$direction. Due to the already performed image alignment (see Sect. \ref{align}), each pixel at a given frequency $\left[j(\nu_u),i(\nu_u),k(\nu_u)\right]$ corresponds to its counterpart at the other frequencies $\left[j(\nu_p),i(\nu_p),k(\nu_p)\right]$, where $j(\nu_{u/p}),i(\nu_{u/p})$ are the array indices of each VLBA map in $x-$ and $y-$direction and $k(\nu_{u/p})$ is the frequency {index}. 
iii) For each position (x,y) a frequency vector $\vec{\nu_{x,y}}=\left[\nu_{0}\cdots \nu_{n}\right]$ and a flux density vector $\vec{S_{\nu,x,y}}=\left[S_{\nu,0}\cdots S_{\nu,n}\right]$ vector can be extracted, with $n$ the total number of maps included in the spectral analysis. 

\begin{figure*}
\centering 
\scalebox{1.2}{\includegraphics[width=14cm]{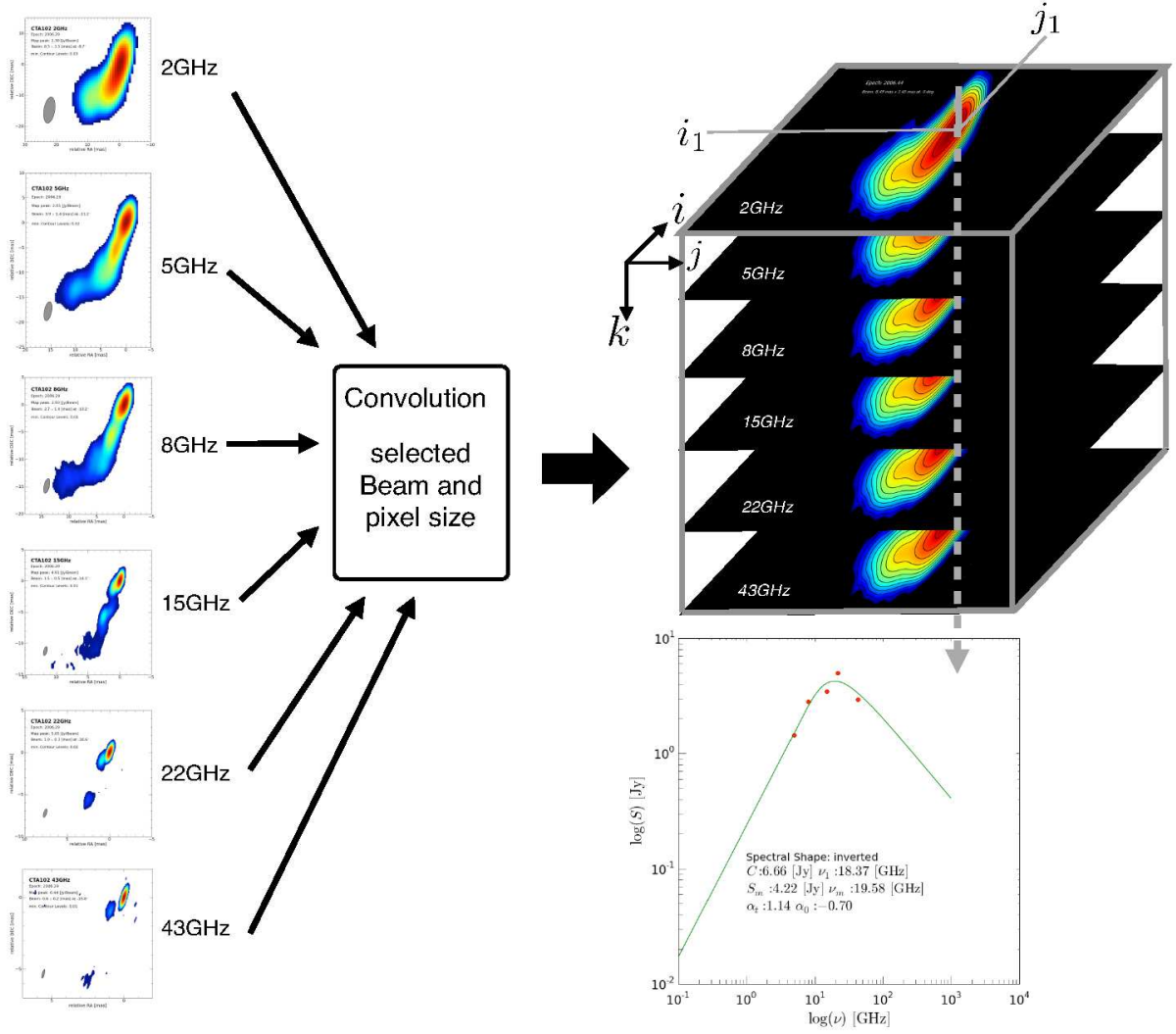}} 
\caption{Illustration of the basic steps within our spectral analysis algorithm: i) image alignment and convolution with common beam and pixel size, ii) stacking of the images, and iii) fitting of the synchrotron-self absorbed spectrum. See text for more details.} 
\label{specanaroutine} 
\end{figure*}

By fitting the approximation of the synchrotron-self absorbed spectrum to $\vec{\nu_{x,y}}$ and $\vec{S_{\nu,x,y}}$, the turnover frequency, $\nu_{m,x,y}$, the turnover flux density $S_{m,x,y}$, and the optically thin spectral index, $\alpha_{0,x,y}$, can be obtained, while fixing $\alpha_{t,x,y}=5/2$ (assuming homogeneous synchrotron emission feature). Extending this technique to the overall source structure results in a 2D-distribution of the turnover values for each multi-frequency epoch. Figure \ref{specanaroutine} illustrates our spectral analysis algorithm. 

We estimated the uncertainties on the derived spectral parameters including the uncertainties of the measured parameters and the sensitivity of the computing method to the shifts between images, as described in Appendix D. In short, first we assumed that the uncertainties in the flux density vary between 5\% and 15\%. We included this frequency-dependent flux density uncertainty in our error analysis and, additionally, we incorporated the flux density errors for each pixel, based on the image SNR at each of them. To estimate the influence of the uncertainties of the opacity shifts we assumed that the uncertainties of the obtained shift positions are within the used pixel size. Therefore, we randomly shifted each map within the allowed position, extracted the spectral parameters and calculated the scatter of the obtained values.  This was done by performing a Monte Carlo simulation for which we used up to $10^4$ random shifts for each epoch to derive the uncertainties for the spectral parameters. We rejected turnover values which are not well constrained within our data set, i.e., spectra for which we obtained turnover frequencies, $\nu_{m}$, lower than 75\% percent of the minimum frequency in the data set. This procedure provides the most reliable turnover values and is incorporated in presented evolution of the spectral parameters.

\section{Core-shift analysis}
\label{coreshiftres}
We used the technique presented in Sect. \ref{align} to align the VLBI observations of CTA\,102. Due to the difference in the image parameters, especially in the convolving beam size, we used only adjacent frequency maps in the alignment. The shift values obtained relatively to the initial position are presented in Table~\ref{2dshifts}.

\begin{figure*}
\centering 
\includegraphics[width=17cm]{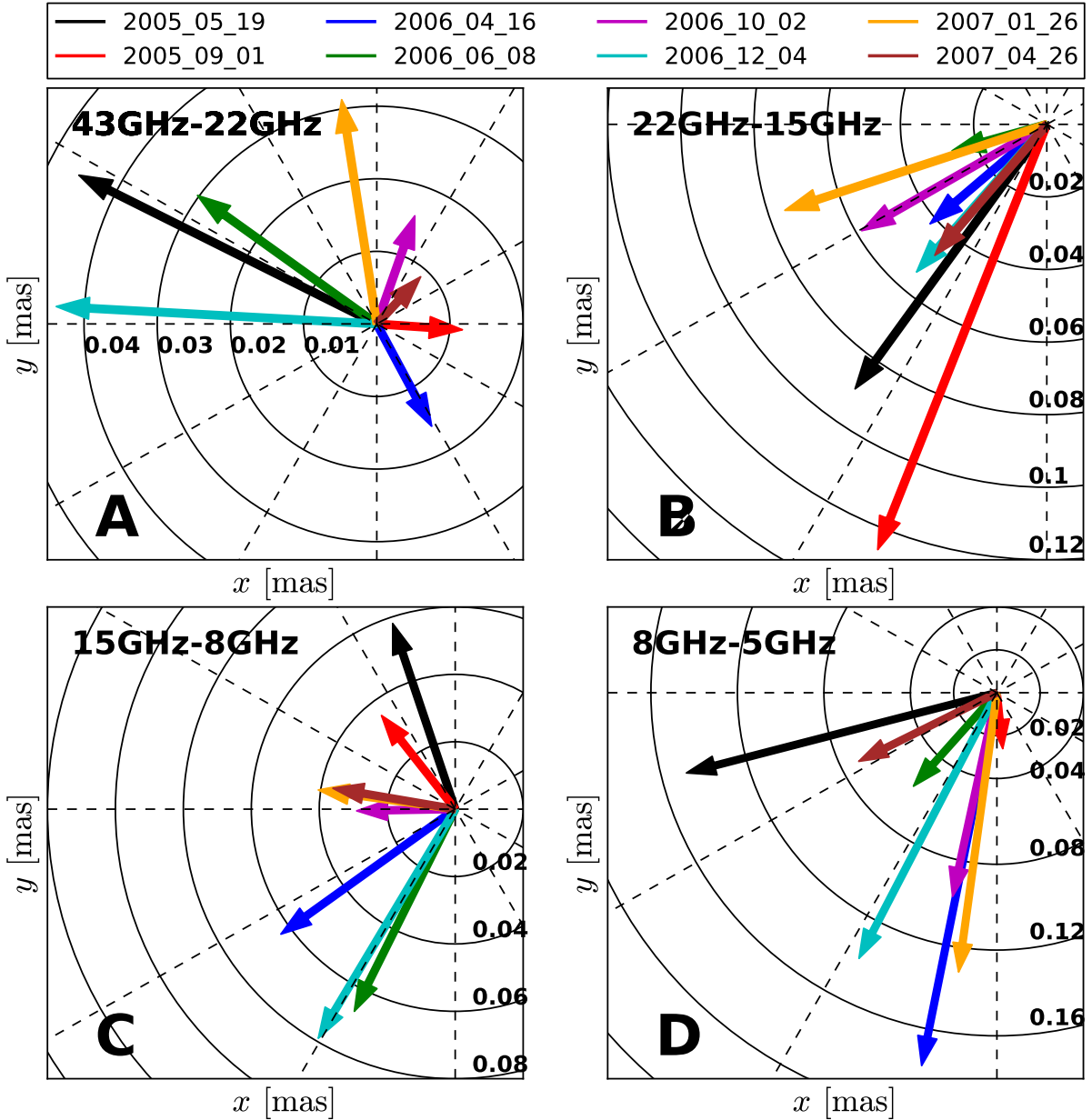} 
\caption{Variation of the core-shift vector with time for different frequency pairs (panel A: $(43-22)\,\mathrm{GHz}$, panel B: $(22-15)\,\mathrm{GHz}$, panel C: $(15-8)\,\mathrm{GHz}$ and panel D: $(8-5)\,\mathrm{GHz}$). The solid circles correspond to the radial distance from the reference core and the dashed lines are drawn in intervals of $30^\circ$. The color of the arrow indicated the different observational epochs. (See text for more details).} 
\label{coreshiftvec} 
\end{figure*}

There are two different shift directions towards the north-east (positive $x-$ and $y-$ directions) for frequencies larger than 22\,$\mathrm{GHz}$ (see panel A in Fig.~\ref{coreshiftvec}) and towards the south-east direction for lower frequencies (relative to the highest frequency involved in the analysis, see panels B, C, and D in Fig.~\ref{coreshiftvec}). An exception of this observed behavior appears in the April 2006 epoch, possibly due to the large flux density outburst during this time. 

We derived the core-shift vector from the correction of the opacity shift (see Table~\ref{2dshifts}) and the position of the VLBI core (see Paper II for more details on the jet model fitting). Figure~\ref{coreshiftvec} shows the variation of the core-shift vector for the different frequency pairs with time. The solid circles indicate the distance from the reference core in $\mathrm{mas}$, the dashed lines are drawn in intervals of $30^\circ$ and the different colors of the arrow correspond to the different observational epochs. 

\begin{table*}
\caption{Correction for the opacity shifts obtained by 2D cross correlation between adjacent frequencies}  
\label{2dshifts}
\centering 
\small
\begin{tabular}{@{}c c c c c c c c c c c c c@{}} 
\hline\hline
Epoch & \multicolumn{2}{c}{$(86-43)\,\mathrm{GHz}$} &  \multicolumn{2}{c}{$(43-22)\,\mathrm{GHz}$} &  \multicolumn{2}{c}{$(22-15)\,\mathrm{GHz}$} &  \multicolumn{2}{c}{$(15-8)\,\mathrm{GHz}$} &  \multicolumn{2}{c}{$(8-5)\,\mathrm{GHz}$} &  \multicolumn{2}{c}{$(5-2)\,\mathrm{GHz}$}\\
	   &	 x	& 	y 	&	x	&	y	&	x	&	y	&	 x	& 	y 	&	x	&	y	&	x	&	y\\
$[$yyyy-mm-dd] & [mas] & [mas] &[mas] & [mas] & [mas] &[mas] & [mas] & [mas] &[mas] & [mas] & [mas] &[mas] \\
\hline
2005-05-19 & 0.02	& 0.02 	& 0.04 	& 0.01 	& 0.04 	& $-$0.04 &0.12 	& $-$0.12	& 0.38 		&$-$0.28 		& -- 		& --\\
2009-09-01 & -- 	& --		& 0.01 	&$-$0.02	& 0.06 	& $-$0.11	& 0.10 	&0.02	& 0.32 		& $-$0.52 	&-- 		& -- \\
2006-04-16 & -- 	& --		& 0.02	&$-$0.03	& 0.03	&$-$0.01	& 0.09	&$-$0.09	& 0.24		& $-$0.52		& 0.80	&	$-$2.25 \\
2006-06-08 &--		& --		&0.01	& 0.02	& 0.01	&$-$0.01  &0.03	&$-$0.05	&0.12		& $-$0.32		& 0.80	&	$-$2.54 \\
2006-10-02 &--		& --		&0.03	&$-$0.03	& 0.05	&$-$0.02	&0.03	&$-$0.03	&0.26		& $-$0.45		& --		& -- \\
2006-12-04 &--		& --		&0.08	&$-$0.05	& 0.06	&$-$0.04	&0.06	&$-$0.09	&0.35		& $-$0.52		& --		& -- \\
2007-01-26 &--		& --		&0.03	&0.00 	&0.10	&$-$0.05	&0.04	&$-$0.03	&0.30		& $-$0.50		& --		& -- \\
2007-04-26 &--		& --		&0.01	&$-$0.02	&0.03	&$-$0.02	&0.09	&$-$0.06	&0.48		& $-$0.64		& --		& -- \\	
\hline

\end{tabular} 
\end{table*}

\begin{table*}[t!]
\caption{Core shift values relative to the reference frequency (given in the second column) for different epochs}  
\label{coreshiftstab1}
\centering 
\small
\begin{tabular}{@{}c c c c c c c c c c c c c c@{}} 
\hline\hline
Epoch & $\nu_\mathrm{ref}$ &\multicolumn{2}{c}{$43\,\mathrm{GHz}$} &  \multicolumn{2}{c}{$22\,\mathrm{GHz}$} &  \multicolumn{2}{c}{$15\,\mathrm{GHz}$} &  \multicolumn{2}{c}{$8\,\mathrm{GHz}$} &  \multicolumn{2}{c}{$5\,\mathrm{GHz}$} &  \multicolumn{2}{c}{$2\,\mathrm{GHz}$}\\
	   &	 & r	& 	PA 	&	r	&	PA	&	r	&	PA	&	 r	& 	PA 	&	r	&	PA	&	r	&	PA\\
$[$yyyy-mm-dd] &[GHz]& [mas] & [$^\circ$] &[mas] & [$^\circ$] & [mas] &[$^\circ$] & [mas] & [$^\circ$] &[mas] & [$^\circ$] & [mas] &[$^\circ$] \\
\hline
2005-05-19 &86	& 0.03$\pm$0.01	& 85 	& 0.10$\pm$0.02 	& 50 	& 0.20$\pm$0.05 	& $-$13 	&0.31$\pm$0.1 	& 16	& 0.57$\pm$0.14 		&$-$1 		& -- 		& --\\
2009-09-01 &43	& -- 	& --		& 0.02$\pm$0.02 	&$-$177	& 0.14$\pm$0.05 	& $-$74	& 0.24$\pm$0.08 	&$-$58	& 0.37$\pm$0.14 		& $-$65 	&-- 		& -- \\
2006-04-16 &43	& -- 	& --		& 0.02$\pm$0.02	&$-$118	& 0.06$\pm$0.05	&$-$60	& 0.17$\pm$0.08	&$-$46	& 0.45$\pm$0.14		& $-$66		& 1.1$\pm$0.3	&	$-$47 \\
2006-06-08 &43	&--		& --		&0.03$\pm$0.02	& 35		& 0.08$\pm$0.05	&12  	&0.18$\pm$0.20	&$-$31	&0.33$\pm$0.30		& $-$38		& 0.5$\pm$0.4	&	0 \\
2006-10-02 &43	&--		& --		&0.02$\pm$0.02	&109	& 0.06$\pm$0.05	&$-$18	&0.14$\pm$0.08	&$-$11	&0.28$\pm$0.14		& $-$49		& --		& -- \\
2006-12-04 &43	&--		& --		&0.04$\pm$0.03	&3		& 0.13$\pm$0.05	&$-$25	&0.29$\pm$0.08	&$-$42	&0.59$\pm$0.14		& $-$52		& --		& -- \\
2007-01-26 &43	&--		& --		&0.03$\pm$0.02	&81		& 0.11$\pm$0.05	&5		&0.23$\pm$0.08	&6		&0.40$\pm$0.14		& $-$41		& --		& -- \\
2007-04-26 &43	&--		& --		&0.01$\pm$0.02	&133	&0.05$\pm$0.05	&$-$50	&0.12$\pm$0.08	&$-$21	&0.25$\pm$0.14		& $-$24		& --		& -- \\	
\hline
\end{tabular} 
\end{table*}

\begin{table*}
\caption{Physical parameters from the core-shift analysis for three different models (see text for more details).}  
\label{coreb1n1}
\centering  
\begin{tabular}{@{}c c c c c c c c c c@{}} 
\hline\hline
Epoch		&$k_{r}$ 	&$\alpha_0$ 	&  $\Delta r_{(22-15)\mathrm{GHz}}$	&$\Omega_{r\nu}$	&	$B_1$	& $N_1^\star$		&$r_\mathrm{core,22GHz}$	&	$B_\mathrm{core,22GHz}$	 &$N_\mathrm{core,22GHz}^\star$\\
{[yyyy-mm-dd]}	&	[1]	   	& [1]		&	[mas] 						& [pc GHz]		&	[G]		& $[10^3\mathrm{cm}^{-3}]$	&		[pc]				&	[G]		& $[\mathrm{cm}^{-3}]$		\\
\hline
\multicolumn{10}{l}{model i): $k_r=1=\mathrm{const}$, $\alpha_0=-0.5=\mathrm{const}$, $\vartheta=\vartheta_\mathrm{crit}$ and $\beta_\mathrm{app}=13$} \\
2005-05-19 & 1.0 & $-$0.5 & 0.05 & 19.1 & 1.04 & 4.7 & 11.3 &0.09 &37 \\
2005-09-01 & 1.0 & $-$0.5 & 0.01 & 3.8 & 0.31 & 0.4 & 2.3 &0.14 &82\\
2006-04-14 & 1.0 & $-$0.5 & 0.02 & 7.7 & 0.52 & 1.2 & 4.5 &0.12 &58\\
2006-06-08 & 1.0 & $-$0.5 & 0.03 & 11.5 & 0.71 & 2.2 & 6.8 &0.10 &47\\
2006-10-02 & 1.0 & $-$0.5 & 0.02 & 7.7 & 0.52 & 1.2 & 4.5 &0.12 &58\\
2006-12-02 & 1.0 & $-$0.5 & 0.04 & 15.3 & 0.88 & 3.3 & 9.1 &0.10 &41\\
2007-01-26 & 1.0 & $-$0.5 & 0.03 & 11.5 & 0.71 & 2.2 & 6.8 &0.10 &47\\
2007-04-26 & 1.0 & $-$0.5 & 0.01 & 3.8 & 0.31 & 0.4 & 2.2 &0.14 &82\\
\multicolumn{10}{l}{model ii): $k_r=\mathrm{var.}$, $\alpha_0=-0.5=\mathrm{const}$, $\vartheta=2.6^\circ$, $\varphi=0.5^\circ$ and $\delta=17$} \\
2005-05-19 & 1.0 & $-$0.5 & 0.05 & 19.1 & 1.22 & 6.4 & 19.1 &0.06 &17\\
2005-09-01 & 1.0 & $-$0.5 & 0.01 & 3.8 & 0.36 & 0.6 & 3.8 &0.09 &39\\
2006-04-14 & 1.0 & $-$0.5 & 0.02 & 7.7 & 0.61 & 1.6 & 7.7 &0.08 &28\\
2006-06-08 & 1.0 & $-$0.5 & 0.03 & 11.5 & 0.83 & 3.0 & 11.5 &0.07 &23\\
2006-10-02 & 1.0 & $-$0.5 & 0.02 & 7.7 & 0.61 & 1.6 & 7.7 &0.08 &28\\
2006-12-02 & 1.0 & $-$0.5 & 0.04 & 15.3 & 1.03 & 4.6 & 15.3 &0.07 &20\\
2007-01-26 & 1.0 & $-$0.5 & 0.03 & 11.5 & 0.83 & 3.0 & 11.5 &0.07 &23\\
2007-04-26 & 1.0 & $-$0.5 & 0.01 & 3.8 & 0.36 & 0.6 & 3.8 &0.09 &39\\
\multicolumn{10}{l}{model iii): $k_r=\mathrm{var.}$, $\alpha_0=-0.5=\mathrm{const}$, $\vartheta=2.6^\circ$, $\varphi=0.5^\circ$ and $\delta=17$} \\
2005-05-19 & 0.9 & $-$0.5 & 0.05 & 23.7 & 0.91 & 3.5 & 16.8& 0.05& 12\\
2005-09-01 & 0.8 & $-$0.5 & 0.01 & 6.3 & 0.26 & 0.3 & 2.9& 0.09& 33\\
2006-04-14 & 0.8 & $-$0.5 & 0.02 & 12.6 & 0.39 & 0.6 & 5.8& 0.07& 19\\
2006-06-08 & 1.2 & $-$0.5 & 0.03 & 8.5 & 1.48 & 9.2 & 14.3& 0.10& 45\\
2006-10-02 & 0.7 & $-$0.5 & 0.02 & 18.4 & 0.31 & 0.4 & 4.9& 0.06& 17\\
2006-12-02 & 0.8 & $-$0.5 & 0.04 & 25.2 & 0.59 & 1.5 & 11.6& 0.05& 11\\
2007-01-26 & 0.8 & $-$0.5 & 0.03 & 18.9 & 0.50 & 1.0 & 8.7& 0.06& 14\\
2007-04-26 & 0.6 & $-$0.5 & 0.01 & 15.7 & 0.18 & 0.1 & 2.0& 0.09& 36\\
\hline
\multicolumn{10}{l}{$^\star$: $\gamma_\mathrm{min}=1$ and $\gamma_\mathrm{max}=1\cdot10^5$ were used for the calculation of the particle density} \\
\end{tabular}

\end{table*}

The variation of the $43\,\mathrm{GHz}-22\,\mathrm{GHz}$ core-shift vector around $0.01\,\mathrm{mas}$ reflects the uncertainties in the alignment method used (see panel A in Fig. \ref{coreshiftvec}). The $22\,\mathrm{GHz}-15\,\mathrm{GHz}$ and $8\,\mathrm{GHz}-5\,\mathrm{GHz}$ shift-vectors are mainly oriented in south-east direction (see panel B and D in Fig. \ref{coreshiftvec}) whereas the $15\,\mathrm{GHz}-8\,\mathrm{GHz}$ ones show some variation in the north-east to south-east direction. This variation in the direction of the core-shift vector for this frequency pair may reflect the slightly different orientation of the radio jet axis between $15\,\mathrm{GHz}$ and $8\,\mathrm{GHz}$. For the other frequency pairs, the difference in the jet axis is not as pronounced as for the $15\,\mathrm{GHz}-8\,\mathrm{GHz}$ pair (see contour plots in Fig.\ref{allcont}).

In order to derive the parameter $k_r$ (see Eq. ~\ref{coreshift}), we used the highest frequency during each observation as a reference frequency and computed the relative radial core-shift. Since CTA\,102 has a curved structure, clearly visible in Fig.~\ref{allcont}, we would have underestimated the core-shift by using only the one-dimensional radial distance between two VLBA cores. Therefore, we calculated the core-shift along a curved trajectory given by the $x$- and $y$-position of the image alignment. Table~\ref{coreshiftstab1} presents the calculated core-shift values relative to the highest frequency during each observation epoch.

Since we measure the core shift relative to a reference frequency, $\nu_{\mathrm{ref}}$, Eq. ~\ref{coreshift} can be re-written in the following form \citep{OSullivan:2009p1877}:

\begin{equation}
r_{\mathrm{core}}=A\left(\nu^{-1/k_{r}}-\nu_{\mathrm{ref}}^{-1/k_{r}}\right).
\label{cshift}
\end{equation}

We applied  Eq. ~\ref{cshift} to the core-shifts and the results are presented in Table~\ref{krfit}. The weighted average of the different $k_r$ and $A$ values obtained is $k_r=0.8\pm0.1$ and $A=3.4\pm1.6$. All values are compatible with $k_r\sim1$ but for the October 2006 and April 2007 observations.

\begin{table}[h!]
\caption{Parameters for the core-frequency dependency from the core-shift analysis}  
\label{krfit}
\centering  
\begin{tabular}{@{}c c c c@{}} 
\hline\hline
Epoch		&	$\nu_{\mathrm{ref}}$& $k_{r}$ 	&$A$ \\	
{[yyyy-mm-dd]}	& 	[GHz] 			 &	[1]	   	& [1]	  \\
\hline
2005-05-19 &86& 0.9$\pm$0.1		& 3.3$\pm$0.7\\
2005-09-01 &43& 0.8$\pm$0.4		& 3.7$\pm$4.4\\
2006-04-14 &43& 0.8$\pm$0.1		& 2.9$\pm$0.6 \\
2006-06-08 &43& 1.2$\pm$0.3		& 1.0$\pm$0.3 \\
2006-10-02 &43& 0.7$\pm$0.1		& 3.2$\pm$1.4\\
2006-12-02 &43& 0.8$\pm$0.1		& 5.2$\pm$1.5\\
2007-01-26 &43& 0.8$\pm$0.2 	& 3.4$\pm$1.6\\
2007-04-26 &43& 0.6$\pm$0.1 	& 3.9$\pm$2.0 \\
\hline
\end{tabular} 
\end{table} 

Assuming that the jet is in equipartition and has a conical geometry, which is a valid assumption for region C as seen from the value obtained for $k_r$, the magnetic field can be obtained using Eqs.~ \ref{b1coreshiftg} -- \ref{b1rcore} {(see Sect. \ref{jetridge} for further details on the jet width in region C)} . In general, the radial evolution of the magnetic field in the jet is given by $B\propto r^{-\epsilon b}$, where $\epsilon$ is the jet opening index ($R\propto r^{\epsilon}$) and $b$ parametrizes the evolution of the magnetic field ($B\propto R^{-b}$). For a jet in equipartition the particle density has to decrease as $N\propto R^{-2b}$ or, in terms of distance along the jet $N\propto r^{-2b\epsilon}$. Assuming a conical jet (i.e., $\epsilon=1$) and a toroidal magnetic field ($b=1$), an estimate can be derived for the magnetic field, $B_\mathrm{core}=B_1r_\mathrm{core}^{-1}$, and the relativistic particle density at the core, $N_\mathrm{core}=N_1r_\mathrm{core}^{-2}$. 
For the calculation of the magnetic field we used the shifts obtained between $22\,\mathrm{GHz}$ and $15\,\mathrm{GHz}$ because we obtained a small variation in the core-shift orientation for this frequency pair. We used an viewing angle of $\vartheta=2.6^\circ$, a half opening angle $\varphi=0.5^\circ$, a Doppler factor $\delta=17$ and an apparent speed, $\beta_\mathrm{app}=13$\,\textit{c}, derived from the components C1 and C2 (see Paper~II). These components were ejected during our observations and are the ones that best reflect the velocities in the core region.

Table~\ref{coreb1n1} presents values for the magnetic field and the particle density computed from the core-shift results using i) $k_r=1$, $\alpha_0=-0.5$, $\vartheta=\vartheta_\mathrm{crit}$ and $\beta_\mathrm{app}=13$ and ii) $k_r\neq1$, $\alpha_0=-0.5$, $\vartheta=2.6^\circ$, $\varphi=0.5^\circ$ and $\delta=17$, and iii) $k_r\neq1$ and $\alpha_0=-0.5$, $\vartheta=2.6^\circ$, $\varphi=0.5^\circ$ and $\delta=17$. For the calculation of the particle density we assumed a fixed ratio of $\gamma_\mathrm{max}/\gamma_\mathrm{min}=10^5$ and a lower electron Lorentz factor $\gamma_\mathrm{min}=1$.

\begin{figure}[h!]
\resizebox{\hsize}{!}{\includegraphics{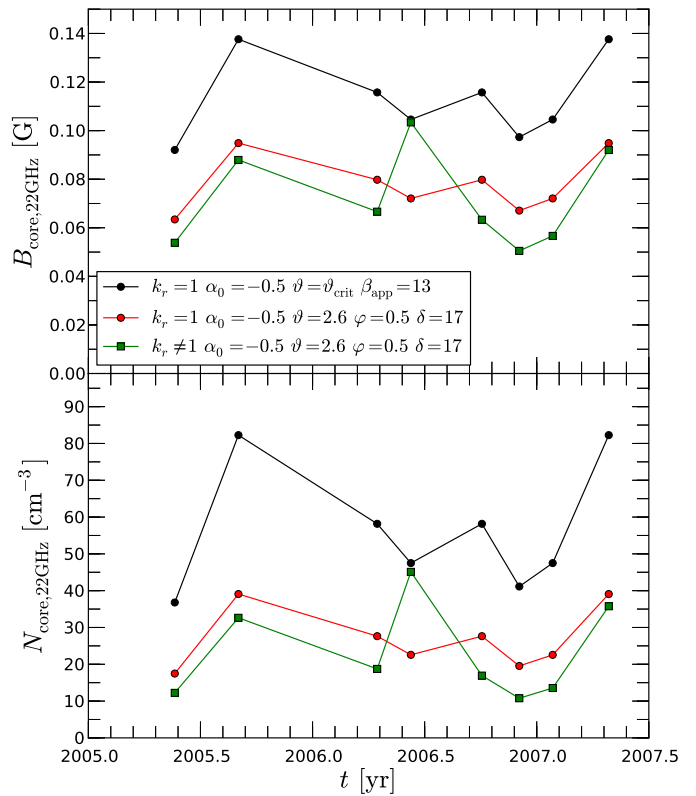}} 
\caption{Evolution of the core magnetic field, $B_\mathrm{core}$ (top), and the relativistic particle density at the core, $N_\mathrm{core}$ (bottom), as derived from the core shift analysis for three different models (For details see text).}
\label{csplot} 
\end{figure}

Figure~\ref{csplot} shows the evolution of $B_\mathrm{core,22GHz}$ and $N_\mathrm{core,22GHz}$ for the three models. For model i), we obtained the values around $B_\mathrm{core,22GHz}\sim0.1\,\mathrm{G}$ and $N_\mathrm{core,22GHz}=60\,\mathrm{cm^{-3}}$ (black points in Fig. \ref{csplot}). Using a more detailed jet parameters, while keeping $k_r=1$ and $\alpha_0=-0.5$ leads to smaller values of $B_\mathrm{core,22GHz}\sim0.08\,\mathrm{G}$ and $N_\mathrm{core,22GHz}\sim30\mathrm{cm^{-3}}$ as compared to model i. On average, slightly smaller values are computed for $B_\mathrm{core,22GHz}$ and $N_\mathrm{core,22GHz}$ if we use the obtained $k_r$ values. The average $B_\mathrm{core,22GHz}=0.07\,\mathrm{G}$ and $N_\mathrm{core,22GHz}=23\,\mathrm{cm^{-3}}$.

\section{Transversal jet structure and jet ridge-line}
\label{jetridge}
A conical jet geometry was assumed in the modelling presented in the previous section. We studied the transversal jet structure of CTA~102 to test this assumption. For this, we derived the jet ridge line, i.e., the line connecting the local flux density maxima along the jet, and fitted a Gaussian to the flux density profiles perpendicular to the jet ridge-line \citep[see, e.g.,][]{Pushkarev:2009p5426}. 
Since we were interested in the average jet width and ridge line, we used stacked VLBA images, which include all the observations at a given frequency, here from 2005 until 2007. The advantage of using stacked images is that the values obtained for the width and the ridge line are less affected by individual calibration uncertainties. For the convolution of the individual images {we} used the average beam size for each frequency (see Table~\ref{averpara}).

\begin{figure}[h!]
\resizebox{\hsize}{!}{\includegraphics{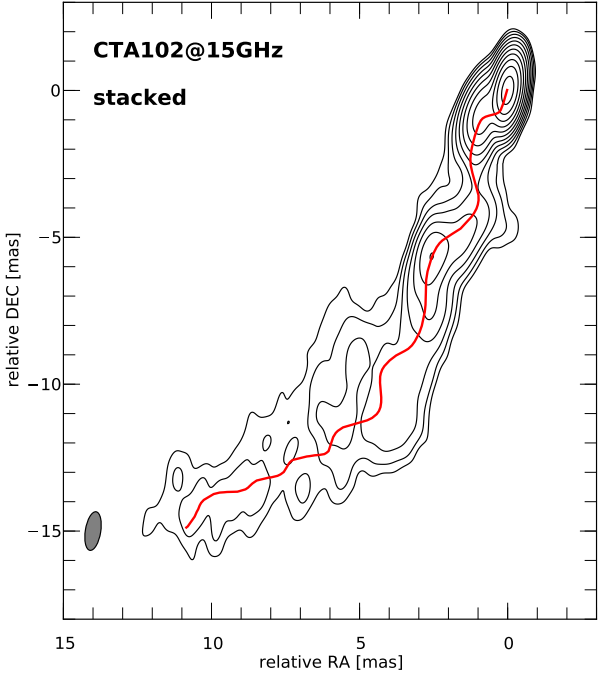}} 
\caption{Stacked $15\,\mathrm{GHz}$ contour map of CTA\,102 obtained from observations between 2005 and 2007 convolved with a common beam size of $1.33\,\mathrm{mas}\times0.52\,\mathrm{mas}$ and a P. A. of $-9^\circ$ with over-plotted jet ridge line. The lowest contour level is drawn at 5$\times$ the average off-source rms ($4\,\mathrm{mJy}$) and the contours increase with steps of 2.} 
\label{15stacked} 
\end{figure}

\begin{figure}[h!]
\resizebox{\hsize}{!}{\includegraphics{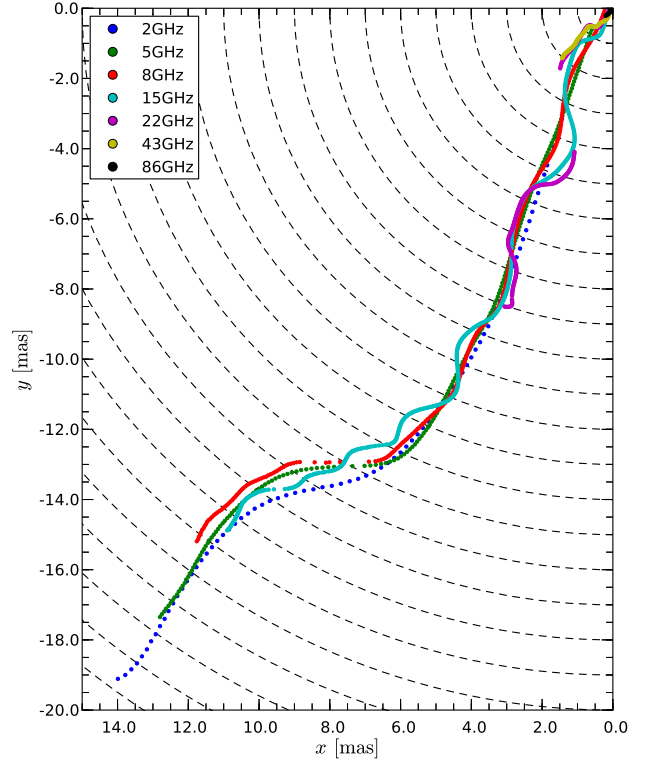}} 
\caption{Jet ridge line for CTA\,102 obtained from stacked VLBA images at different frequencies after re-alignment. The dashed lines correspond to the radial distances and are plotted every $1\,\mathrm{mas}$.} 
\label{ridgeline} 
\end{figure}

Figure~\ref{15stacked} shows a $15\,\mathrm{GHz}$ stacked VLBI image of CTA\,102 over-plotted with the obtained jet ridge-line. We corrected the starting point of the jet ridge line at each frequency for the opacity shift using the average image shift obtained as explained in Sect.~\ref{coreshiftres}. The results are presented in Fig.~\ref{ridgeline}.

After applying the opacity correction to the ridge lines derived at different frequencies, we obtained a good spatial agreement between them (see Fig.~\ref{ridgeline}). The jet ridge line at low frequencies $\nu<8\,\mathrm{GHz}$ shows an oscillating pattern with an observed wavelength $\lambda_\mathrm{obs,1}\sim20\,\mathrm{mas}$. Within the first $13\,\mathrm{mas}$ from the core, the ridge line is oriented along an observed angle of $-65^\circ$. At higher frequencies, a second oscillating pattern developing on top of the first one is observed, as reported in \citet{Perucho:2012p5917} for the case of the jet in the quasar S5~0836+710. This second pattern is best visible at $15\,\mathrm{GHz}$ and $22\,\mathrm{GHz}$ and has an observed wavelength, $\lambda_\mathrm{obs,2}\sim5\,\mathrm{mas}$. There are also indications for an additional pattern with $\lambda_\mathrm{obs,3}<1\,\mathrm{mas}$ visible at the highest frequencies, $43\,\mathrm{GHz}$ and $86\,\mathrm{GHz}$. A full analysis on the jet ridge lines as described in \citet{Perucho:2012p5917} could lead to additional parameters of the three dimensional helical patterns, which are potentially connected to fluid instabilities, but this is beyond the scope of this paper.

\begin{figure}[h!]
\resizebox{\hsize}{!}{\includegraphics{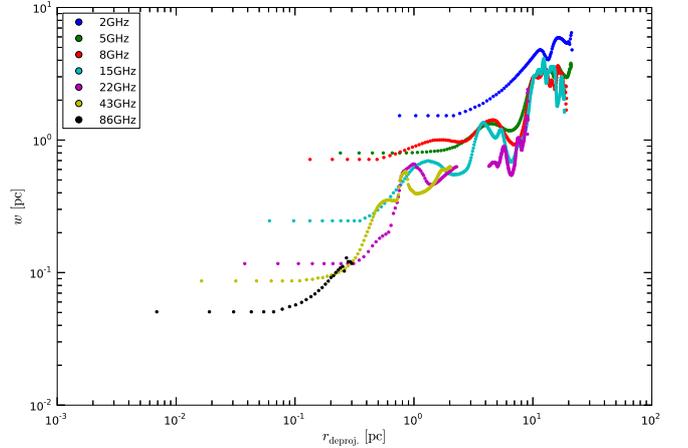}} 
\caption{Jet width of CTA\,102 obtained from stacked VLBA images at different frequencies after re-alignment} 
\label{jetwidth} 
\end{figure}

Figure \ref{jetwidth} shows the variation of the de-convolved jet width, $w=\sqrt{d^2-b_\phi^2}$, with distance, where $d$ is the FWHM of the fitted Gaussian and $b_\phi$ is the size of the beam transversal to the jet ridge line along the jet. As we did for the jet ridge line, we corrected the starting position using the average core-shift value, $r=3.4\cdot\nu^{-1.25}$. The constant horizontal points for each frequency reflect the resolution limit where we can not resolve the transversal structure of the jet.
The position of the first observed rise of the jet width increases with decreasing frequency and is in good agreement with the position found at neighboring frequencies, which supports the validity of the measurement. Tracing this point back to the highest frequency available in our data set ($86\,\mathrm{GHz}$) leads to a distance of $0.1\,\mathrm{mas}$ from the jet nozzle and to a jet width of $0.05\,\mathrm{mas}$, which corresponds to one half of the convolving beam size transversal to the jet ridge line. The jet width is varying, with several indications for collimation and recollimation along the jet. This behavior is best visible at $15\,\mathrm{GHz}$. For this frequency our analysis reveals three local maxima of the jet width at a core distance along the ridge line of $r=1.2\,\mathrm{mas}$, $r=5.4\,\mathrm{mas}$ and $r=13.3\,\mathrm{mas}$. At the higher frequencies $\nu=22\,\mathrm{GHz}$ and $\nu=43\,\mathrm{GHz}$ we also find indications of recollimation of the jet at the same positions.

{We approximated the jet width with a power law in each of the four different regions C, D, B, and A, in which we divided the jet.} The jet width can be parametrized as a power law, $R\propto r^{\epsilon}$ where $\epsilon$ is the jet opening index. The jet opening index obtained for each of these regions is presented in Table \ref{opening}. A single power law fit applied to the jet width gives $\epsilon=0.6\pm0.1$ for the four regions. 

\begin{table}[h!]
\caption{Estimates for the exponent $\epsilon$, defining the jet geometry, $R\propto r^{\epsilon}$.}  
\label{opening}
\centering  
\begin{tabular}{c c c c c}
\hline\hline
region &C	&	D	&	B	&	A\\
\hline
$\epsilon$& $0.8\pm0.1^a$	&	$0.8\pm0.1^b$	&	$-1.0\pm0.1$	&	$1.3\pm0.2$ \\
\hline
\multicolumn{5}{l}{$^a$ fitted together with region D, otherwise $\epsilon=1.2$}\\
\multicolumn{5}{l}{$^b$ fitted together with region C, otherwise $\epsilon=0.4$}\\
\end{tabular}
\end{table} 

The results of the jet width analysis show that CTA\,102 embeds a pinching jet width, i.e., it consists of regions where the jet cross-section is opening or closing. At $r<1\,\mathrm{mas}$ the jet shows a nearly conical geometry $\epsilon=0.8\pm0.1$. Farther downstream $(r>4\,\mathrm{mas})$ the jet geometry differs significantly from conical.

\section{Spectral analysis}
\label{resspecana}
Here we present the result of the spectral analysis for the different jet regions as introduced in Paper~II (see the previous section). We show the spectral parameters along the jet ridge-line and averaged values transversally to it, which reflect possible transversal gradients in the spectral parameters across the jet. The 2D distribution of the spectral parameters is presented in Appendix C

  Since the spectral turnover is outside our frequency range for regions B and A, we fitted a power law $S_\nu\propto\nu^{\alpha}$ to the measured flux densities. The uncertainties on the spectral parameters were calculated using a Monte-Carlo simulation taking into account the uncertainties of the image alignment and the flux densities of the individual pixels. Since the uv-coverage is changing with frequency we only used those uv-radii covered throughout the selected frequency range. For more details see Appendix B and Fromm (in preparation)  for the detailed treatment of the capabilities and the limitations of the pixel based spectral analysis applied {to} multi-frequency VLBI images.

\subsection{Region C $(r<1\,\mathrm{mas})$}
\label{cregion}
The core region is characterized by an inverted spectrum (increasing flux density with increasing frequency) throughout our data set, except for the May 2005 observations. Therefore, we applied a power-law fit to obtain the 2D distribution of the spectral index $\alpha$.  The region could be best studied using the high-frequency VLBI maps ($\nu>8\,\mathrm{GHz}$), a common beam size of $0.95\times0.33\,\mathrm{mas}$ with a P.A. of $-13^\circ$, which corresponds to the average beam of the $22\,\mathrm{GHz}$ observations, and a pixel size of $0.03\,\mathrm{mas}$. These image settings allowed us to extract the spectral index without generating image artifacts in the low-frequency maps or losing structural information in the high frequency maps.

\begin{figure}[b!]
\resizebox{\hsize}{!}{\includegraphics{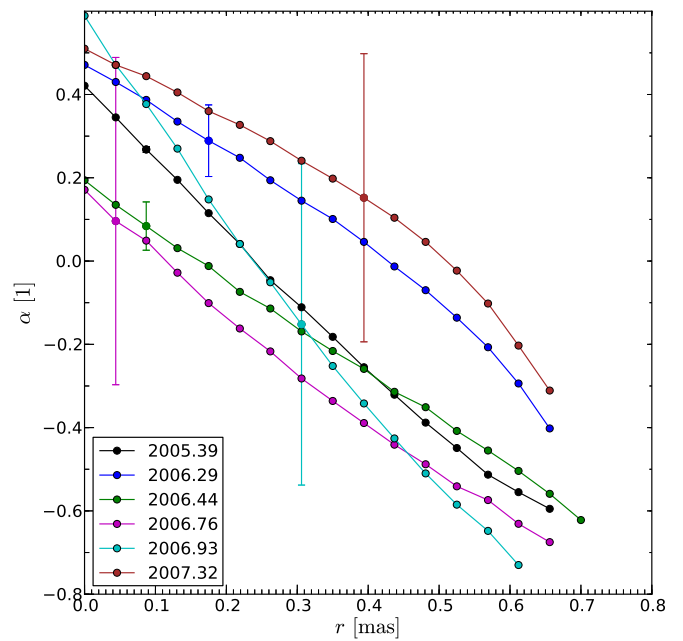}} 
\caption{Evolution of the spectral index along the jet ridge line, $\alpha$  $(S_\nu\propto \nu^{\alpha})$ for region C $(r<1\,\mathrm{mas})$, where we used a beam size of $0.95\times0.33\,\mathrm{mas}$ with a P.A. of $-13^\circ$. For reasons of readability only one error bar per epoch is shown. } 
\label{Calpha1d} 
\end{figure}

The spectral index along the jet axis is presented in Fig. \ref{Calpha1d}, where we used the $43\,\mathrm{GHz}$ core of each observation as the origin of coordinates. The core region of the September 2005 and January 2007 observations of CTA\,102 could not be explained with a single power law due to the rise of a flare i.e., the increase of the high frequency flux densities, here the $43\,\mathrm{GHz}$ flux densities (see Fig. \ref{lc}). Therefore, we excluded these two epochs from the analysis of the core region.  Within $r<0.3\,\mathrm{mas}$ from the core, the spectral index is positive $(\alpha>0)$, which corresponds to an inverted spectrum and farther downstream $\alpha$ decreases with distance. There is a variation of $\alpha$ with distance and time. The latter is best visible between May 2005 and October 2006, when the slope of the spectral index along the jet changes from $-$1.6 (May 2005) to $-$1.1 (June 2006) and back to $-$1.3 (October 2006). This could be an indication for the injection of relativistic particles and the connected energy losses during the propagation of a relativistic shock wave. 

\begin{figure}[b!]
\resizebox{\hsize}{!}{\includegraphics{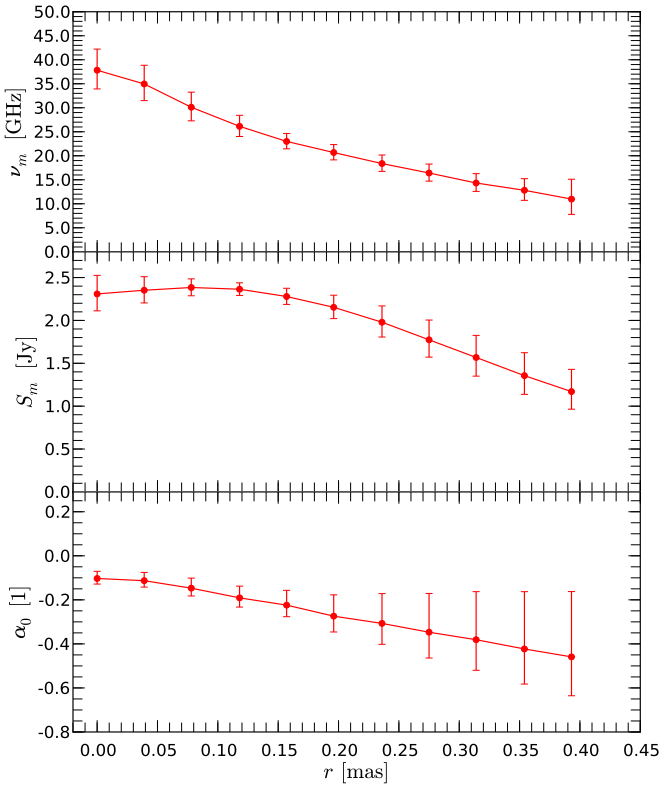}} 
\caption{Spatial evolution of the turnover values along the jet ridge line for the May 2005 observation (2005.39).  We used beam size of $0.95\times0.33\,\mathrm{mas}$ with a P.A. of $-13^\circ$ and pixel size of $0.03\,\mathrm{mas}$. The error estimates were derived from a Monte Carlo simulation (see text for more details). Top panel: Turnover frequency, $\nu_{m}$; middle panel: turnover flux density, $S_{m}$; bottom panel: optically thin spectral index, $\alpha_0$}
\label{1dvmsm} 
\end{figure}

Since our frequency coverage for the May 2005 observations spans from $5\,\mathrm{GHz}$ to $86\,\mathrm{GHz}$, estimates on the values at the spectral turnover could be derived. For this epoch, we only present axial turnover values, since we could hardly resolve the transversal structure of the source with the selected beam size. Figure \ref{1dvmsm} shows the spatial evolution of the turnover frequency, $\nu_m$, the turnover flux density, $S_m$, and the optically thin spectral index, $\alpha_0$ along the jet ridge line. The optically thin spectral index decreases from $-0.1$ at  $r=0.15\,\mathrm{mas}$ to $-0.45$ at $r=0.5\,\mathrm{mas}$. At the same distance the turnover frequency changes from $\nu_m=30\,\mathrm{GHz}$ to $\nu_m=15\,\mathrm{GHz}$ at $r=0.5\,\mathrm{mas}$. A drop in the turnover flux density, $S_m$, occurs at $r=0.2\,\mathrm{mas}$ from $S_m=2.3\,\mathrm{Jy}$ to $S_m=1.2\,\mathrm{Jy}$.

\begin{figure}[b!]
\resizebox{\hsize}{!}{\includegraphics{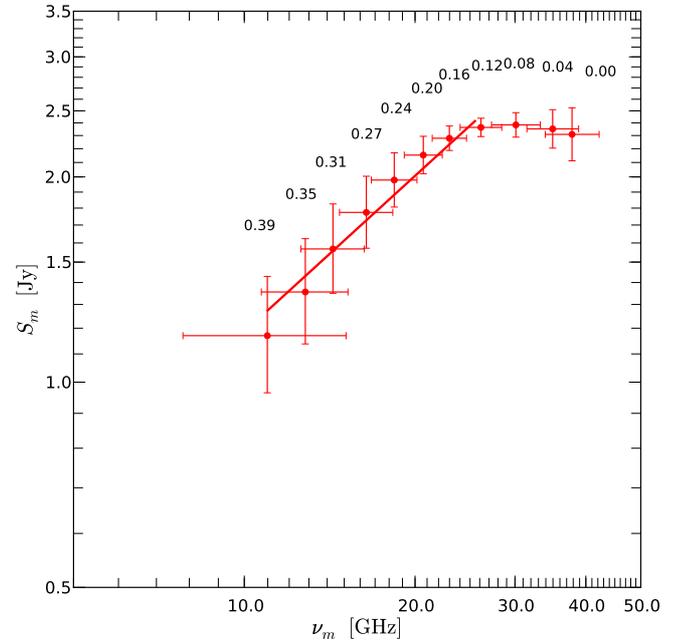}} 
\caption{Turnover frequency--turnover flux density plane for region C for the May 2005 observations. The number correspond to the distance from the core in $\mathrm{mas}$ and the solid red line to a power law fit $(S_m\propto \nu_m^{f})$.}
\label{Cvmsm} 
\end{figure}

In the turnover frequency--turnover flux density plane, the spatial evolution shows a constant turnover flux density $S_m=2.3\,\mathrm{Jy}$ for $r<0.15\,\mathrm{mas}$, while the turnover frequency decreases (see fig. \ref{Cvmsm}). For $r>0.15\,\mathrm{mas}$ both, the turnover frequency and the turnover flux density decline. This decrease in the turnover position is in agreement with the expected behaviour, when the (adiabatic) expansion losses are the dominant energy loss mechanism \citep{Marscher:1985p50}. The constant turnover flux density measured across the region at $r<0.1\,\mathrm{mas}$ could be caused by the smearing of the flux density during the convolution. 

\subsection{Region D $(1\,\mathrm{mas}<r<4\,\mathrm{mas}$)}
\label{regionD}
For the spectral analysis of region D we used a beam size of $1.33\times0.52\,\mathrm{mas}$ and a pixel size of $0.04\,\mathrm{mas}$ and included the $5\,\mathrm{GHz}$ and $8\,\mathrm{GHz}$ VLBI maps into the analysis. Due to this extended frequency range we could derive the turnover values by applying Eq. ~\ref{snuapprox}.  We rejected turnover frequencies below $5\,\mathrm{GHz}$ and optically thin spectral indices, $\alpha_0>-0.2$.
 
\begin{figure}[h!]
\resizebox{\hsize}{!}{\includegraphics{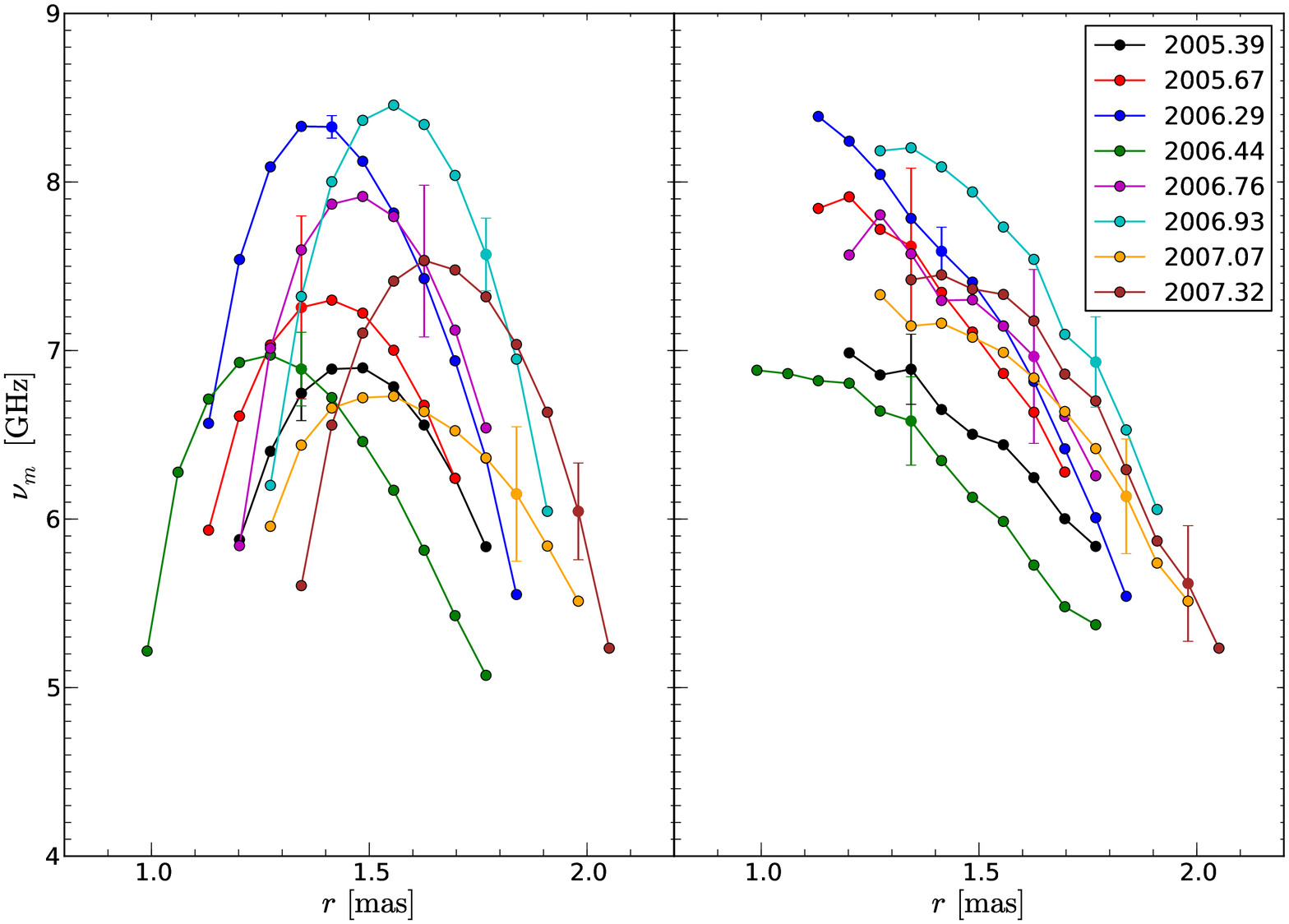}} 
\caption{Evolution of the turnover frequency, $\nu_m$,  for region D $(1\,\mathrm{mas}<r<4\,\mathrm{mas})$ along the jet axis, where we used a beam size of $1.33\times0.52\,\mathrm{mas}$ with a P.A. of $-9^\circ$. For reasons of readability only one error bar per epoch is shown. Left: values along the jet ridge line. Right: average values transversal to the jet ridge line.} 
\label{Dvm1d} 
\end{figure}

The values obtained for the turnover frequency in this region are between $5.0\,\mathrm{GHz}$ and $10.0\,\mathrm{GHz}$. The axial turnover frequencies exhibit a curved distribution, where the value of the peak and its position vary in time (see left panel in Fig.~\ref{Dvm1d}). This curvature could be an artifact of the convolution. However, the transversally averaged values show a monotonic decrease in the turnover frequency (see the right panel in Fig.~\ref{Dvm1d}).

\begin{figure}[h!]
\resizebox{\hsize}{!}{\includegraphics{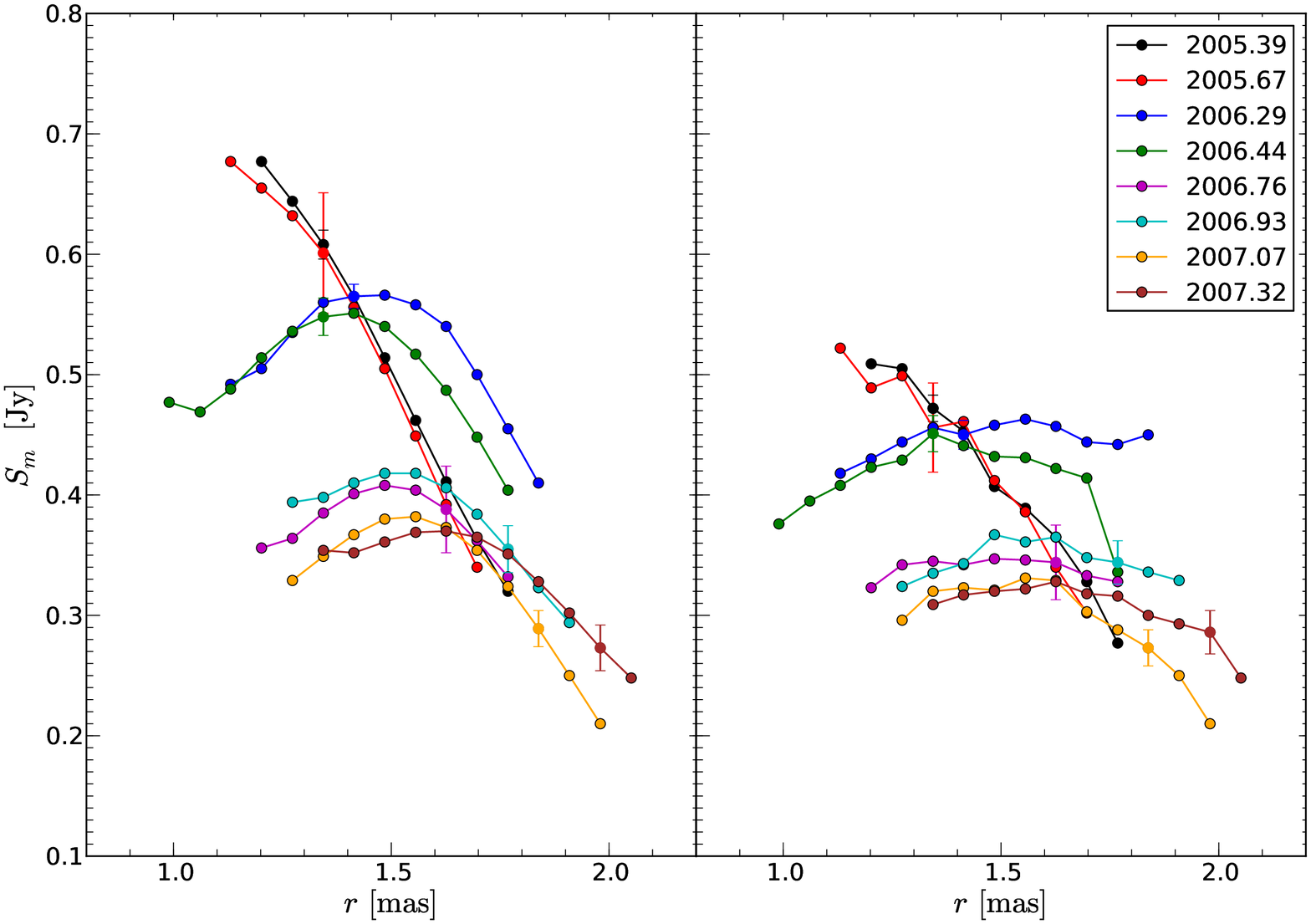}} 
\caption{Same as Fig. \ref{Dvm1d} for the turnover flux density, $S_m$.} 
\label{Dsm1d} 
\end{figure}

\begin{figure}[h!]
\resizebox{\hsize}{!}{\includegraphics{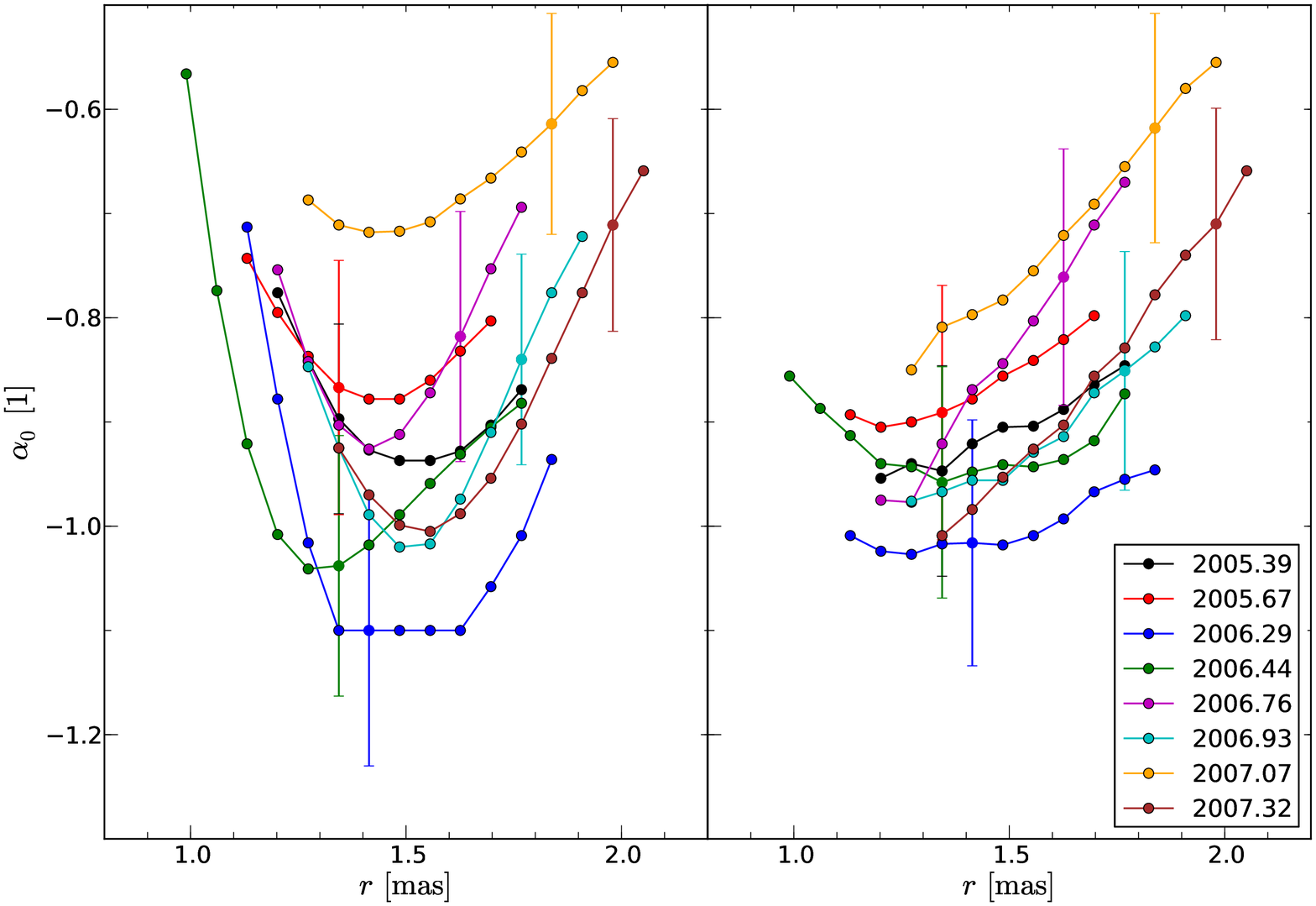}} 
\caption{Same as Fig. \ref{Dvm1d} for the optically thin spectral index, $\alpha_0$.} 
\label{Da01d} 
\end{figure}

The evolution of the turnover flux density, $S_m$, shows a clear decreasing trend with time and a slight shift in the position of the flux density maxima with distance from the core (see Fig. \ref{Dsm1d}). The optically thin spectral index shows an increasing trend with distance from $\alpha_0\sim-1.0$ to $\alpha_0\sim-0.5$, but the large uncertainties in the optically thin spectral index do not allow us to detect a significant temporal variation (see Fig. \ref{Da01d}).

\begin{figure}[h!]
\resizebox{\hsize}{!}{\includegraphics{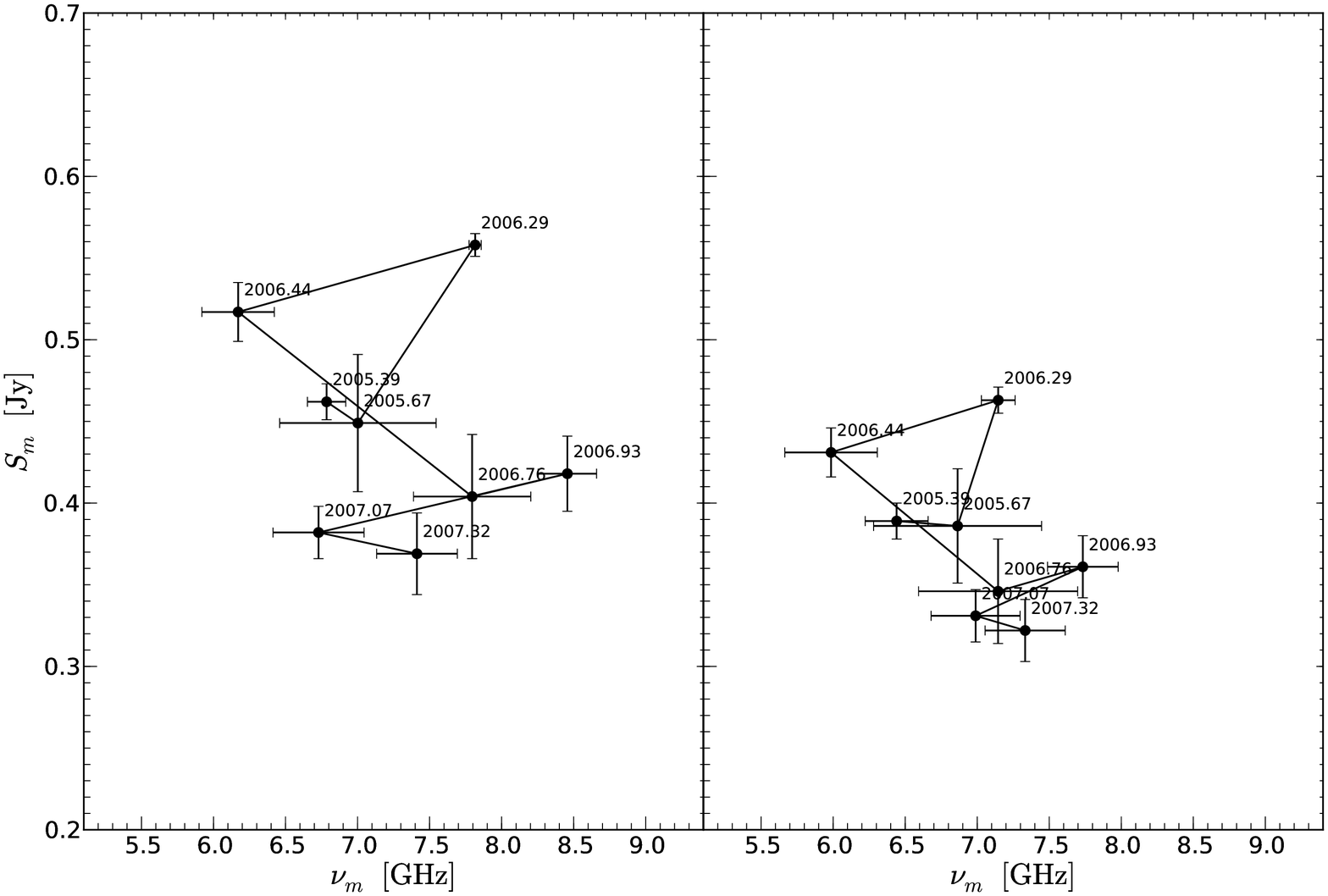}} 
\caption{Turnover frequency--turnover flux density plane for region D at a fixed position of $r=1.5\,\mathrm{mas}$, where the numbers represent the temporal evolution. Left: axial values Right: transversal averaged values (for more details see text).} 
\label{Dvmsm} 
\end{figure}

We can see significant changes in the turnover frequency and turnover flux density with time and position in region D. Figure \ref{Dvmsm} shows the variation of the turnover values in the turnover frequency - turnover flux density $(\nu_m-S_m)$ plane for a fixed position at $r=1.5\,\mathrm{mas}$. The kinematic analysis of CTA\,102 revealed a traveling component, labeled as D2, within this region and a possible standing component at $r=1.5\,\mathrm{mas}$ (see region D in Paper~II). The evolution in the $\nu_m-S_m$ plane shows an increase in both $\nu_m$ and $S_m$ between 2005.3 and 2006.3 followed by an increase in 2006.44. For $t>2006.44$, the turnover flux density is smaller than in 2005.3 and the turnover frequency varies slightly.

The kinematical results (Paper~II) and the present spectral study point towards a situation in which component D2 crosses region D during the time span of our observations, possibly interacting with a standing emission region at $r\simeq 1.5$~mas. Following the classical shock-in-jet model, we would expect a continuous decrease in turnover flux density and turnover frequency, since at this position, $r=1.5\,\mathrm{mas}$ (de-projected $86\,\mathrm{pc}$), the main energy loss mechanism should be adiabatic expansion losses. However, we detect an increase in $S_m$ at this position, which could be due to an re-acceleration of relativistic electrons or an increase in the particle density (see also Paper~I). From the spectral analysis of region C we know that the turnover flux density and turnover frequency are monotonically decreasing (see Fig.~\ref{Cvmsm}). Therefore, the increase in the $\nu_m-S_m$ plane requires a region of locally increased density and/or magnetic field. Such a local increase in the physical parameters could be created by a recollimation shock. The results of the kinematic analysis in region D (Paper~II) and the study of the transversal jet width (Sect.~\ref{jetridge}) support this assumption.

\subsection{Region B $(4\,\mathrm{mas}<r<10\,\mathrm{mas}$)}
\begin{figure}[h!]
\resizebox{\hsize}{!}{\includegraphics{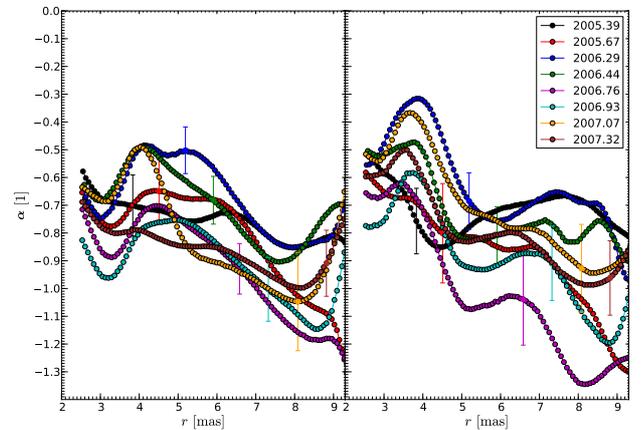}} 
\caption{Evolution of the spectral index, $\alpha$  $(S_\nu\propto \nu^{\alpha})$ for region B $(4\,\mathrm{mas}<r<10\,\mathrm{mas})$  along the jet ridge line. For reasons of readability only one error bar per epoch is shown. Left: values along the jet ridge line. Right: average values transversal to the jet ridge line.} 
\label{Ba01d} 
\end{figure}

This region is characterized by three nearly stationary components (see Paper~II) and by a constant {jet} width (see Sect.~\ref{jetridge}). Figure~\ref{Ba01d} shows the value of the spectral index along the jet. This region could be best studied by using a frequency range from $5\,\mathrm{GHz}$ to $15\,\mathrm{GHz}$, a beam size of  $2.32\times0.97\,\mathrm{mas}$, a P.A of $-7^\circ$, and a pixel size of $0.10\,\mathrm{mas}$. The spectral index at $r\sim2.0\,\mathrm{mas}$ lies between $-0.8<\alpha<-0.6$, which is consistent with the continuation of the observed evolution in region D (see Fig.~\ref{Da01d}). For $r>2.0\,\mathrm{mas}$ an increase in the spectral index can be observed, followed by a plateau of nearly constant value at $r=4.0\,\mathrm{mas}$ (left panel in Fig.~\ref{Ba01d}). For larger distances, the spectral index decreases with distance until $r\sim7.0\,\mathrm{mas}$. Farther downstream there is an additional rise of the spectral index, although less pronounced than the one at $r=4\,\mathrm{mas}$.

\subsection{Region A $(8\,\mathrm{mas}<r<14\,\mathrm{mas}$)}
\begin{figure}[h!]
\resizebox{\hsize}{!}{\includegraphics{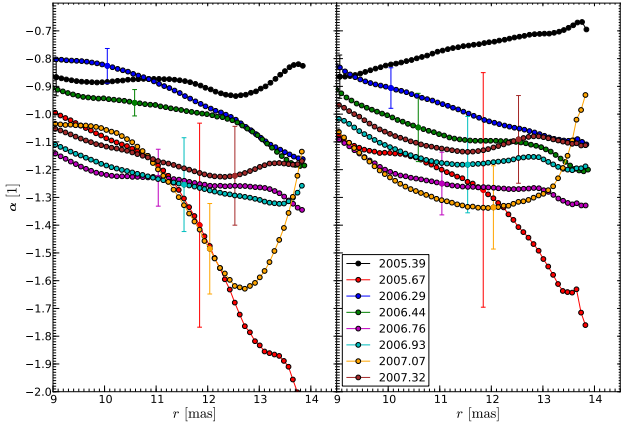}} 
\caption{Evolution of the spectral index, $\alpha$  $(S_\nu\propto \nu^{\alpha})$ for region A $(8\,\mathrm{mas}<r<14\,\mathrm{mas})$  along the jet ridge line. For reasons of readability only one error bar per epoch is shown. Left: values along the jet ridge line. Right: average values transversal to the jet ridge line.} 
\label{A2a1d} 
\end{figure}

For region A we used the VLBI images at $5\,\mathrm{GHz}$,  $8\,\mathrm{GHz}$, and  $15\,\mathrm{GHz}$ and a beam of $3.65\times1.52\,\mathrm{mas}$, a P. A. of $-8^\circ$, and a pixel size of $0.15\,\mathrm{mas}$. In contrast to region B, the spectral index decreases from $\alpha\sim-1.0$ to $\alpha\sim-1.2$ without major variations, best visible in the transversally averaged values (see Fig.~\ref{A2a1d}). The increase of the averaged values for the May 2005 observations (2005.39) is due to a region of high spectral indices at the edges of the jet (see Fig.~\ref{A1alpha2d} in Appendix B). The sparse sampling of the  uv coverage, especially on short ranges leads to strong variation on the outermost structure. Additional to the uv-sampling, antenna or weather problems (e.g., rain), which mainly affects the low frequencies and especially the outer structure of the source, {lead to strong flux variations and therefore variations in the spectral index}. {Therefore, we restricted ourself} to distances $r<14\,\mathrm{mas}$ in the spectral analysis of the outermost structure.

\section{Spatial and temporal evolution of physical parameters}
\label{physpara}

\subsection{Position of the jet nozzle}
{For the core region C we could only derive the turnover frequency, $\nu_m$, and the turnover flux density, $S_m$ for the May 2005 observations due to extended frequency range up to $86\,\mathrm{GHz}$.}
The spectral turnover values allow us to compute the magnetic field and the normalization coefficient of the relativistic electron distribution using Eqs.~\ref{bfield} and \ref{knorm}, respectively. Furthermore, we could provide some estimates on the number density of the relativistic particles and on the magnetization, $\sigma$, of the plasma using the first order model presented in Sect. \ref{gammaevo} and the jet width (see Sect. \ref{jetridge}).

Notice that in this section we use the opacity corrected core position, i.e., radial origin corresponds to the location where $\tau=1$ for $\nu\rightarrow\infty$ using the results of the core-shift analysis (see Table \ref{krfit} in Sect. \ref{coreshiftres}). In Table \ref{nozzle} we present the offset corrections for all epochs.
 
\begin{table}[h!]
\caption{Results of the core-shift analysis.}  
\label{nozzle}
\centering  
\begin{tabular}{c l} 
\hline\hline
Epoch	&	$\Delta r_\mathrm{nozzle}$ $[\mathrm{mas}]$\\
\hline
2005-05-19 & 	0.02$\pm$ 0.01\\
2005-09-01 &	0.03$\pm$0.08\\
2006-04-14 & 	0.03$\pm$0.02  \\
2006-06-08 & 	0.04$\pm$0.03  \\
2006-10-02 & 	0.01$\pm$0.01\\
2006-12-02 & 	0.05$\pm$0.03\\
2007-01-26 & 	0.04$\pm$0.04\\
2007-04-26 &   0.01$\pm$0.01\\
\hline
\end{tabular}
\end{table} 

\subsection{Core region in May 2005}
For the calculation of the magnetic field and the normalization coefficient of the relativistic electron distribution  we need to know the evolution of the jet width and an estimate for the Doppler factor along the jet. Since the turnover values for region C were obtained using the average beam size of the $22\,\mathrm{GHz}$ VLBI maps (see Sect.~\ref{cregion}), we used the jet width calculated for this frequency (see Sect.~\ref{jetridge}).
Based on the flux density evolution of the component labeled as C2 we calculated a variability Doppler factor $\delta_\mathrm{var}=17\pm3$. A parameter space study based on the gradients in the brightness temperature, $T_b$, revealed that the Doppler factor increases within region C (see Paper~II). Accordingly with the numbers {obtained} in Paper~II, we adopted an exponent of $d=-0.6$ ($\delta=R^{-d}$) and an upper value of $\delta=17$ for the following analysis. 

\begin{figure}[h!]
\resizebox{\hsize}{!}{\includegraphics{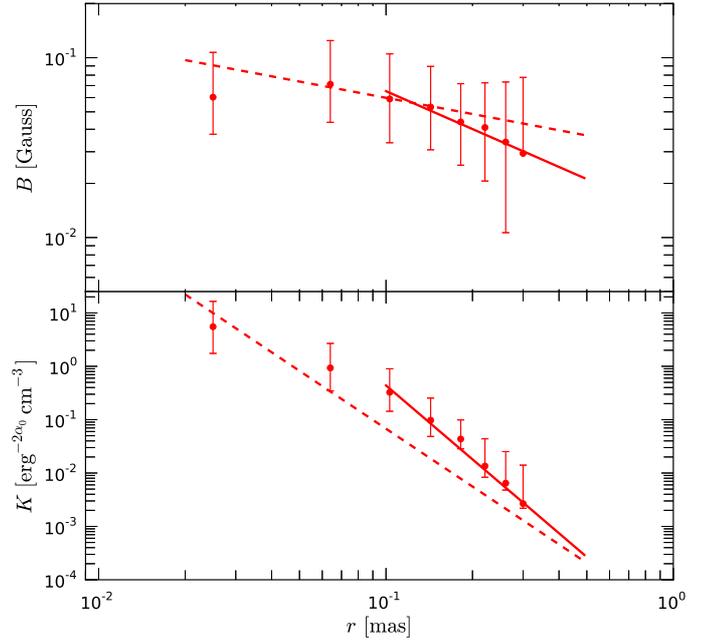}} 
\caption{Evolution of the magnetic field $B$ (top panel) and the normalization coefficient, $K$ (bottom panel) for the May 2005 observation. The solid and dashed red lines correspond to a power law fits (see text for details).} 
\label{BKevo} 
\end{figure}

The magnetic field computed from the aforementioned parameters is presented in the upper panel of Fig.~\ref{BKevo}. The magnetic field intensity decreases from $60\,\mathrm{mG}$ at $r=0.1\,\mathrm{mas}$ to $30\,\mathrm{mG}$ at $r=0.4\,\mathrm{mas}$. The results of an approximation of the evolution with a power law $(B\propto r^{b^\prime})$ gives an exponent $b^\prime=-0.3\pm0.1$ for the whole range ($0.02\,\mathrm{mas}<r<0.4\,\mathrm{mas}$, dashed red line in the upper panel of Fig.~\ref{BKevo}). The flat spectral index for $r<0.1\,\mathrm{mas}$ could be due to convolution effects. Therefore, we additionally fitted the evolution of the magnetic field intensity for $r>0.1\,\mathrm{mas}$ and obtained an exponent of $b^\prime=0.7\pm0.1$ (solid red line in the upper panel of Fig. \ref{BKevo}).
Similar values for the magnetic field, $B$, were obtained using the core-shift {(see magnetic field values in Sect.~\ref{coreshiftres})}. Since both methods are independent, the result of the magnetic field can be regarded as meaningful estimate.\\

The normalization coefficient of the relativistic electron distribution spans from $5.0\,\mathrm{erg^{-2\alpha_0}cm^{-3}}$ to $10^{-3}\,\mathrm{erg^{-2\alpha_0}cm^{-3}}$. We used a power law $(K\propto r^{k^\prime})$ to fit the evolution of $K$ with distance along the jet and obtained $k^\prime=-3.6\pm0.3$ for the entire region (dashed red line in the lower panel of Fig. \ref{BKevo}). As in the case of the magnetic field, the values for $K$ steepen for $r>0.1\,\mathrm{mas}$. A power law fit for this region results in $k^\prime=-4.6\pm0.2$ (solid red line in the lower panel of Fig. \ref{BKevo}). Notice that the values of the normalization coefficient can not be directly compared due to the varying spectral index along the jet (see bottom panel in Fig.~\ref{1dvmsm} and Eq. \ref{knorm}). 

The slopes obtained from the variation of the magnetic field, $B$, and the normalization coefficient, $K$, (see Fig.~\ref{BKevo}) correspond to the evolution along the jet and include the influence of the jet geometry (e.g., the magnetic field along the jet is given by $B\propto r^{\epsilon b}$). We thus have to take into account the evolution of the jet radius in order to derive the geometry of the magnetic field. The evolution of the transversal jet size with distance in region C can be fitted with an exponent $\epsilon=0.8\pm0.1$ ($R\propto r^{\epsilon}$). This results in $b=-0.4\pm0.1$ for the overall region and to $b=-0.9\pm0.2$ for $r>0.1\,\mathrm{mas}$. For the evolution of the normalization coefficient K we computed an exponent of $k=-4.5\pm0.7$ for $0.02\,\mathrm{mas}<r<0.45\,\mathrm{mas}$ and $k=-5.7\pm0.8$ for $r>0.1\,\mathrm{mas}$.

Assuming adiabatic expansion losses as the main energy loss mechanism, and following \citet{Marscher:1985p50, Lobanov:1999p2299}, the slope in the $\nu_m-S_m$ plane is given by:
\begin{equation}
\eta_\mathrm{adi}=-\frac{2s+13-5k-b(2s+3)-d(3s+7)}{2(k-1)+(b+d)(s+2)},
\label{etaadi}
\end{equation}

\noindent with $b$ the exponent for the evolution of the magnetic field intensity $(B\propto R^{-b})$, $k$ the exponent for the evolution of the normalization coefficient of the relativistic electron distribution $(K\propto R^{-k})$, $d$ the exponent for the evolution of the Doppler factor $(\delta\propto R^{-d})$, and $s$ the spectral slope $(N=K\gamma^{-s})$. A power law fit to the variation of the turnover flux density with respect to the turnover frequency provides an exponent $\eta=0.8\pm0.1$ (see Fig. \ref{Cvmsm}). Together with estimates on $b$, $k$ and $d$, we obtained the average spectral slope $\langle s \rangle=2.3\pm0.6$, by solving Eq. \ref{etaadi} for $s$. This value  for the spectral slope is, within the uncertainties, in agreement with the results of the parameter space study based on the observed brightness temperature gradients for the adopted value of $d=-0.6$ (see panel 2 in Fig.~16 of Paper~II).\\

\subsection{Region D}
For the calculation of the magnetic field and the normalization coefficient for region D we used the jet width obtained from the stacked $15\,\mathrm{GHz}$ VLBI maps (see Fig.~\ref{jetwidth} in Sect.~\ref{jetridge}). The jet expands until $r\sim1.5\,\mathrm{mas}$ and collimates between $1.5\,\mathrm{mas}<r<2.4\,\mathrm{mas}$. Power law fits applied to the expansion and collimation regions give $\epsilon=0.8\pm0.1$ and $\epsilon=-0.5\pm0.1$ (where $\epsilon$ is the jet opening index, $R\propto r^{\epsilon}$). The kinematic analysis of region D shows a decreasing trend for the Doppler factor, which is based only on two data points, due to the lack of traveling components within this region. Because of this, we used for the calculation of the magnetic field and the normalization coefficient the average of $\langle\delta\rangle=7$.

\begin{figure}[h!]
\resizebox{\hsize}{!}{\includegraphics{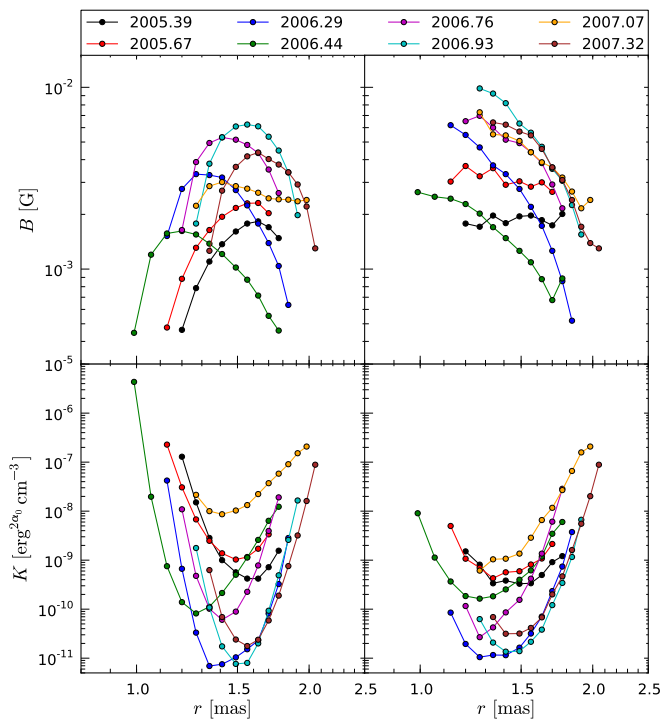}} 
\caption{Evolution of the magnetic field $B$ (top panel) and the normalization coefficient, $K$ (bottom panel) for region D. The left hand side correspond to the axial values and the right hand side transversally averaged value along the jet} 
\label{BKevoD} 
\end{figure}

The upper panels in Fig.~\ref{BKevoD} show the spatial and temporal evolution of the axial (left panel) and transversally averaged (right panel) magnetic field intensity. The axial magnetic field varies between $6\,\mathrm{mG}$ and $0.5\,\mathrm{mG}$ and the transversal averaged magnetic field spans from $10\,\mathrm{mG}$ to $0.5\,\mathrm{mG}$. Like the turnover frequency for region D, the magnetic field intensity on-axis shows a parabola shaped distribution (there is a rise and a decrease within the region) whereas the transversally averaged magnetic field decreases with distance. However, both show an increase between 2005.39 and 2006.29, which could be due to the passage of a traveling feature through this region (see component D2 in Paper~II).

The evolution of the normalization coefficient, $K$, with distance to the core is presented in the lower panels of Fig.~\ref{BKevoD}. The left panel shows the on-axis values and the right panel shows the transversally averaged ones. In both cases, $K$ decreases in the region  $1.0\,\mathrm{mas}<r<1.5\,\mathrm{mas}$ and increases farther downstream. The variation in $K$ is larger for the axial values (from $5\times 10^{-5}\,\mathrm{erg^{2\alpha_0}cm^{-3}}$ to $10^{-11}\,\mathrm{erg^{2\alpha_0}cm^{-3}}$) than  for the transversally averaged values (between $10^{-11}\,\mathrm{erg^{2\alpha_0}cm^{-3}}$ and $10^{-7}\,\mathrm{erg^{2\alpha_0}cm^{-3}}$). The values for $K$ depend on the optically thin spectral index and cannot therefore be directly compared. In contrast to the magnetic field intensity, the shape of the distribution is comparable for both values, and the increase in $K$ could reflect a reaccleration of relativistic particles.

\begin{figure}[h!]
\resizebox{\hsize}{!}{\includegraphics{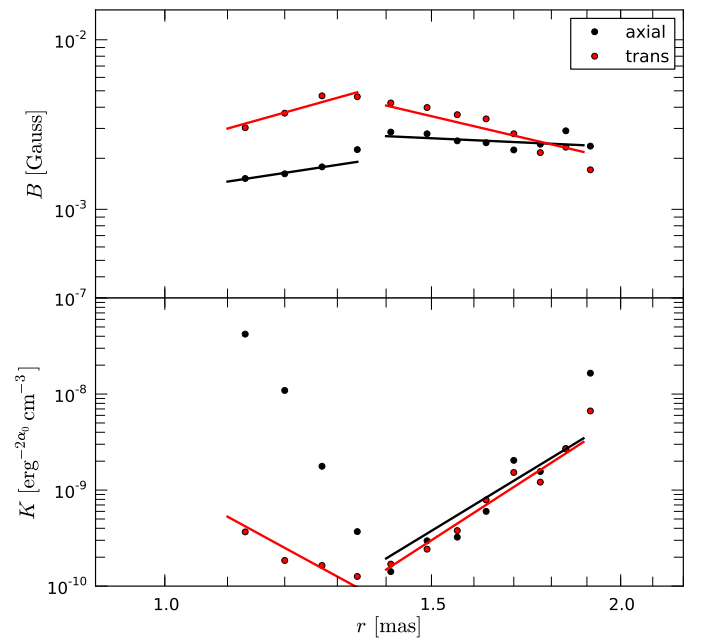}} 
\caption{Evolution of the time averaged magnetic field $B$ (top panel) and the normalization coefficient, $K$ (bottom panel) for region D. The black points and lines correspond to averaged axial values and the red points an lines to the transversally averaged values along the jet.}
\label{BKevoDaver} 
\end{figure}

Due to the large scatter and uncertainties in $B$ and $K$ we use in the following the time-averaged value at each position. The evolution of these time-averaged values with distance could be approximated by a power law and the results for the exponents are presented in Fig.~\ref{BKevoDaver} and Table~\ref{kbfit}.

\begin{table}[h!]
\caption{Results of the power law approximations for $B$ and $K$ (as seen in Fig.~\ref{BKevoDaver}) in region D.}  
\label{kbfit}
\centering  
\begin{tabular}{@{}c c c c c @{}} 
\hline\hline
		&	\multicolumn{4}{c}{$r<1.4\,\mathrm{mas}$ $(\epsilon=0.8\pm0.1)$}  \\	
		& 	b$^\prime$	& k$^\prime$  & b &k  \\
\hline
axial 	&	1.4$\pm$0.2	&	--			&	1.8$\pm$0.3	& ---			 \\
trans. 	&	2.5$\pm$0.7	&	$-9.6\pm3$	&	3.1$\pm$1.0	&$-12\pm4$	 \\
\hline
		&	\multicolumn{4}{c}{$r>1.4\,\mathrm{mas}$ $(\epsilon=-0.5\pm0.1)$} \\	
		& 	b$^\prime$	& k$^\prime$ & b & k   \\
\hline
axial 	&	$-0.4\pm0.3$	&	9.6$\pm$3	&0.8$\pm$0.6	& $-19\pm7$ \\
trans. 	&	$-2.1\pm0.3$	&	10.2$\pm$3	&4.2$\pm$1.8	&$-20\pm8$ \\
\hline
\end{tabular} 
\end{table} 

For $r<1.4\,\mathrm{mas}$, $b^\prime$ is for both, on-axis and transversally averaged values in Fig.~\ref{BKevoDaver}, that expected for a poloidal field ($b^\prime\sim2$), within the uncertainties. However, there is a significant difference between the value on-axis and the averaged one, indicating that the field is poloidal on average, but could have a toroidal component closer to the axis. The strong decrease in the value of $K$ on-axis results in an exponent $k^\prime\sim30$ (not given in the Table), which reflects the huge scatter in the temporal evolution of $K$ (see Fig. \ref{BKevoD}). For the transversally averaged normalization coefficient, $K$, we obtained an exponent of $k=-12\pm4$. 

   For $r>1.4\,\mathrm{mas}$ (within region D), the difference in the exponent between the on-axis values and the transversally averaged ones is larger. On-axis, the exponent indicates that the magnetic field is not organized in poloidal geometry, whereas the transversally averaged value of the field would keep the poloidal structure from the previous region. For both the on-axis and the transversally averaged normalization coefficients the exponent is $k\sim20$. The unclear geometry of the magnetic field close to the jet axis in region D, together with the kinematic analysis for this region, could be interpreted in terms of an additional shock-shock interaction, as already suggested in Paper~II. Such an interaction could lead to strong variations in magnetic field intensity and orientation. In particular, an increase in the transversal component (radial) is expected at the shock. Numerical simulations show that, at the position of recollimation shocks, there is an adiabatic compression that causes a (symmetric) bump in the spectral index \citep{Mimica:2009p42}.\\

\subsection{Derivation of the evolution of $\gamma_\mathrm{min}$ and $\gamma_\mathrm{max}$}
\label{gammaevo}

\citet{Marscher:1987p2165} pointed out that the energy limits, $E_{\mathrm{min}}=\gamma_\mathrm{min}m_ec^2$ and $E_{\mathrm{max}}=\gamma_\mathrm{min}m_ec^2$ in the expression describing the synchrotron spectrum are difficult to estimate. From the results obtained in this paper, we can now put constraints on the evolution of the minimum and maximum Lorentz factors of the non-thermal population along the jet in CTA~102. However, here we present a first-order approximation for the evolution of the lower, $\gamma_\mathrm{min}$, and upper, $\gamma_\mathrm{max}$ electron Lorentz factors. We consider two different jet geometries, i) a conical pressure matched jet ($d_k=p_0/p_a=1$) as a reference model and ii) an over-pressured jet ($d_k\neq1$). Based on these two different assumptions, and using some of the parameters obtained from the core-shift, ridge-line and spectral analysis, we computed the evolution of the electron Lorentz factor assuming iii) only adiabatic losses and iv) synchrotron and adiabatic losses. For this calculation we used an adiabatic index $\hat{\gamma}=4/3$, which corresponds to an ultra-relativistic $e^-e^+$ plasma. 
\paragraph{Conical jet:}
The first step in the calculation of the evolution of the electron Lorentz factor is to define the properties of the jet. In the case of a pressure-matched jet, the profile is conical and it is characterized by its radius at the jet nozzle, $R_0$, and its opening angle, $\varphi$. The opening angle, $\varphi$, can be obtained either from the measured bulk Lorentz factor $\varphi\sim1/\Gamma$ \citep[e.g.,][]{Konigl:1981p15} or from the transversal size of the jet \citep[e.g.,][]{Pushkarev:2009p5426}. The jet width, $R_0$, at the nozzle can be calculated in the following way:
\begin{equation}
R_0=R_j-r_{j-0}\tan{\varphi},
\end{equation} 
with $R_j$ the jet width obtained at a position, $r_j$, where the jet can be transversally resolved and $r_{j-0}$ the distance to the jet nozzle including the opacity shift correction (see Sect. \ref{coreshift1}). Finally the jet width for a conical geometry can be written as a power law:
\begin{equation}
R(r)=R_0\left(\frac{r}{r_0}\right)^\epsilon,
\label{conjet}
\end{equation}
where $\epsilon=1$ and $r_0$ is a normalization distance. 

\paragraph{Overpressured jet:} An initial over-pressure at the jet nozzle leads to the formation of recollimation shocks farther downstream \citep[e.g.,][]{Daly:1988p3}. Such stationary features can be detected and identified with VLBI observations. 
For the calculation of the jet width at the nozzle we used the approximation presented by \citet{Daly:1988p3}:
\begin{eqnarray*}
\nonumber
R_0\sim\frac{r_\mathrm{max}}{3.3\Gamma_0 d_k},
\end{eqnarray*}
where $r_\mathrm{max}$ is the de-projected distance between the jet nozzle and the recollimation shock, $\Gamma_0$ is the bulk Lorentz factor of the fluid and $d_k=p_0/p_\mathrm{ext}$ is the initial overpressure.

Assuming a locally homogeneous ambient medium, the distance between the jet nozzle and the location of maximal jet width is roughly $r\left(R_\mathrm{max}\right)\sim0.5\, r_\mathrm{max}$ \citep{Komissarov:1997p5504}. Using $R_0$, the jet opening index, $\epsilon$, can be derived for $r<r\left(R_\mathrm{max}\right)$. For $r>\left(R_\mathrm{max}\right)$, the jet is collimated by the ambient pressure, reaching a local minimum in the radius at the position of the standing shock. As in the case of the jet opening region, and given the jet width at the position of the standing shock, $\epsilon$ can be computed for this region. If there are additional recollimation shocks with known distance from the previous shock and jet width, one can proceed as in the case of the first recollimation shock and define the jet geometry.

For each case (conical and over pressured jet), the evolution of the maximum and minimum Lorentz factors of the non-thermal population considering adiabatic and synchrotron losses, and adiabatic losses alone, can be computed as follows:

\paragraph{Adiabatic and synchrotron losses:}
In order to provide an estimate for the variation of the electron Lorentz factor $\gamma$ along the jet, the energy-loss equation, which includes both, radiative and adiabatic losses \citep[see, e.g.,][]{Georganopoulos:1998p5775,Mimica:2009p42}, has to be solved:
\begin{equation}
\frac{d\gamma}{dr}=-\left(\frac{d\gamma}{dr}\right)_\mathrm{syn}-\left(\frac{d\gamma}{dr}\right)_\mathrm{adi},
\end{equation}
where $\gamma$ corresponds to the electron Lorentz factor and $r$ to the distance along the jet. The synchrotron and adiabatic losses are given by:
\begin{eqnarray}
\left(\frac{d\gamma}{dr}\right)_\mathrm{syn}&=&\left(\frac{2}{3}\right)^2\frac{e^4}{m_e^3c^6}\gamma^2B^2\\
\left(\frac{d\gamma}{dr}\right)_\mathrm{adi}&=&\frac{\gamma}{3}\frac{d\ln\rho}{dz}
\end{eqnarray}
Parametrizing the magnetic field, $B=B_0(r/r_0)^{-b}$, and the jet geometry, $R=R_0(r/r_0)^\epsilon$, the energy loss equation can be simplified to:
\begin{equation}
\frac{d\gamma}{dr}=-\left(\frac{2}{3}\right)^2\frac{e^4}{m_e^3c^6}\gamma^2B_0^2\left(\frac{r}{r_0}\right)^{-2\epsilon b}-\frac{\gamma\epsilon}{r}
\end{equation} 
This differential equation has the following analytical solution:
\begin{equation}
\gamma(r)=\frac{\gamma_0\left(\epsilon+2\epsilon b-1\right)r_0^{\epsilon}}{-C\gamma_0 r r_0^\epsilon+r^\epsilon\left(\frac{r}{r_0}\right)^{2\epsilon b}\left(\epsilon+2\epsilon b+ C\gamma_0 r_0-1\right)}\left(\frac{r}{r_0}\right)^{2\epsilon b},
\end{equation}
where $C=\left(\frac{2}{3}\right)^2(e^4)/(m_e^3c^6)B_1^2$, and $\gamma_0=\gamma\left(r_0\right)$ corresponds to the initial value.\\

An estimate of the jet parameters at the jet nozzle is done by assuming that they must be similar to those at the first recollimation shock, which we locate at $r\simeq0.1$~mas. This is justified by the fact that the conditions (pressure and density) at the position of the standing shock wave are roughly the same as those at the jet nozzle \citep[in a locally homogeneous ambient medium, see e.g.,][]{Daly:1988p3, Komissarov:1997p5504}. These parameters, as obtained from this work and paper~II, are listed in Table~\ref{gmodel}.

\begin{table}[h!]
\caption{Parameters used in the calculation of $\gamma_\mathrm{min}$ and $\gamma_\mathrm{max}$ presented in Fig.~\ref{losses}.}  
\label{gmodel}
\centering  
\begin{tabular}{@{}l l l l@{}} 
\hline\hline
Parameter &Symbol & Value & Source\\
\hline
normalization of B-field		&  	$B_0$			&	$0.1\,\mathrm{G}$	&	VLBI$^\dag$	\\
exponent B-Field evolution	& 	$b$				&	$1.0$			&	VLBI$^\dag$	\\
normalization distance		& 	$r_0$			&	$1.0\,\mathrm{pc}$	&	VLBI$^\dag$	\\
viewing angle				&	$\vartheta$		&	$2.6^\circ	$		&	VLBI$^\dag$	\\
Bulk Lorentz factor			&	$\Gamma_0$		&	$13$				&	VLBI$^\ddag$	\\
pressure matched jet 		&	$d_k$			&	$3$				&	VLBI$^\star$	\\
jet opening index			&	$\epsilon$			&	0.6, 0.8,$-$0.5,1.2	&	VLBI$^\dag$   	\\
electron Lorentz factor		&	$\gamma_\mathrm{min,max}$	&100, $1\cdot10^5$	& Theory$^\ast$\\
\hline
\multicolumn{4}{l}{$^\dag$ taken from this paper}\\
\multicolumn{4}{l}{$^\ddag$ taken from paper II}\\
\multicolumn{4}{l}{$^\star$ taken from \citet{Jorstad:2005p4121}}\\
\multicolumn{4}{l}{$^\ast$ taken from \citet{Mimica:2009p42}}\\
\end{tabular} 
\end{table}

\begin{figure*}[t!]
\centering 
\includegraphics[width=17cm]{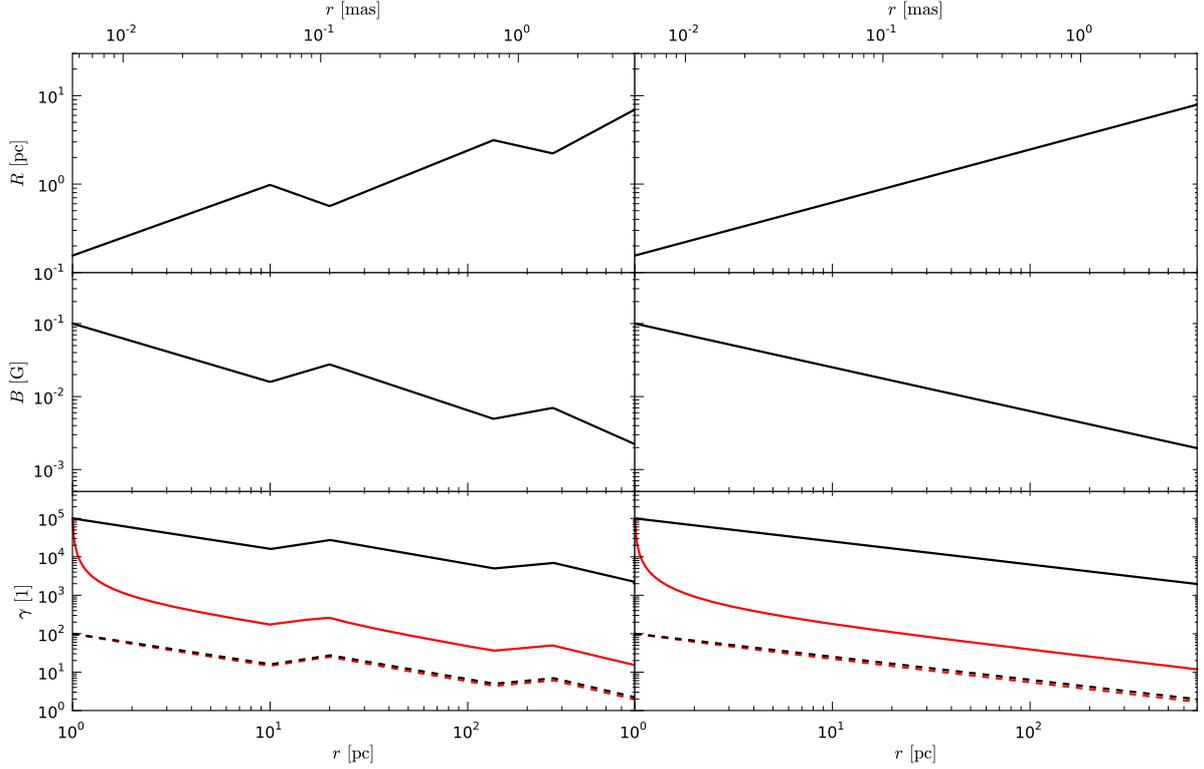} 
\caption{Evolution of source intrinsic parameters for an over-pressured jet, $d_k\neq1$, (left panels) and for a conical jet, $d_k=1$, (right panels). The upper panels show the variation of the jet width, $R$, in $\mathrm{pc}$ along the jet, the middle panels the evolution of the magnetic field, $B$, in $\mathrm{G}$ and the bottom panels the variation of the upper and lower electron Lorentz factors, $\gamma_\mathrm{min,max}$. The solid and dashed red lines correspond to a model where we take both, synchrotron and adiabatic losses into account. In order to demonstrate the influence of radiative losses we computed additionally the evolution of the electron Lorentz factor only for adiabatic losses (solid and dashed lines in bottom panels). See text for more details. Notice, that the bottom x-axis is drawn in $\mathrm{pc}$ and to allow for the direct comparison with the observations we plotted the upper x-axis in $\mathrm{mas}$.}
\label{losses} 
\end{figure*}

Figure~\ref{losses} shows the results of our calculations for an over-pressured (left panels) and a conical jet (right panels) using the parameters presented in Table \ref{gmodel}. Notice that the x-axes are given in pc (bottom) and in mas (top) to facilitate the comparison with observations. The panels show (from top to bottom) the evolution of the jet width with distance along the jet (fitted from the evolution of the jet radius with distance in Fig.~\ref{jetwidth}), the evolution of the magnetic field intensity (from Figs.~\ref{BKevo} for the core and \ref{BKevoDaver} for region D), and computed evolution of the maximum and minimum electron Lorentz factor, $\gamma_\mathrm{min,max}$. The red solid and dashed lines in the bottom panels correspond to the variation of the electron Lorentz factor taking synchrotron and adiabatic losses into account and the black solid and dashed lines show the evolution of the electron Lorentz factor with only adiabatic losses.

For both jet models, the evolution of the upper electron Lorentz factor, $\gamma_\mathrm{max}$, decreases fast if the synchrotron losses are taken into account. The adiabatic losses are the dominant energy loss mechanism for $r>3\,\mathrm{pc}$, which is visible in the similar slopes of the solid black and red lines. In contrast, the evolution of the lower electron Lorentz factor, $\gamma_\mathrm{min}$, is hardly affected by synchrotron losses.\\

The calculations show that in the case of an over-pressured jet, the magnetic field increases at the position of the recollimation shocks (at $r\sim20\,\mathrm{pc}$ and $r\sim270\,\mathrm{pc}$) and at the same position, both the minimum and the maximum electron Lorentz factor increase. This increase of the Lorentz factor corresponds to the local increase of the particle density, i.e., adiabatic compression \citep[see, e.g.,][]{Mimica:2009p42}. Farther downstream, the jet expands and the magnetic field and the Lorentz factor decrease again. At the second recollimation shock, there is again a local increase in both, magnetic field and electron Lorentz factor. This behavior is also visible in the case of adiabatic losses alone.\\

As opposed to the over-pressured jet, the evolution of the magnetic field and the electron Lorentz factor are continuously decreasing in the case of a conical jet (see right panels in Fig. \ref{losses}).

\subsection{Particle density, $N_\mathrm{tot}$ and relativistic energy density, $U_e$}
\begin{figure}[h!]
\resizebox{\hsize}{!}{\includegraphics{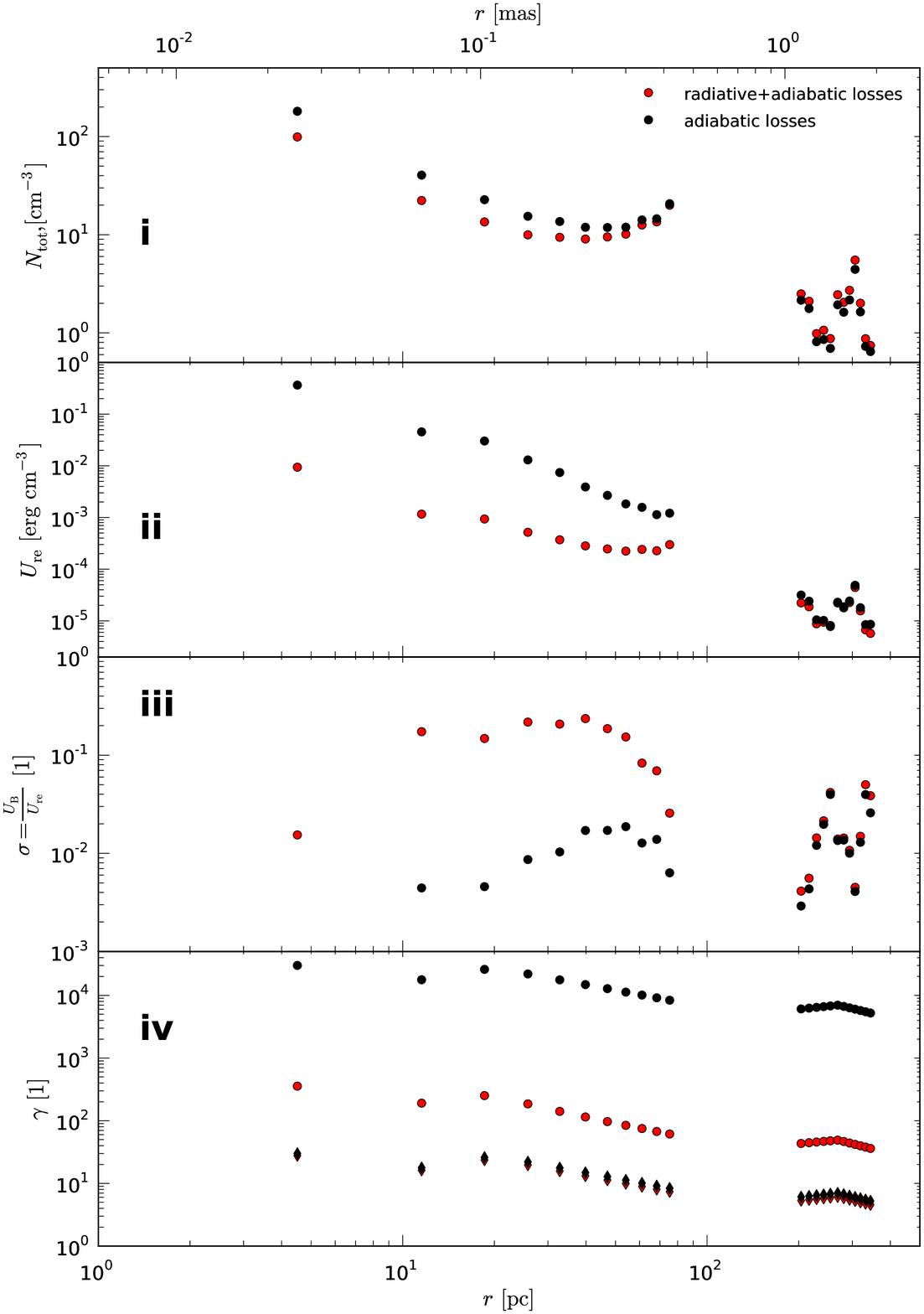}} 
\caption{Evolution of the source intrinsic parameters taking radiative and adiabatic losses into account (red points) and only adiabatic losses (black points). The {different} panels show the variation of the total particle density, $N_\mathrm{tot}$ (panel i), the relativistic energy density, $U_\mathrm{re}$ (panel ii), the magnetization $\sigma=U_\mathrm{B}/U_\mathrm{re}$ (panel iii) and the electron Lorentz factor (panel iv). The circle and diamond marker in {panel iv} correspond to the upper and lower electron Lorentz factors. For more details see text.} 
\label{Ntotal} 
\end{figure}

Using the values computed for $\gamma_\mathrm{min}$ and $\gamma_\mathrm{max}$, together with the magnetic field, $B$, and the normalization coefficient, $K$, we could compute estimates for the evolution of the total particle density, $N_\mathrm{tot}$, and the relativistic energy density, $U_\mathrm{re}$ for regions C and D. The values of $N_\mathrm{tot}$ and $U_\mathrm{re}$ in region D were obtained from the average magnetic field and normalization coefficient (see Fig.~\ref{BKevoDaver}). Figure~\ref{Ntotal} presents the results of our calculations taking radiative and adiabatic losses, and only adiabatic losses into account. The panels show the evolution of the total particle density, $N_\mathrm{tot}$ (panel i), the relativistic energy density, $U_\mathrm{re}$ (panel ii), the magnetization $\sigma=U_\mathrm{B}/U_\mathrm{re}$ (panel iii) and the electron Lorentz factor (panel iv). The circle and diamond markers in panel iv correspond to the upper and lower electron Lorentz factors. Since our model for the evolution of the electron Lorentz factor does not take into account traveling shock waves, the average values of $B$ and $K$ in region D reflect a calculated steady state. However, the possible passage of a traveling component through this region was reported in Paper~II, so this may be the reason for differences between the model and the observed behavior.

The total particle density for region C ($r<1\,\mathrm{mas}$) takes values between (200--15)~cm$^{-3}$ if we consider only adiabatic losses and in the range of (100--10)~cm$^{-3}$ if we additionally take radiative, here synchrotron losses, into account (panel i in Fig. \ref{Ntotal}). Farther downstream in region D ($1\,\mathrm{mas}<r<4\,\mathrm{mas}$) the particle density drops to roughly 1~cm$^{-3}$ and increases at $r\sim1.5\,\mathrm{mas}$ to 5~cm$^{-3}$ (location of a possible standing shock). The values for both models lead to similar values since $N_\mathrm{tot}\propto\gamma_\mathrm{min}^{1-s}$ (see Eq. ~\ref{Ntot}) and the evolution of the lower electron Lorentz factor is not affected by radiative losses (see panel iv).\\

The influence of {radiative} losses is clearly visible in the evolution of the relativistic energy density, $U_\mathrm{re}$, for region C (panel ii in Fig. \ref{Ntotal}). The energy density takes values in the range of ($5\times10^{-1}$--$5\times10^{-3}$)\,erg\,cm$^{-3}$ if only adiabatic losses are considered and from ($10^{-2}$-$10^{-3}$)\,erg\,cm$^{-3}$ if both, synchrotron and adiabatic losses are included. Since the spectral slope for region C is $s<2$, the relativistic energy density is mainly proportional to $\gamma_\mathrm{max}$ (see Eq. \ref{ue}). As shown in Sect. \ref{gammaevo}, radiative losses strongly affect the variation of $\gamma_\mathrm{max}$. The difference between the two models nearly disappears in region D, where we obtain values between $10^{-4}$\,erg\,cm$^{-3}$ and $10^{-5}$\,erg\,cm$^{-3}$. The dependence of $U_\mathrm{re}$ changes from $U_\mathrm{re}\propto \gamma_\mathrm{max}^{2-s}$ to $U_\mathrm{re}\propto \gamma_\mathrm{min}^{2-s}$ since $s>2$ within region D (see Eq. ~\ref{ue}). As in the case of $N_\mathrm{total}$, the relativistic energy density increases by a factor 6 at $r\sim1.5\,\mathrm{mas}$.\\

For both models, the jet is particle dominated ($\sigma<1$) for $r>0.1\,\mathrm{mas}$ (see panel iv in Fig.~\ref{Ntotal}). However, the mentioned differences in $U_\mathrm{re}$ lead to a difference of a factor 50 in the value of $\sigma$ between the two models. The magnetization increases from a calculated value around $\sigma\sim0.01$ in region C to $\sigma\sim0.1$ in region D (see the third panel in Fig.~\ref{Ntotal}) due to the increase of the magnetic field close to the location where we think that a standing shock is located, and drops to $\sigma\sim0.01$ at $r=1.5\,\mathrm{mas}$ due to the increase of the particle density (first panel in Fig.~\ref{Ntotal}) and the related rise of the relativistic energy density (second panel in Fig.~\ref{Ntotal}).

\section{Discussion}
\label{disc}
\subsection{Region C (core)}

  Region~C covers the first milliarcsecond of the radio-jet.
The component ejected after the 2006 flare (C2) evolves during the time spanned by the observations
(2005-2007) within this region. This component has a velocity of $0.25\pm0.04$~mas/yr,
which translates into an apparent velocity of $13\pm2\,c$. In two
years the component has covered an apparent distance of 0.5~mas. Unfortunately,
we do not have enough resolution to identify C2 with accuracy in the
spectral index maps of the region (see Fig.~\ref{Calpha1d} and Fig.~\ref{Calpha2d} in Appendix C).

  However, following the temporal evolution of the spectral index
distributions (both the axial and the transversally averaged values, see
Fig.~\ref{Calpha1d}), we observe that the first significant increase (beyond errors)
of the spectral index in the region occurs at $\sim 2006.4$ (blue dots).
At this time, C2 should be crossing the region around 0.1~mas from the position of
the 43~GHz core (0~mas in the plot, see Fig. A.25 in Paper II), which
is the expected location of a stationary feature from our results in
Papers~I and II.

  In Paper I, we deduced that the flare started around 2005.6 from the spectral evolution of the source as derived from the single-dish data. This evolution was interpreted in terms of
the shock-in-jet model \citep{Marscher:1985p50}. The $\nu_m,\, S_m$ plot after this flare revealed a
Compton stage, which lasted until 2005.8, followed by an adiabatic stage. 
However, in the period 2006.0--2006.3, there was a reversal of the expected evolution, with an increase in the peak flux ($S_m$), while the peak frequency ($\nu_m$) stayed basically constant. Finally, after 2006.3 the component returned to the expected spectral evolution dominated by adiabatic losses. This second and unexpected peak in $S_m$ could be understood as due to injection of relativistic particles into the system, or to the existing
particles going through a compression \citep{Mimica:2009p42}. It 
was discussed in terms of the interaction of the component ejected after
the flare with a standing shock close to the core region. The authors initially interpreted this stage as a new Compton stage due to its similar behavior in the $\nu_m,\, S_m$ plot. However,
it could well be explained as a break within the adiabatic stage. This
possibility is being tested using numerical simulations (Fromm et al.,
in preparation). 

{The observed increase of the spectral index in region C (Fig.~\ref{Calpha1d}) from the 2005 epoch (black dots)
 to that in 2006.3 (blue dots) fits well into the description given in the previous paragraph. After the passage of C2 through the innermost region, it should arrive to the expansion
region between $\simeq 0.1$~mas and $\simeq 1$~mas (see Fig.~\ref{jetwidth}). Then, the observed decrease in the spectral index can also be understood within the evolution of C2 inside the
core. The rises of the spectral index in 2006.9 (light blue in Fig.~\ref{Calpha1d}) is compatible with no changes within errors. However, the increase in 2007.4 (brown
dots in Fig.~\ref{Calpha1d}) is clearly beyond errors and very
similar to that in 2006.3.} Although there is no evidence for a
component crossing the core or the $0.1$~mas standing feature at this
epoch, component C3 is identified between 0 and 0.1~mas from the 43~GHz core
position (see Fig.~16 in Paper II). Therefore this new increase could
be related with the ejection of a new component.

    The observed decrease of the spectral index with distance at each
epoch can be easily understood in terms of synchrotron and adiabatic
cooling times, which are shorter in the case of higher frequencies
\citep[see, e.g.,][]{Blandford:1979p33}, so a steepening of the spectral distribution with distance
is expected. In summary, the spectral evolution of region C until 2007 is consistent with the propagation of a shock produced by the perturbation ejected in the 2006 radio-flare.

\subsection{Region D}

   Figures~\ref{Dvm1d}, \ref{Dsm1d}, and \ref{Da01d} show the peak frequency, peak flux
and the spectral index, respectively of the spectral distribution of the jet along region
D for the different epochs. The whole picture fits well with the passage
of component D2 through the region, as shown by both the axial
and averaged values of $S_m$. The peak in emission propagates from
$r\simeq 1$~mas to $r\simeq 1.5$~mas in two years, which is in agreement
with the velocity of the component D2 ($0.16\pm0.01$~mas/yr, see Table~3 in Paper~II).

   There is a discrepancy between the axial and averaged profile shapes
of $\alpha_0$ and $\nu_m$ in the 1.0--1.5~mas interval, which could be related to
edge effects (i.e., at the edges of the studied region) in the axial values produced by the alignment procedure.
The result is compatible with a smooth decrease of the peak frequency
with distance in the region. The large errors in the derivation of the
spectral index make it difficult to make any statement regarding its
evolution in space and time within the region, but the result seems to
indicate that it does not undergo strong changes.        

       Figure~\ref{jetwidth} indicates that the jet is recollimating
between 1~mas and 2~mas and this is confirmed in Fig.~8 of Paper~II,
which shows the continuous presence of a component at
$\simeq$1.5~mas from the core, first identified as D1 and later as D2 in this Figure.
This was also reported by \citet{Jorstad:2005p4121}. The interaction between
D2 and this standing feature could be responsible for the bump in
emission at epochs 2005.39 to 2006.44 seen in the $S_m$ plot (black,
red, dark blue, and dark green points) and for the spectral index being
possibly constant or slightly increasing within the region. A reconfinement region
should generate a compression of the emitting particle distribution and be seen
as a symmetric bump in the spectral index profile along the jet \citep{Mimica:2009p42}. 

\subsection{Regions B and A}        

     The spectral index profile within region~B (see Fig.~\ref{Ba01d}) shows
a clear indication for one or two bumps at 4-5~mas. The fairly symmetric
shape of the bumps is an indication of a compression of the particles
\citep{Mimica:2009p42}. This could be also the case between 1.5 and 2.5~mas
(region~B, see below and Fig.~\ref{Ba01d}), but we do not have data to probe this hypotheses. 
In addition, the kinematics of the source between 4 and 5~mas
indicate that the components are stationary. Finally, the jet
width is constant between $r\simeq 4$~mas and $r\simeq
8$~mas. These facts favor the standing shock scenario.  From the data,
it is difficult to say whether the whole region includes several of such
shocks (as indicated by the fitted components), or whether it is a
single, long reconfinement region similar to that observed in numerical
simulations \citep{Perucho:2007p9}. In this respect, the maximum
of emission in the radio maps corresponds to component B2 (see
Fig.~\ref{allcont}), which is at $r\simeq6$~mas from the core, but the first
bump in the spectral index is observed around $r\simeq 4$~mas.

     Regarding region~A, Fig.~\ref{A2a1d} shows the profile of the spectral
index at the different epochs. A new plateau in the jet width, starting
around 10~mas (see Fig.~\ref{jetwidth}), could have an influence on the little
change observed across this region at most epochs. However, within
errors, the result is compatible with a smooth decrease of the spectral
index with distance, as expected from adiabatic losses \citep[see, e.g.,][]{Mimica:2009p42}.

\subsection{The jet in CTA~102. A global perspective}     

 From the accumulated evidences in Papers~I, II, and this work, we can
discuss the general nature of the jet in CTA~102. First of all, the jet
shows kinks and helical structures with different wavelengths showing up
at different observing frequencies \citep[see also][]{Perucho:2012p5917}. The
regions in the stacked image (0--1~mas, 1--2~mas, 5--8~mas, and
$>$10~mas) coincide with the transversal displacement of the ridge line
in East--West direction at 15~GHz, which is a central frequency to the
whole analysis. The latter is the main reason why the jet viewing angle
at this frequency plays a crucial role in selecting regions for
our analysis.

  Following the conclusions derived in Paper~II, the jet flow should be
accelerating in order to explain the observed increase in the Doppler factor at the core region
(0.1 to 1~mas, see Fig.~13 in Paper~II). Moreover, acceleration of the flow is also
expected in a hot flow after compression in a standing shock, as it has
been claimed to be the case at 0.1~mas.

   Farther downstream, the changes in the brightness temperature along
the jet include discontinuous jumps at $r\simeq1-2$~mas and
$r\simeq4-5$~mas (see Fig.~14 in Paper~II), which cannot be accounted
for in terms of a continuous change in the viewing angle. At the same
positions (1-2~mas and 4~mas), the jet-width profile flattens (see
Fig.~\ref{jetwidth}) and the spectral index shows 'bumps' in
some cases (e.g., $r\simeq4-5$~mas, Fig.~\ref{Ba01d}). Both facts point to
the presence of reconfinement shocks. Therefore, the jet
morphology at, e.g., 15~GHz, indicates that helical motion is 
responsible for the highest relative flux in regions D, B, and A. However, the
spectral properties and, to some extent also the kinematics, favor the
over-pressured jet scenario and the existence of recollimation regions
at $r\simeq 0.1$~mas (de-projected 16~pc from the core), at $r\simeq
1-2$~mas (de-projected 160-320~pc from the core), at $r\simeq 5$~mas
(de-projected 800~pc from the core), and possibly at $r\simeq 10$~mas
(de-projected 1600~pc from the core). The increasing distance between
the positions of the subsequent shock candidates suggest that the
jet propagates in a decreasing density ambient medium. However, it is
difficult to know whether, for instance, there is only one or more
standing shocks in region~B (components B1, B2 and B3 at $4\,\mathrm{mas}\leq r
\leq8$~mas, see Paper~II). The lack of correspondence of the plateaus in the jet radius
(including stationary components and bumps in the spectral index
profiles) with the boosted regions of the jet can be taken as an
evidence of their independent nature.

   Interestingly, the width of the jet at 2~GHz increases in region D
whereas it remains unchanged at higher frequencies. Only at
$r\simeq10$~mas the jet-width at 2~GHz becomes flatter. This result
should be understood in terms of transversal jet structure and different
internal and external dynamics in the jet, which can be compared to
simulations or theoretical models of over-pressured jets  \citep{Begelman:1984p75,Perucho:2007p9,Nalewajko:2012p6040}.
At the largest scales, the 2~GHz jet-width is
$\simeq 6$~mas ($\simeq 50$~pc), whereas at 5~GHz the jet-width is
$\simeq 3.5$~mas ($\simeq 30$~pc), which can give an idea of the width
of the shear/mixing layer surrounding the jet \citep[see, e.g.,][]{Perucho:2007p9,Wang:2011p6258}. 
A deeper study of the ridge-line
and jet transversal structure will be presented elsewhere.

 In conclusion, from the present data, we claim that a complex interaction between different structures
is probably taking place in CTA~102.

\section{Conclusions}
\label{sum}
In this paper we studied in detail the radio spectral variations in the jet of CTA\,102 during the 2006 major flare.
The core-shift analysis revealed that, on average, the observed VLBI core position behaves like $r\propto\nu^{-1}$.
We calculated the magnetic field at the core to be $B_\mathrm{core}\sim100\,\mathrm{mG}$ and the particle density
$N_\mathrm{core}\sim40\,\mathrm{cm^{-3}}$.

The spectral analysis of the core region showed a significant increase in the spectral index, $\alpha$, during the period of the flare with a steepening of $\alpha$ for $r>0.1\,\mathrm{mas}$. The increase of the spectral index is in agreement with the possible interaction of a traveling component C2 and a standing shock at this position. The location coincides with the first detection of the jet expansion. 

Within this region, adiabatic losses are dominant and the evolution of the magnetic field intensity indicates a toroidal geometry, with values of (9--50)$\,\mathrm{mG}$.\\
Farther downstream, we found evidence for the possible crossing of the feature D2 through another possible standing feature at r$\simeq$1.5\,mas. The analysis of the transversal structure showed evidence for recollimation of the jet. The spectral evolution of this region revealed an increase in the turnover frequency and turnover flux density during the period of the interaction between the components. The evolution of the magnetic field intensity in this region is compatible with a mixture of toroidal and poloidal structure between 10~mG and 1~mG.

Between $r$=3\,mas and $r$=8\,mas there are additional regions of recollimation, as shown by the jet width and the spectral index, which further support our hypothesis of an over-pressured jet.

Summarizing the results of this series of papers, we conclude that the jet of CTA~102 cannot be simply described by an over-pressured or a helical jet, since we found evidence for both during our analysis. The morphology of VLBI images and the jet ridge line {show} a clear helical pattern. The kinematic analysis reveals several standing components and the jet width exhibits regions with jet collimation, both being indications of an over-pressured jet. The symmetric variations around the locations of this stationary components is an additional proof for the existence of recollimation shocks in CTA~102. All these facts lead to the conclusion that the jet of CTA~102 is best described by helical patterns or flow motion developing within an over-pressured jet.{Whether the observed helical structure corresponds to a pattern, as reported in \citet{Perucho:2012p5917}, or the flow follows this helical path should be studied in detail in future.}\\

Due to the limited resolution and frequency range, we could not derive the spectral evolution of the VLBA core and for jet features farther downstream ($r>1\,\mathrm{mas}$). Moreover, the interaction between a traveling shock-wave and a recollimation shock is a highly non-linear process which requires a more detailed investigation. Such a study is only possible by performing numerical simulations and the full radiative transfer calculations including a self-consistent treatment of the relativistic electron distribution \citep{Mimica:2009p42}. We plan to further study shock-shock-interaction in parsec-scale jets using relativistic hydrodynamic simulations and extend our analysis to other sources.

\begin{acknowledgements}
C.M.F. was supported for this research through a stipend from the International Max Planck Research School (IMPRS) for Astronomy and 
Astrophysics at the Universities of Bonn and Cologne. Part of this work was supported by the COST Action MP0905 Black Holes in a 
violent Universe through the scientific stays, STSM-MP0905-140211-004292, STSM-MP0905-300711-008633 and STSM-MP0905-110711-008634.\\
E.R. acknowledges partial support from MINECO grant AYA2009-13036-C02-02 and the Generalitat Valenciana grant PROMETEO-2009-104.\\ 
M.P. acknowledges financial support from the MINECO grants AYA2010-21322-C03-01, AYA2010-21097-C03-01 and CONSOLIDER2007-00050, and from the Generalitat Valenciana grant PROMETEO-2009-103. M.P. acknowledges partial support from MINECO through a  Juan de la Cierva contract.\\
P. M. acknowledges the support from the European Research Council (grant CAMAP-259276).\\
This work is based on observations with VLBA. The VLBA is operated by the NRAO, a facility of the NSF under cooperative agreement by Associated Universities Inc.
\end{acknowledgements}

\bibliographystyle{aa} 
\bibliography{biblio2}
\clearpage
\section*{Appendix A}
\label{specana}
In this section we review the basic equations of synchrotron self-absorption and present the relations needed for the performed spectral analysis \citep[see e.g.,][]{Pacholczyk:1970p1583,Marscher:1987p2165,Lobanov:1998p2310,Turler:1999p342}.
\newline The emission at frequency $\nu$, $\epsilon_{\nu}$, and absorption coefficients, $\kappa_{\nu}$,  of a power law distribution of relativistic electrons, $N(E)=KE^{-s}$, where $K$ is the 
normalization coefficient of the distribution and $s$ the spectral slope of the relativistic electron distribution, can be written as \citep[for details see][]{Pacholczyk:1970p1583}:

\begin{eqnarray}
\epsilon_\nu&=&c_{\epsilon}(s)K \left(B\sin\varphi\right)^{(s+1)/2}\nu^{-(s-1)/2}\\
\kappa_\nu&=&c_{\kappa}(s)K \left(B\sin\varphi\right)^{(s+2)/2} \nu^{-(s+4)/2},
\end{eqnarray}
where $B$ is the magnetic field, $\varphi$ the pitch angle and $\nu$ the frequency. The constants $c_{\epsilon}(s)$ and $c_{\kappa}(s)$ are given by:

\begin{eqnarray}
c_{\epsilon}(s)&=&\frac{\sqrt{3}e^{3}}{16\pi m_{e}c^{2}}\left(\frac{3e}{2\pi m_{e}^{3}c^{5}}\right)^{\frac{s-1}{2}}\left(\frac{s+\frac{7}{3}}{s+1}\right)\hat{\Gamma}\left(\frac{3s-1}{12}\right) \nonumber \\
&&\hat{\Gamma}\left(\frac{3s+7}{12}\right) \label{epss} \\
c_{\kappa}(s)&=&\frac{\sqrt{3}\pi}{72}e m_{e}^{5}c^{10}\left(\frac{3e}{2\pi m_{e}^{3}c^{5}}\right)^{\frac{s+4}{2}}\left(\frac{s+10}{3}\right)\hat{\Gamma}\left(\frac{3s+2}{12}\right) \nonumber\\
&&\hat{\Gamma}\left(\frac{3s+10}{12}\right) \label{kaps},
\end{eqnarray}
being $e$ the electron charge, $m_{e}$ the electron rest-mass, $c$ the speed of light and $\hat{\Gamma}$ the complete Euler-Gamma function. For a random magnetic field, the constants above have to be averaged over the pitch angle $\varphi$, i.e., multiplied by $c_{\epsilon,b}$ and $c_{\kappa,b}$, respectively, with

\begin{eqnarray}
c_{\epsilon,b}(s)&=&\frac{\sqrt{\pi}}{2}\,\hat{\Gamma}\left(\frac{s+5}{4}\right)\left(\hat{\Gamma}\left(\frac{s+7}{4}\right)\right)^{-1} \label{epsb}\\
c_{\kappa,b}(s)&=&\frac{\sqrt{\pi}}{2}\,\hat{\Gamma}\left(\frac{s+6}{4}\right)\left(\hat{\Gamma}\left(\frac{s+8}{4}\right)\right)^{-1} \label{kapb}.
\end{eqnarray}

The specific intensity, $I_\nu$, can be written as:
\begin{equation}
I_\nu=\frac{\epsilon_\nu}{\kappa_\nu}\left(1-e^{-\tau_{\nu}}\right),
\label{intensity}
\end{equation}
where $\epsilon_\nu$ and $\kappa_\nu$ are the emission and absorption coefficients and $\tau_\nu=\kappa_\nu x$  
is the optical depth, with $x$ the distance along the line of sight. Defining $\nu_{1}$ as the frequency where $\tau_{\nu}=1$, Eq. ~\ref{intensity}  takes the following form \citep{Pacholczyk:1970p1583}:
\begin{equation}
I_{\nu}=I_{\nu_{1}}\left(\frac{\nu}{\nu_1}\right)^{\alpha_t}\left[1-\exp\left(\frac{\nu}{\nu_{1}}\right)^{\alpha_0-\alpha_t}\right],
\label{inu}
\end{equation}
where $\alpha_{t}$ is the optically thick spectral index ($\alpha_{t}=5/2$ for a homogenous source), and $\alpha_{0}<0$ is the optically thin spectral index. The optically thin spectral index is connected to the spectral slope, $s$,  by the following relation:
\begin{equation}
\alpha_{0}=-\frac{(s-1)}{2}.
\end{equation}
Using the transformation from intensities to flux densities Eq. ~\ref{inu} can be expressed in terms of the observed turnover flux density, $S_{m}$, and turnover frequency, $\nu_{m}$ \citep{Turler:1999p342}:
\begin{equation}
S_\nu \approx S_m\left(\frac{\nu}{\nu_m}\right)^{\alpha_t}\frac{1-\exp{\left(-\tau_m\left(\nu/\nu_m\right)^{\alpha_0-\alpha_t}\right)}}{1-\exp{(-\tau_m)}},
\label{snuapprox}
\end{equation}
where $\tau_m\approx3/2\left(\sqrt{1-\frac{8\alpha_0}{3\alpha_t}}-1\right)$ is the optical depth at the turnover. 
Depending on the value of $\nu/\nu_{m}$, Eq. ~\ref{snuapprox} describes an optically thick $(\nu<\nu_{m})$ or 
optically thin $(\nu>\nu_{m})$ spectrum with their characteristic shapes  $S_{\nu}\propto \nu^{5/2}$ and  
$S_{\nu}\propto \nu^{-(s-1)/2}$, respectively.\\

\subsection*{Magnetic field, $B$, and particle density, $K$}
Once the turnover frequency, $\nu_{m}$, and the turnover flux density, $S_{m}$ are obtained (see Sect. \ref{spectralana}), estimates for the magnetic field, $B$, and the normalization coefficient, $K$, \citep[see, e.g.,][]{Marscher:1987p2165} can be derived. Following \citet{Lind:1985p3025}, the emission, $\epsilon_{\nu}$, and absorption coefficient, $\kappa_{\nu}$, have to be corrected for relativistic and cosmological effects. In the following, primed variables correspond to the observers frame and the equations will be derived for a random magnetic field (isotropic pitch angle, $\varphi$), with all parameters in $\mathrm{cgs}$ units. Introducing these corrections we obtain:

\begin{eqnarray}
\epsilon^{\prime}_{\nu^{\prime}}&=&\delta^{2}\epsilon_{\left({\nu^{\prime}(1+z)/\delta}\right)}\\
\kappa^{\prime}_{\nu^{\prime}}&=&\delta^{-1}\kappa_{\left({\nu^{\prime}(1+z)/\delta}\right)}
\label{kappa}
\end{eqnarray}

\noindent with $\delta=\Gamma^{-1}\left(1-\beta\cos\vartheta\right)^{-1}$ the Doppler factor, {with $\beta=v/c$, $\vartheta$ the viewing angle}, and $z$ the redshift. The optically thin flux, $S^\prime_{\nu^{\prime},\mathrm{thin}}=\Omega\epsilon_\nu^\prime R$ (with $\Omega$ the solid angle and $R$ the size of the emission region), is given by:

\begin{equation}
S_{\nu^{\prime},\mathrm{thin}}=\frac{\pi}{D_{l}^{2}}c_{\epsilon}(s)c_{\epsilon,b}(s)(1+z)^{-\frac{s-3}{2}}R^{3}\delta^{\frac{s+5}{2}}K B^{\frac{s+1}{2}}\nu^{\prime-\frac{s-1}{2}}
\label{snu}
\end{equation}

\noindent and the optical depth, $\tau_{\nu}$, by:

\begin{equation}
\tau^{\prime}_{\nu^{\prime}}=c_{\kappa}(s)c_{\kappa,b}(s)(1+z)^{-\frac{s-4}{2}}R\delta^{\frac{s+4}{2}}K B^{\frac{s+2}{2}}\nu^{\prime-\frac{s+4}{2}}.
\label{tau}
\end{equation}

Using the obtained turnover values, the flux density, $S_\nu^\prime$, in Eq. ~\ref{snu} and the frequency, $\nu^\prime$, in Eq. ~\ref{tau} can be replaced by the turnover flux density, $S_{m}^\prime$, and the turnover frequency, $\nu_{m}^\prime$:
\begin{eqnarray}
S^{\prime}_{m}&=&\pi D_{l}^{-2}c_{\epsilon}(s)c_{\epsilon,b}(s)(1+z)^{-\frac{s-3}{2}}R^{3}\delta^{\frac{s+5}{2}}K B^{\frac{s+1}{2}}\nu_{m}^{\prime-\frac{s-1}{2}}\label{smeq}\\
\tau^{\prime}_{m}&=&c_{\kappa}(s)c_{\kappa,b}(s)(1+z)^{-\frac{s-4}{2}}R\delta^{\frac{s+4}{2}}K B^{\frac{s+2}{2}}\nu_{m}^{\prime-\frac{s+4}{2}}.
\label{tmeq}
\end{eqnarray}

The equations above can be solved for the magnetic field, $B$, and the normalization coefficient, $K$:

\begin{eqnarray}
B&=&\frac{\pi^{2}}{D_{l}^{4}}\left[\frac{c_{\epsilon}(s)c_{\epsilon,b}(s)}{c_{\kappa}(s)c_{\kappa,b}(s)}\right]^{2}(1+z)^{7}R^{4}\delta\nu^{\prime5}_{m}S_{m}^{\prime-2}\tau_{m}^{\prime2}
\label{bfield}\\
K&=&\frac{D_{l}^{2s+4}}{(\pi)^{s+2}}\left[c_{\epsilon}(s)c_{\epsilon,b}(s)\right]^{-(s+2)}\left[c_{\kappa}(s)c_{\kappa,b}(s)\right]^{s+1} \nonumber \\
&&(1+z)^{-(3s+5)}R^{-(2s+5)}\delta^{-(s+3)}\tau_{m}^{\prime-(s+1)}S_{m}^{\prime s+2}\nu_{m}^{\prime-(2s+3)}.
\label{knorm}
\end{eqnarray}
\subsubsection{Number of particles, $N$, relativistic energy density, $U_e$, and magnetization $\sigma$}
\label{NBfield}

The number of particles, $N$, and the total energy distribution of the relativistic particles, $U_e$, can be calculated by integrating the distribution function $N(E)=KE^{-s}$ within the limits $E_{1}=\gamma_\mathrm{min}m_ec^2$ and $E_{2}=\gamma_\mathrm{max}m_ec^2$:
\begin{equation}
N=\frac{K}{s-1}\left(m_{e}c^{2}\right)^{1-s}\gamma_\mathrm{min}^{1-s}\left[1-\left(\frac{\gamma_\mathrm{max}}{\gamma_\mathrm{min}}\right)^{1-s}\right] \label{Ntot}
\end{equation}

\begin{equation} 
U_e=\left\{ \begin{array}{ll} 
\frac{K}{2-s}\left(m_{e}c^{2}\right)^{2-s}\gamma_\mathrm{max}^{2-s}\left[1-\left(\frac{\gamma_\mathrm{min}}{\gamma_\mathrm{max}}\right)^{2-s}\right]& \textrm{if }1<s<2\\ 
\frac{K}{s-2}\left(m_{e}c^{2}\right)^{2-s}\gamma_\mathrm{min}^{2-s}\left[1-\left(\frac{\gamma_\mathrm{max}}{\gamma_\mathrm{min}}\right)^{2-s}\right] & \textrm{if }s>2\\ 
K\ln\left(\frac{\gamma_\mathrm{max}}{\gamma_\mathrm{min}}\right)  & \textrm{if } s=2
\end{array}\right.  \label{ue}
\end{equation}
Together with the magnetic energy density, $U_{b}=B^{2}/(8\pi)$, we can define the ratio between the magnetic energy density and the energy density of the relativistic particles:
\begin{equation}
\sigma=\frac{U_b}{U_e}.
\end{equation}

\subsection*{Core-shift}
\label{coreshift1}
Assuming that the position of the observed VLBI core is identical with the $(\tau=1)$-surface, \citet{Lobanov:1998p2310} used Eq. ~\ref{tau} to derive the frequency-dependent position of the core, the so-called core-shift. A conical jet geometry was assumed in that work, i.e., $R\propto r$, a decreasing magnetic field, $B=B_1r^{-b}$ and a decreasing particle density, $K=K_1r^{-k}$, where the constants $B_1$ and $K_1$ correspond to the magnetic field and electron normalization coefficient at $1\,\mathrm{pc}$. By inserting these assumptions into Eq. ~\ref{tau} and solving the equation for $r$, one obtain:
\begin{equation}
r\propto \nu^{-1/k_r} \label{coreshift},
\end{equation}
where $k_r=\left[2k+2b\left(3-2\alpha_0\right)-2\right]/\left(5-2\alpha_0\right)$, and $\alpha_0$ is the optically thin spectral index.

Measurements of the core-shift yield estimates of  several physical parameters such as the distance to the central engine and the magnetic field at the core \citep[][]{Lobanov:1998p2310,Hirotani:2005p23,OSullivan:2009p1877,Pushkarev:2012p6264}. The core-shift measure is defined as:
\begin{equation}
\Omega_{r\nu}=4.85\cdot10^{-9}\frac{\Delta r_{\nu_1,\nu_2} D_L\nu_1^{1/k_r}\nu_2^{1/k_r}}{(1+z)^2\left(\nu_2^{1/k_r}-\nu_1^{1/k_r}\right)}\,[\mathrm{pc\cdot GHz}],
\end{equation}

\noindent with $\Delta r_{\nu_1,\nu_2}$ the core shift between the frequencies $\nu_1$ and $\nu_2$ in $\mathrm{mas}$ and $D_l$ the luminosity distance in $\mathrm{pc}$. Following \citet{Hirotani:2005p23}, the magnetic field at $1\,\mathrm{pc}$ is given by:

\begin{eqnarray}
B_1\approx\frac{2\pi m_e^2 c^4}{e^3}\left[\frac{e^2}{m_e c^3}\left(\frac{\Omega_{r\nu}}{r_1\sin\vartheta}\right)^{k_r}\right]^{\frac{5-2\alpha_0}{7-2\alpha_0}}\cdot\left[ \pi C\left(\alpha_0\right)\frac{r_1 m_e c^2}{e^2}\right. \nonumber \\
\left. \frac{-2\alpha_0}{\gamma_\mathrm{min}^{2\alpha_0+1}} \frac{\varphi}{\sin{\vartheta}}K\left(\gamma,\alpha_0\right) \left(\frac{\delta}{1+z}\right)^{\frac{3}{2}-\alpha_0}\right]^{\frac{-2}{7-2\alpha_0}}, \label{b1coreshiftg}
\end{eqnarray}
where $C\left(\alpha_0\right)$ and $K\left(\gamma,\alpha_0\right)$ are defined as:
\begin{eqnarray}
C\left(\alpha_0\right)&=&\frac{3^{1-\alpha_0}}{8}\sqrt{\pi}\hat{\Gamma}\left(\frac{7-2\alpha_0}{4}\right)\hat{\Gamma}\left(\frac{5-6\alpha_0}{12}\right)\hat{\Gamma}\left(\frac{25-6\alpha_0}{12}\right) \nonumber \\ 
&&\left[\hat{\Gamma}\left(\frac{9-2\alpha_0}{4}\right)\right]^{-1} \\
K\left(\gamma,\alpha_0\right)&=&\frac{2\alpha_0+1}{2\alpha_0}\frac{\left[\left(\gamma_\mathrm{max}/\gamma_\mathrm{min}\right)^{2\alpha}-1\right]}{\left[\left(\gamma_\mathrm{max}/\gamma_\mathrm{min}\right)^{2\alpha+1}-1\right]}.
\end{eqnarray}
The distance to the central engine is given by:
\begin{equation}
r_\mathrm{core}(\nu)=\Omega_{r\nu}\sin\vartheta^{-1}\nu^{-1/k_r} \label{b1rcoreg}.
\end{equation}
Assuming a conical jet in equipartition between the magnetic energy density and the kinetic energy density \citep[which implies $k_r=1$, see e.g.][]{Lobanov:1998p2310}, and a spectral index of $\alpha_0=-0.5$, the equations above simplify to the following relations:
\begin{eqnarray}
&B_1&\approx 0.025\left[\frac{\Omega_{r\nu}^{3}(1+z)^{2}}{\varphi \delta^2\sin\vartheta}\right]^{1/4}, \label{b1coreshift} \\
&r_\mathrm{core}(\nu)&=\Omega_{r\nu}\sin\vartheta^{-1}\nu^{-1}. \label{b1rcore}
\end{eqnarray}
The particle density $N_1$ can be written for both cases i) $k_r=1$ and $\alpha_0=-0.5$ and ii) $k_r\neq1$ and $\alpha_0\neq-0.5$, as:
\begin{equation}
N_1=\frac{K\left(\gamma,\alpha_0\right)}{8\pi m_e c^2}\gamma_\mathrm{min}^{-1}B_1^2\, [\mathrm{cm^{-3}}].
\end{equation}
Taking $k_r=1$, $\alpha_0=-0.5$ and fraction of $10^3$ and $10^5$ between the upper and lower electron Lorentz factor, the equation above can be written as:
\begin{eqnarray}
N_1&\approx&7\cdot10^{3}\gamma_\mathrm{min}^{-1}B_1^2\,\quad\,{\gamma_\mathrm{max}}/{\gamma_\mathrm{min}}=10^3\\
N_1&\approx&4.2\cdot10^{3}\gamma_\mathrm{min}^{-1}B_1^2\,\quad\,{\gamma_\mathrm{max}}/{\gamma_\mathrm{min}}=10^5
\end{eqnarray}

\section*{Appendix B}
We performed several tests to investigate the influence of the non-identical uv-range on the spectral indices derived from images at different wavelengths and therefore with different projected baselines in wavelength units. For extracting the spectral parameters we use both a power-law fit $S_\nu\propto \nu^{+\alpha}$ and the approximation of the synchrotron-self-absorbed spectrum (see Eq. \ref{snuapprox}). Table~\ref{uvrange} gives the average uv-ranges for the different frequencies. The image parameters, i.e., convolving beam size and pixel size, used are presented in Table \ref{specanapara}.

\begin{table}[h!]
\caption{VLBA $(u,v)$-ranges for different frequencies in our experiments (see text and Table \ref{specanapara} for further details).}  
\label{uvrange}
\centering  
\begin{tabular}{c c c}
\hline\hline
$\nu$ &$r_{uv,\mathrm{min}}$	&$r_{uv,\mathrm{max}}$\\
$[\mathrm{G}]$ & $\mathrm{M}\lambda$ & $\mathrm{M}\lambda$ \\
\hline
2	&	1	&	66\\
5	&	3	&	144\\
8	&	5	&	250\\
15	&	9	&	450\\
22	&	14	&	640\\
43	&	27	&	1240\\
86	&	66	&	1760\\
\hline
\end{tabular}
\end{table} 

\begin{table}[h!]
\caption{Image parameters used for the spectral analysis}  
\label{specanapara}
\centering  
\begin{tabular}{@{}c c c c c c @{}}
\hline\hline
\small
region & r & $\nu$ & beam & ps & uv-range\\
 & $[\mathrm{mas}]$ & $[\mathrm{GHz}]$& $[\mathrm{mas}]$ & $[\mathrm{mas}]$ & $\mathrm{M}\lambda$ \\
\hline
C			& 0--1	&	15--43	&	0.95$\times$0.33,$-13^\circ$	&0.03	& 27--450\\
C$^{\mathrm{a}}$		& 0--1	&	15--86	& 	0.95$\times$0.33,$-13^\circ$	&0.03	& 66--450\\
D$^{\mathrm{b}}$		& 1--4	&	5--22	&	1.33$\times$0.52,$-9^\circ$	&0.04	& 14--144\\
B			& 4--8	&	5--15	&	2.32$\times$0.97,$-7^\circ$	&0.10	&9--144\\
A2			& 8--14	&	5--15	&	3.65$\times$1.52,$-8^\circ$	&0.15	&9--144\\
A1$^{\mathrm{c}}$		&14--20	&	5--8		&	3.65$\times$1.52,$-8^\circ$	&0.15	&9--144\\
B+A$^{\mathrm{d}}$	& 4--20	&	5--43	&	3.65$\times$1.52,$-8^\circ$	&0.10	&27--144\\
\hline
\multicolumn{6}{l}{$^a$ used to extract turnover values for 2005.39}\\
\multicolumn{6}{l}{$^b$ used to extract turnover values}\\
\multicolumn{6}{l}{$^c$ no fitting, spectral index computed between two frequencies}\\
\multicolumn{6}{l}{$^d$ test of influence of frequency range on $\alpha$}\\
\end{tabular}
\end{table}

\begin{figure*} 
\centering 
\includegraphics[width=17cm]{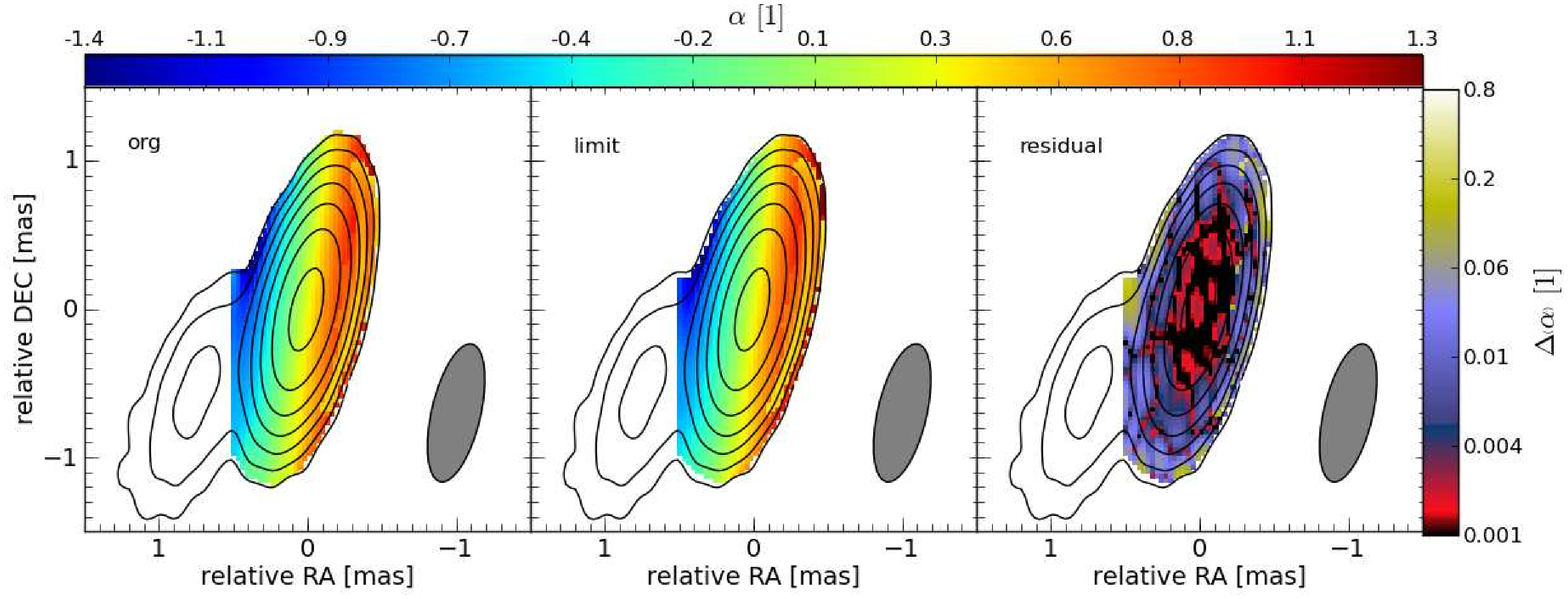} 
\caption{Influence of the uv-range on the 2D distribution of the spectral index, $\alpha$  $(S_\nu\propto \nu^{\alpha})$ for region C $(r<1\,\mathrm{mas})$ using a beam size of  $0.95\times0.33\,\mathrm{mas}$ with a P.A. of $-13^\circ$ and a pixel size of $0.03\,\mathrm{mas}$. The left panel shows the spectral index for a un-limited uv-range, the middle panel for limited uv-range and the right panel the residuals between them. The contours correspond to the $43\,\mathrm{GHz}$ VLBI observations, where the lowest contour is plotted at 10$\times$ the rms value and increase by factors of $2$.} 
\label{uvtest1} 
\end{figure*}

In Fig.~\ref{uvtest1} we show the 2D distribution of the spectral index for region C using a frequency range from $15\,\mathrm{GHz}$ to $43\,\mathrm{GHz}$. The left panel shows the distribution for a non-identical uv-range and the middle panel the distribution for a limited uv-range, here from $27\,\mathrm{M}\lambda$ to $450\,\mathrm{M}\lambda$. The difference in $\alpha$ between the two maps shows that the central region is only marginally affected by the used uv-range ($\Delta\left(\alpha\right)<10^{-3}$) and the discrepancies increase with distance from the centre. The largest discrepancy is found in the edges of the distribution where $\Delta\left(\alpha\right)>0.1$.

\begin{figure*} 
\centering 
\includegraphics[width=17cm]{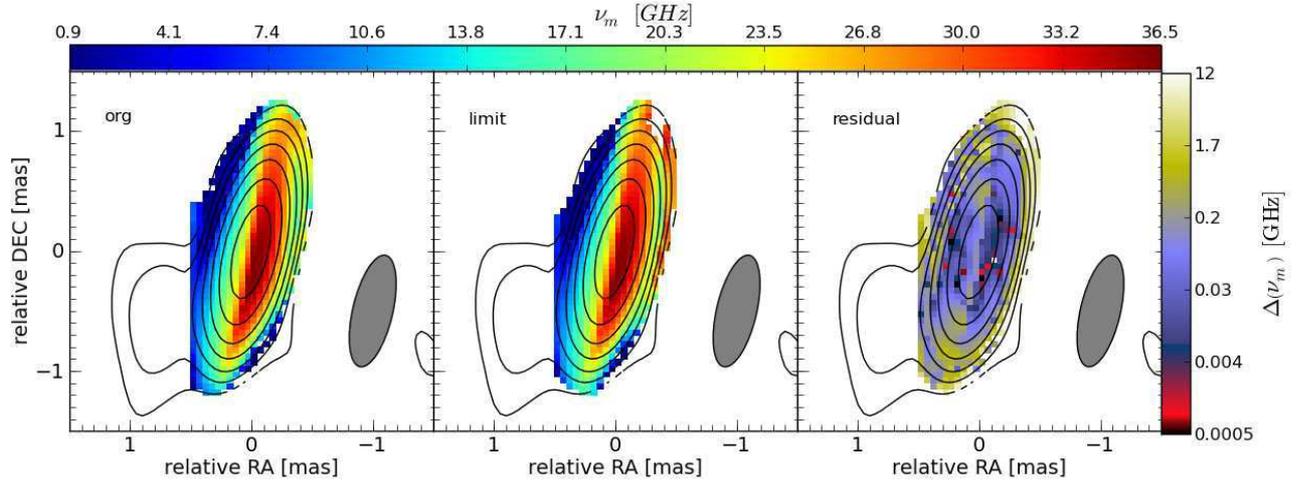} 
\caption{Influence of the uv-range on the 2D distribution of the turnover frequency, $\nu_m$  for region C $(r<1\,\mathrm{mas})$ using a beam size of  $0.95\times0.33\,\mathrm{mas}$ with a P.A. of $-13^\circ$ and a pixel size of $0.03\,\mathrm{mas}$. The left panel shows the turnover frequency for a un-limited uv-range, the middle panel for limited uv-range and the right panel the residuals between them. The contours correspond to the $86\,\mathrm{GHz}$ VLBI observations, where the lowest contour is plotted at 10$\times$ the rms value and increase by factors of $2$.} 
\label{uvtest1vm} 
\end{figure*}

\begin{figure*} 
\centering 
\includegraphics[width=17cm]{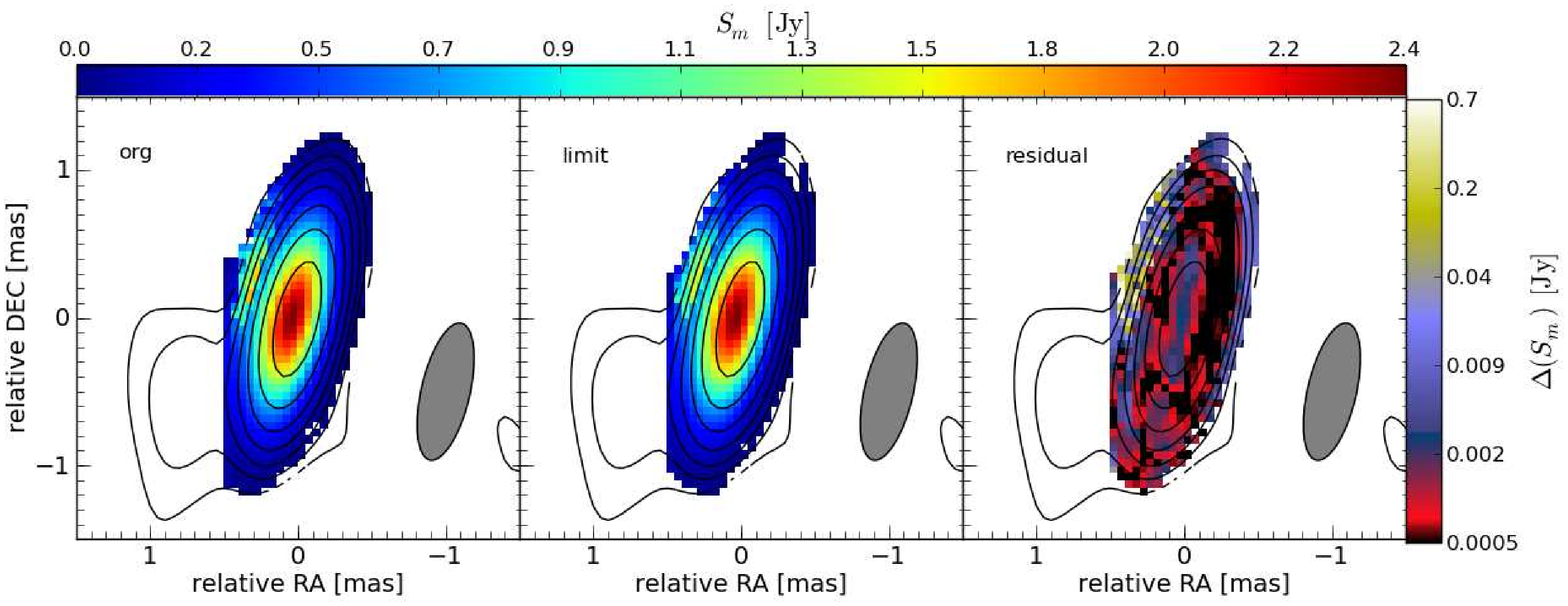} 
\caption{Same as Fig. \ref{uvtest1vm} for the turnover flux density, $S_m$.} 
\label{uvtest1sm} 
\end{figure*}

\begin{figure*} 
\centering 
\includegraphics[width=17cm]{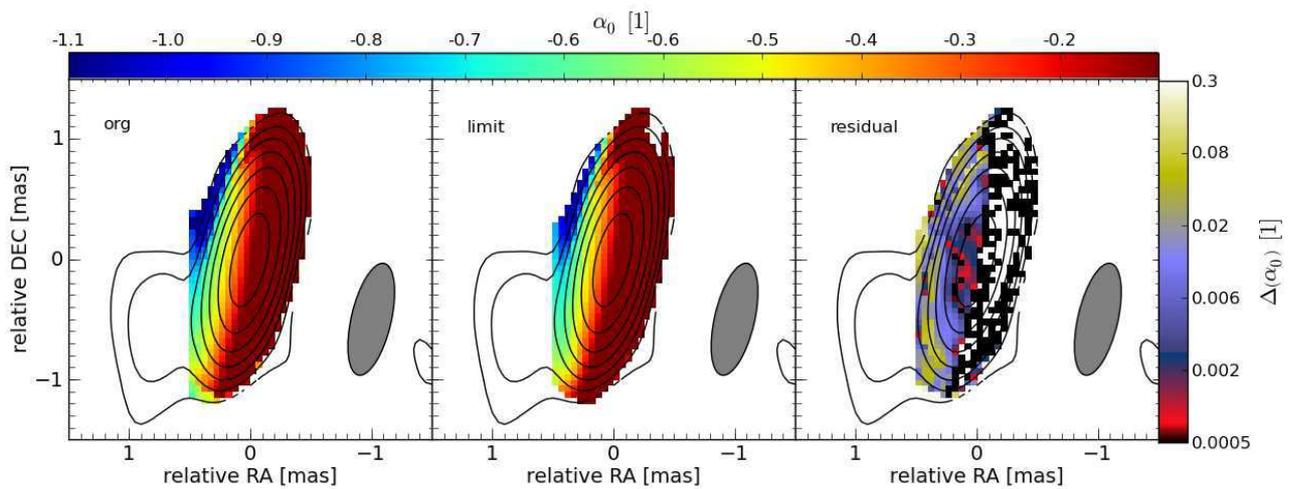} 
\caption{Same as Fig. \ref{uvtest1vm} for the optically thin spectral index, $\alpha_0$} 
\label{uvtest1a0} 
\end{figure*}

For the May 2005 observations of CTA\,102 we could derive the spectral turnover, i.e, $\nu_m$ and $S_m$, using a frequency range from $8\,\mathrm{GHz}$ to $86\,\mathrm{GHz}$. The uv-range is set at $(66-250)\,\mathrm{M}\lambda$ and the difference in the spectral values as compared to the unlimited uv-range is shown in Fig. \ref{uvtest1vm} --\ref{uvtest1a0}. The large frequency range and the short uv-range lead to strong variation of the spectral parameters at the edges of the distribution up to $70\%$ for $\alpha$, $60\%$ for $\nu_m$ and $30\%$ for $S_m$. Despite these large discrepancies at the edges, the values along the jet axis show variations up to $10\%$.

\begin{figure*} 
\centering 
\includegraphics[width=17cm]{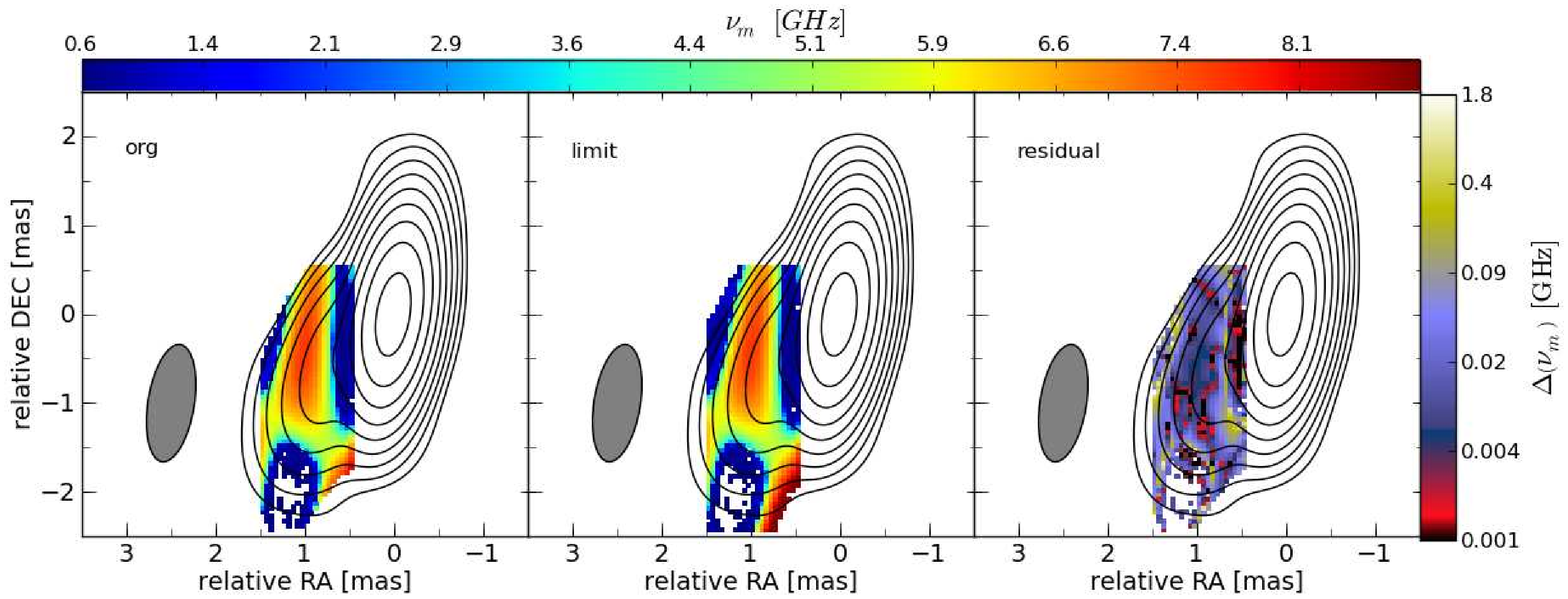} 
\caption{Influence of the uv-range on the 2D distribution of the turnover frequency, $\nu_m$  for region D $(1\,\mathrm{mas}<r<4\,\mathrm{mas})$ using a beam size of  $1.33\times0.52\,\mathrm{mas}$ with a P.A. of $-7^\circ$ and a pixel size of $0.04\,\mathrm{mas}$. The left panel shows the turnover frequency for a un-limited uv-range, the middle panel for limited uv-range and the right panel the residuals between them. The contours correspond to the $43\,\mathrm{GHz}$ VLBI observations, where the lowest contour is plotted at 10$\times$ the rms value and increase by factors of $2$.} 
\label{uvtest2vm} 
\end{figure*}

\begin{figure*} 
\centering 
\includegraphics[width=17cm]{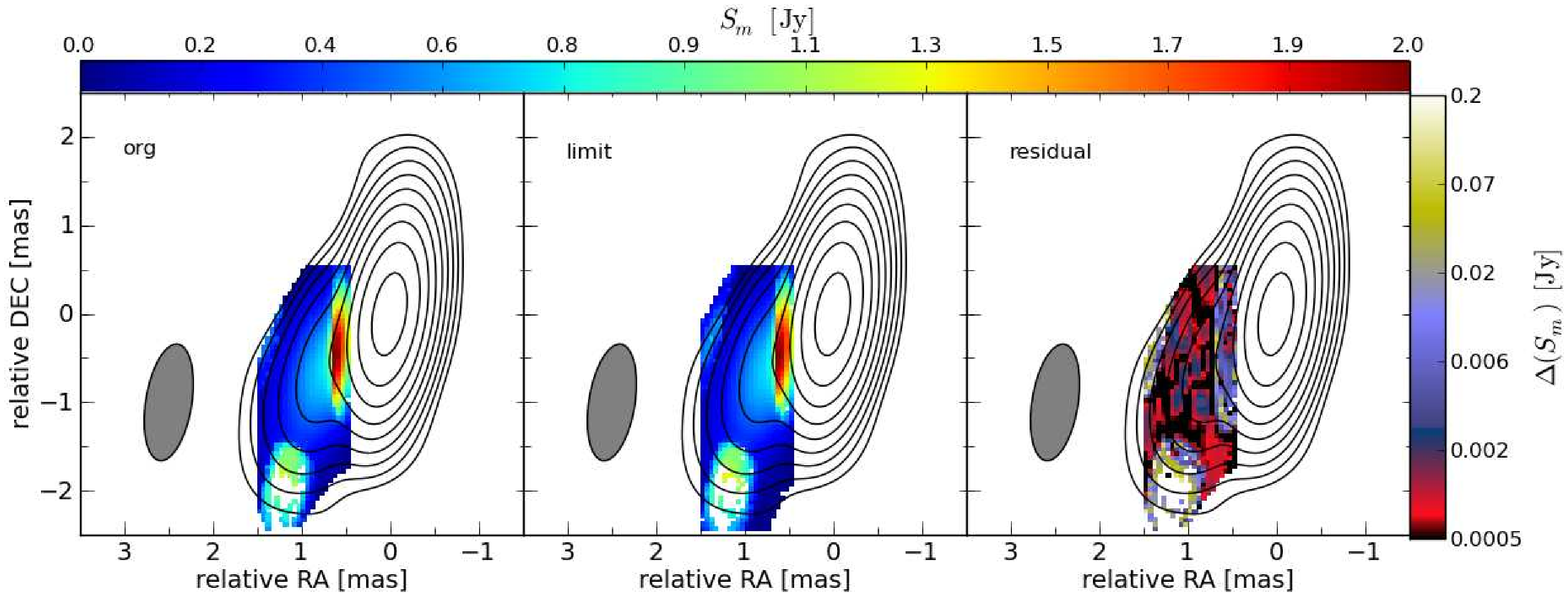} 
\caption{Same as Fig. \ref{uvtest2vm} for the turnover flux density, $S_m$.} 
\label{uvtest2sm} 
\end{figure*}

\begin{figure*} 
\centering 
\includegraphics[width=17cm]{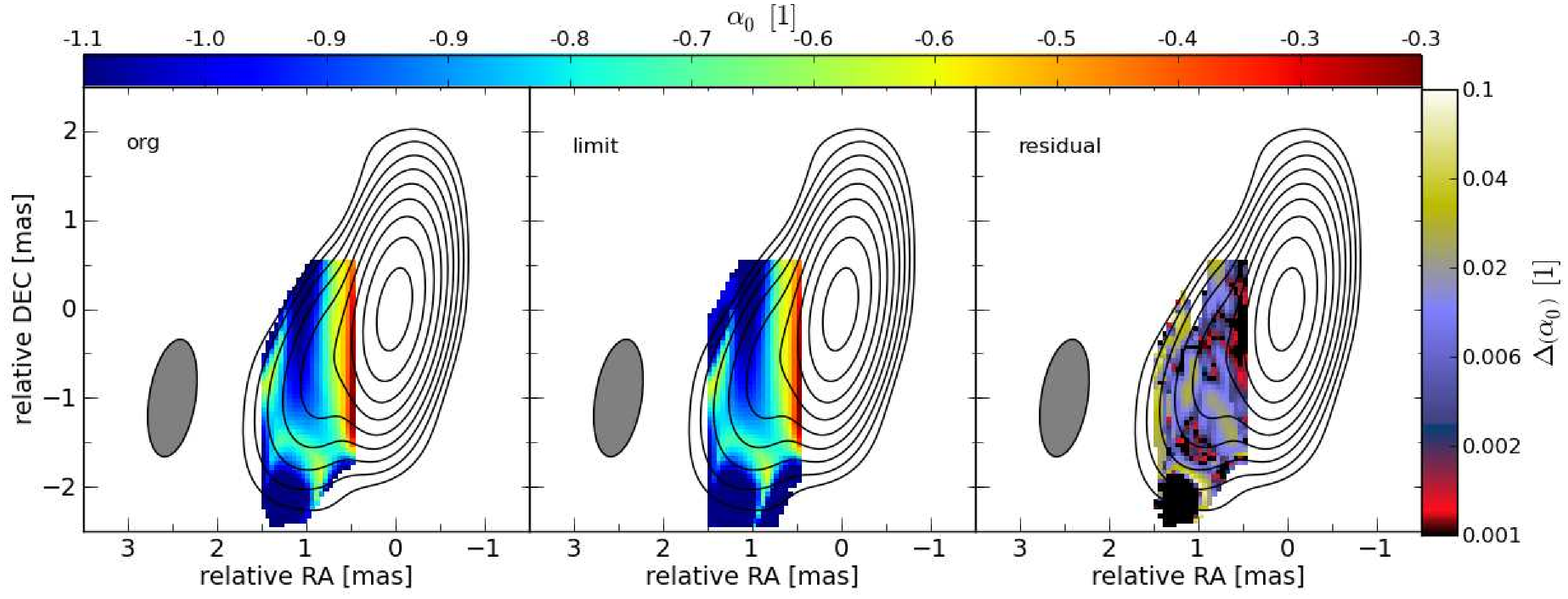} 
\caption{Same as Fig. \ref{uvtest2a0} for the optically thin spectral index, $\alpha_0$.} 
\label{uvtest2a0} 
\end{figure*}

As mentioned in Sect.~\ref{regionD} for region D, the turnover frequency lies within our frequency range and we can derive the turnover frequency, the turnover flux density, and the optically thin spectral index, $\alpha_0$. For the analysis we used a beam size of $(1.33\times0.52)\,\mathrm{mas}$ at a P.A. of $-9^\circ$ and a frequency range from $5\,\mathrm{GHz}$ to $22\,\mathrm{GHz}$. We limited the uv-range to $14\,\mathrm{M}\lambda$ to $144\,\mathrm{M}\lambda$ and compared the results of the spectral analysis to the outcome of the un-limited uv-range (see Fig. \ref{uvtest2vm} -- \ref{uvtest2a0}). In general, the discrepancies between the two methods are largest at the edges of the distribution. However, in contrast to the distribution of $\alpha$ there are also regions in the centre of the distribution that show large differences between maps.\\

The difference between the turnover flux density derived by using limited and full uv-ranges is at most $10\%$ and a similar value is obtained for the optically thin spectral index. For the turnover frequency, the differences are much larger and can reach a $50\%$ at the edges of the distribution. The limiting of the uv-range affects less the distribution of the turnover flux density and more the turnover frequency and optically thin spectral index.

\begin{figure*} 
\centering 
\includegraphics[width=17cm]{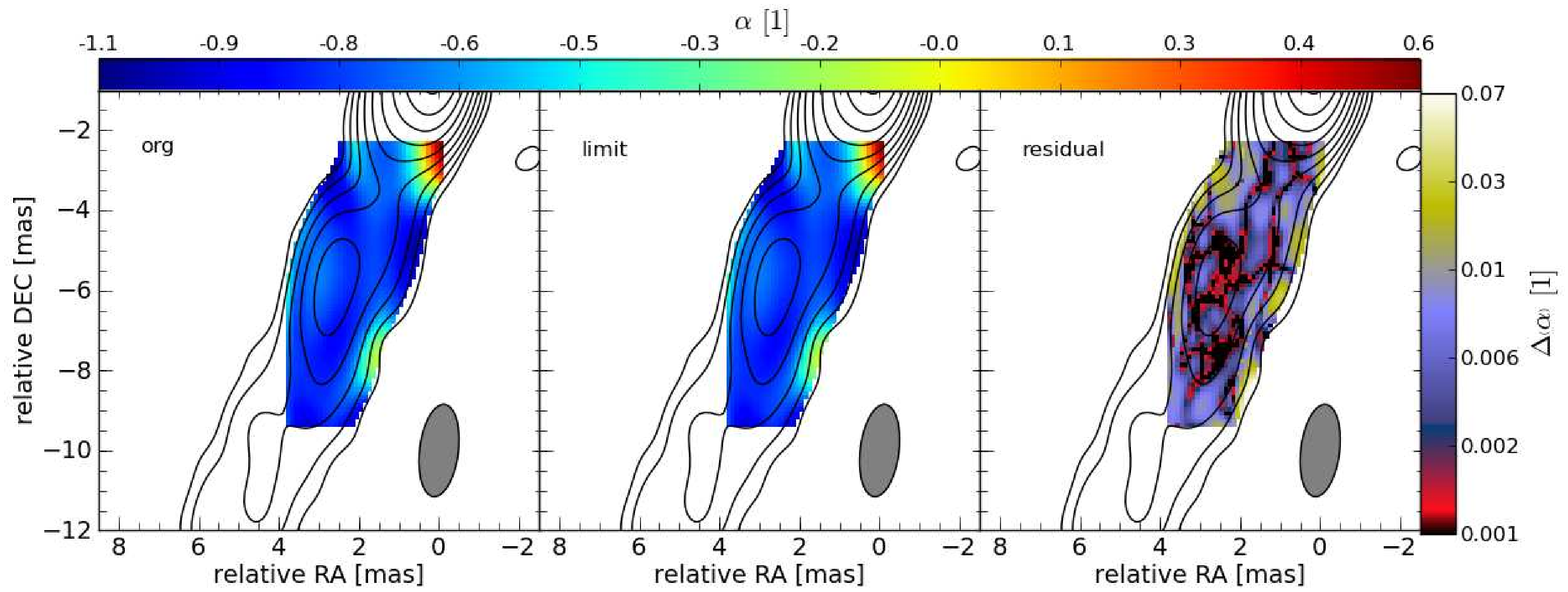} 
\caption{Influence of the uv-range on the 2D distribution of the spectral index, $\alpha$  $(S_\nu\propto \nu^{\alpha})$ for region B $(4\,\mathrm{mas}<r<8\,\mathrm{mas})$ using a beam size of  $2.32\times0.97\,\mathrm{mas}$ with a P.A. of $-7^\circ$ and a pixel size of $0.10\,\mathrm{mas}$. The left panel shows the spectral index for a un-limited uv-range, the middle panel for limited uv-range and the right panel the residuals between them. The contours correspond to the $15\,\mathrm{GHz}$ VLBI observations, where the lowest contour is plotted at 10$\times$ the rms value and increase by factors of $2$.} 
\label{uvtestB} 
\end{figure*}

For region B ($4\,\mathrm{mas}<r<8\,\mathrm{mas}$) we used a frequency range from $5\,\mathrm{GHz}$ to $15\,\mathrm{GHz}$ and the difference in the uv-ranges leads only to small variations in the spectral index. As for region C, the largest differences in $\alpha$ can be found at the edges of the distribution. It is worth to mention that those differences are less than $10\%$ (see Fig. \ref{uvtestB}).

\begin{figure*} 
\centering 
\includegraphics[width=17cm]{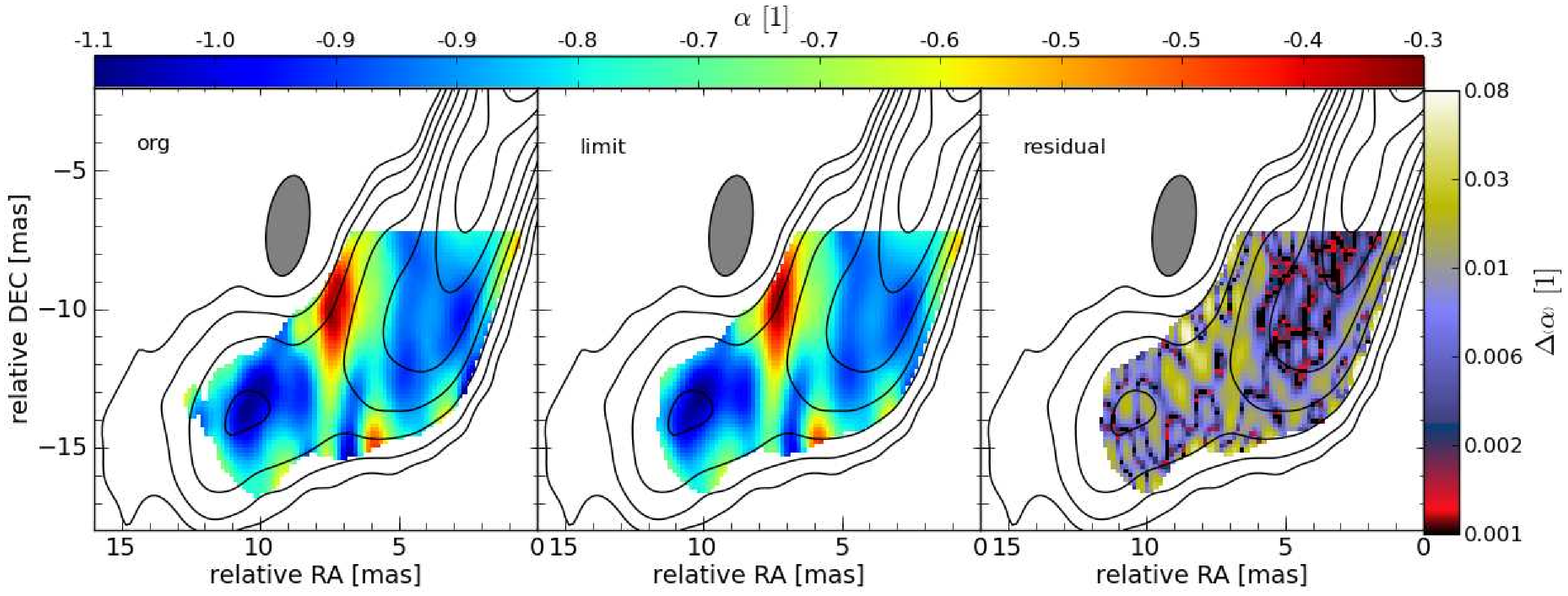} 
\caption{Influence of the uv-range on the 2D distribution of the spectral index, $\alpha$  $(S_\nu\propto \nu^{\alpha})$ for region A and region B $(4\,\mathrm{mas}<r<21\,\mathrm{mas})$ using a beam size of  $3.65\times1.52\,\mathrm{mas}$ with a P.A. of $-8^\circ$ and a pixel size of $0.10\,\mathrm{mas}$. The left panel shows the spectral index for a un-limited uv-range, the middle panel for limited uv-range and the right panel the residuals between them. The contours correspond to the $8\,\mathrm{GHz}$ VLBI observations, where the lowest contour is plotted at 10$\times$ the rms value and increase by factors of $2$.} 
\label{uvtestA} 
\end{figure*}

Besides the influence of the uv-range, we tested the impact of the frequency range on the obtained spectral parameters. Therefore, we enlarged the frequency range from $5\,\mathrm{GHz}$ to $43\,\mathrm{GHz}$ for regions B and A. The result for the spectral index is presented in Fig.~\ref{uvtestA}. The difference in $\alpha$ increased by a factor of 3 as compared to values calculated using a frequency range from $5\,\mathrm{GHz}$ to $15\,\mathrm{GHz}$ (see Fig. \ref{uvtestB}) and regions of increased residuals extend into the central distribution of $\alpha$.

In sum, our test show that:
\newline i) if the frequency range is not larger than a factor 4, the difference in the uv-radii influences mainly the edges of the distribution.
\newline ii) for the calculation of the turnover values the uv-range affects the overall distribution of the spectral parameters significantly, especially at the edges of the analysed jet region. Therefore, the uv-range should be in general be matched.
\clearpage

\section*{Appendix C}
Here we present the 2D distribution of the spectral index, $\alpha$ ($S_\nu\propto\nu^{\alpha}$) or the turnover values ($\nu_m$, $S_m$ and $\alpha_0$) for regions C, D, B, and A. In Table \ref{specanapara} we summarize the used image parameters for the spectral analysis.

\begin{figure*} 
\centering 
\includegraphics[width=17cm]{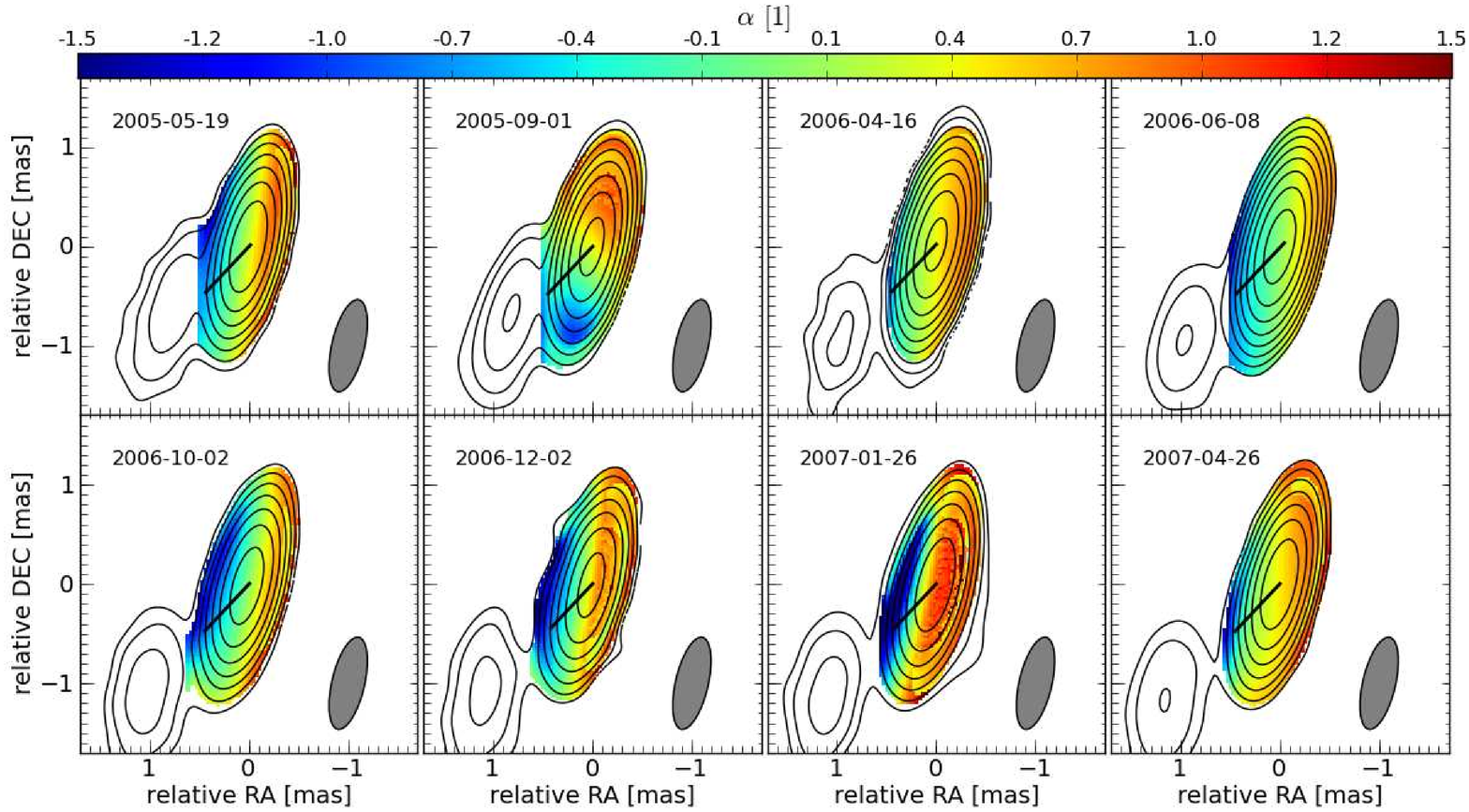} 
\caption{2D distribution of the spectral index, $\alpha$  $(S_\nu\propto \nu^{\alpha})$ for region C $(r<1\,\mathrm{mas})$ using a beam size of  $0.95\times0.33\,\mathrm{mas}$ with a P.A. of $-13^\circ$ and a pixel size of $0.03\,\mathrm{mas}$. The color map in each panel shows for a given epoch (indicated in the top right corner) the distribution of $\alpha$ and the contours correspond to the $43\,\mathrm{GHz}$ VLBI observations, where the lowest contour is plotted at 10$\times$ the rms value and increase by factors of $2$.} 
\label{Calpha2d} 
\end{figure*}

\begin{figure*} 
\centering 
\includegraphics[width=17cm]{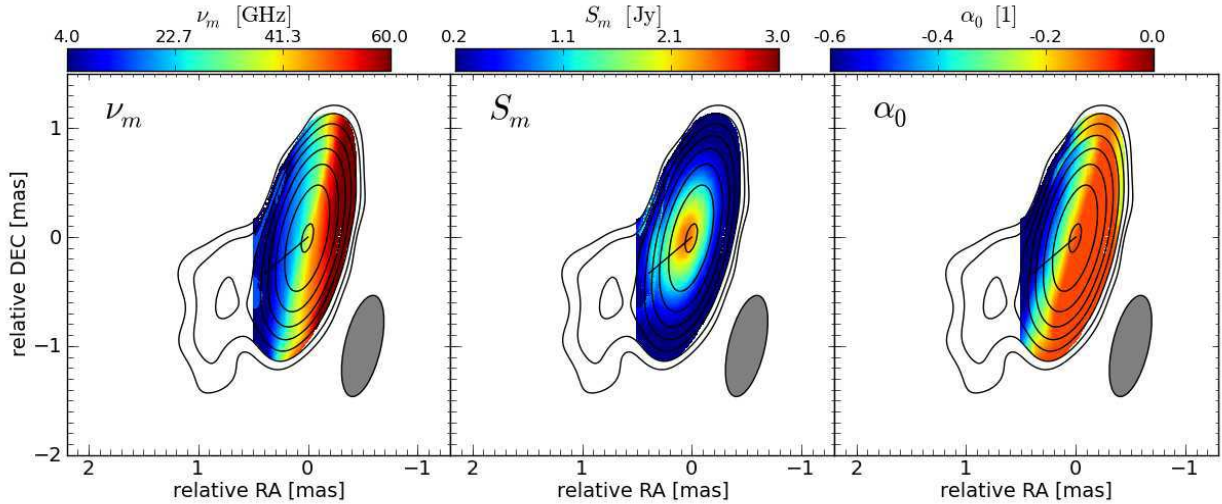} 
\caption{2D distribution of the turnover frequency, $\nu_m$, the turnover flux density, $S_m$, and the optically thin spectral index, $\alpha_0$  for region C $(r<1\,\mathrm{mas})$ for the May 2005 observations using a beam size of  $0.95\times0.33\,\mathrm{mas}$ with a P.A. of $-13^\circ$ and a pixel size of $0.03\,\mathrm{mas}$. The contours correspond to the $86\,\mathrm{GHz}$ VLBI observations, where the lowest contour is plotted at 5$\times$ the rms value and increase by factors of $2$.} 
\label{Calpha2d} 
\end{figure*}
\clearpage

\begin{figure*} 
\centering 
\includegraphics[width=17cm]{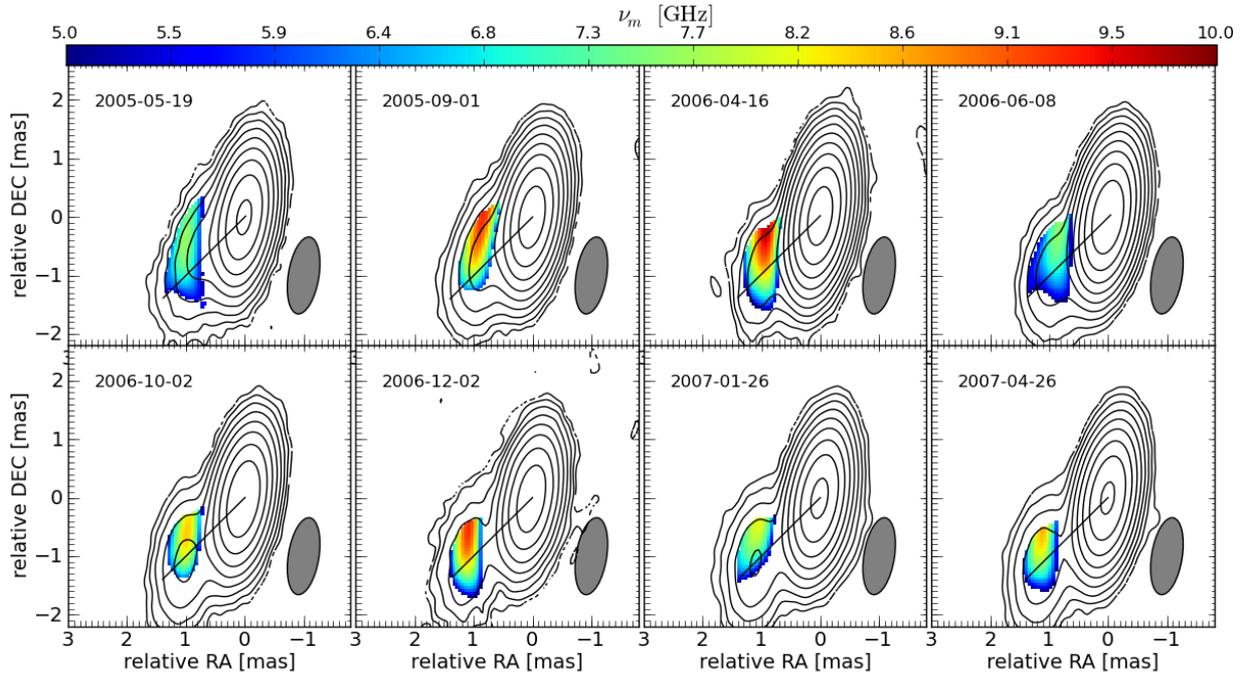} 
\caption{2D distribution of the turnover frequency, $\nu_m$  for region D $(1\,\mathrm{mas}<r<4\,\mathrm{mas})$ using a beam size of  $1.33\times0.52\,\mathrm{mas}$ with a P.A. of $-7^\circ$ and a pixel size of $0.04\,\mathrm{mas}$. The color map in each panel shows for a given epoch (indicated in the top right corner) the distribution of $\nu_m$ and the contours correspond to the $43\,\mathrm{GHz}$ VLBI observations, where the lowest contour is plotted at 5$\times$ the rms value and increase by factors of $2$.} 
\label{Dvm2d} 
\end{figure*}

\begin{figure*} 
\centering 
\includegraphics[width=17cm]{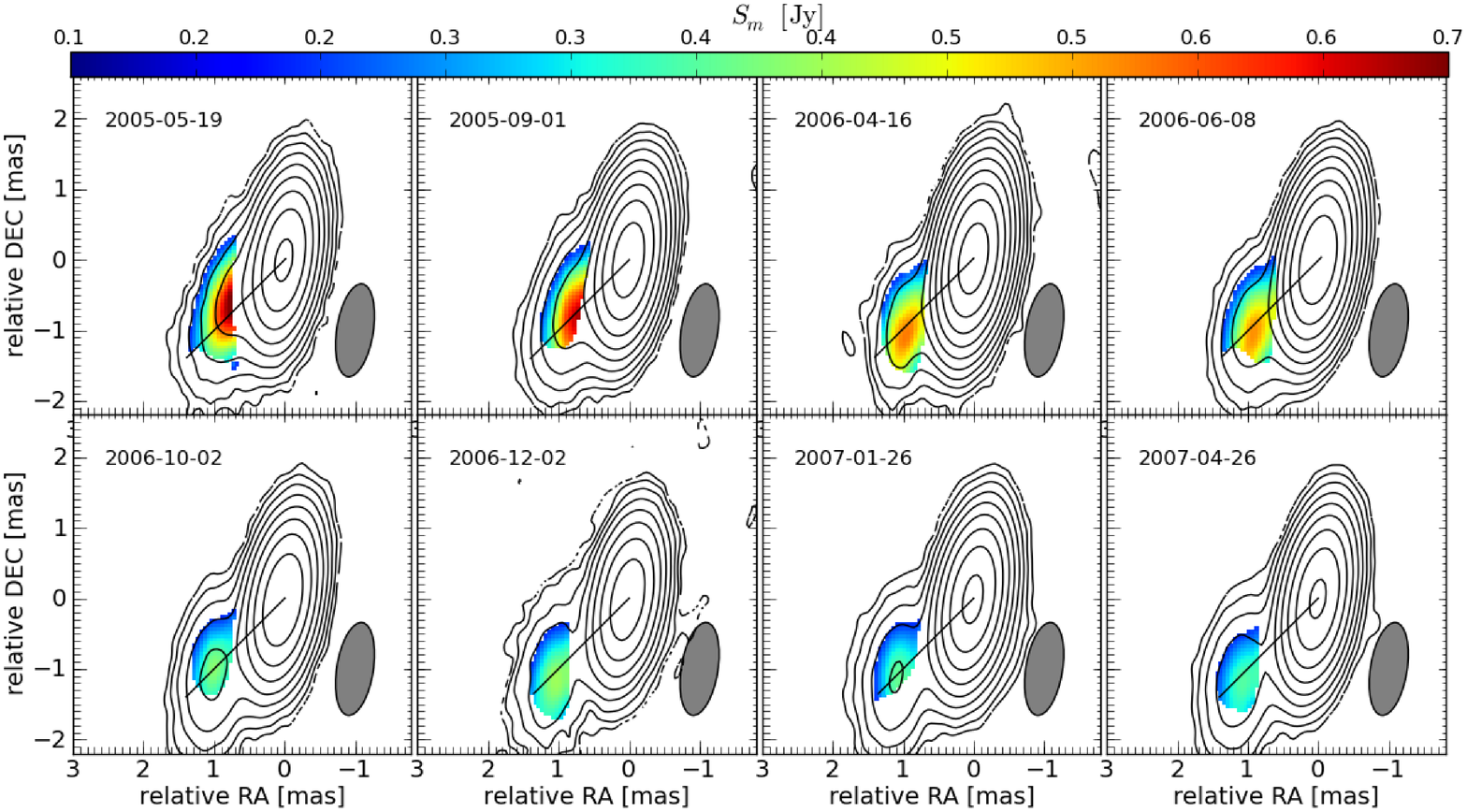} 
\caption{Same as Fig. \ref{Dvm2d} for the turnover flux density, $S_m$.} 
\label{Dsm2d} 
\end{figure*}

\clearpage

\begin{figure*} [h!]
\centering 
\includegraphics[width=17cm]{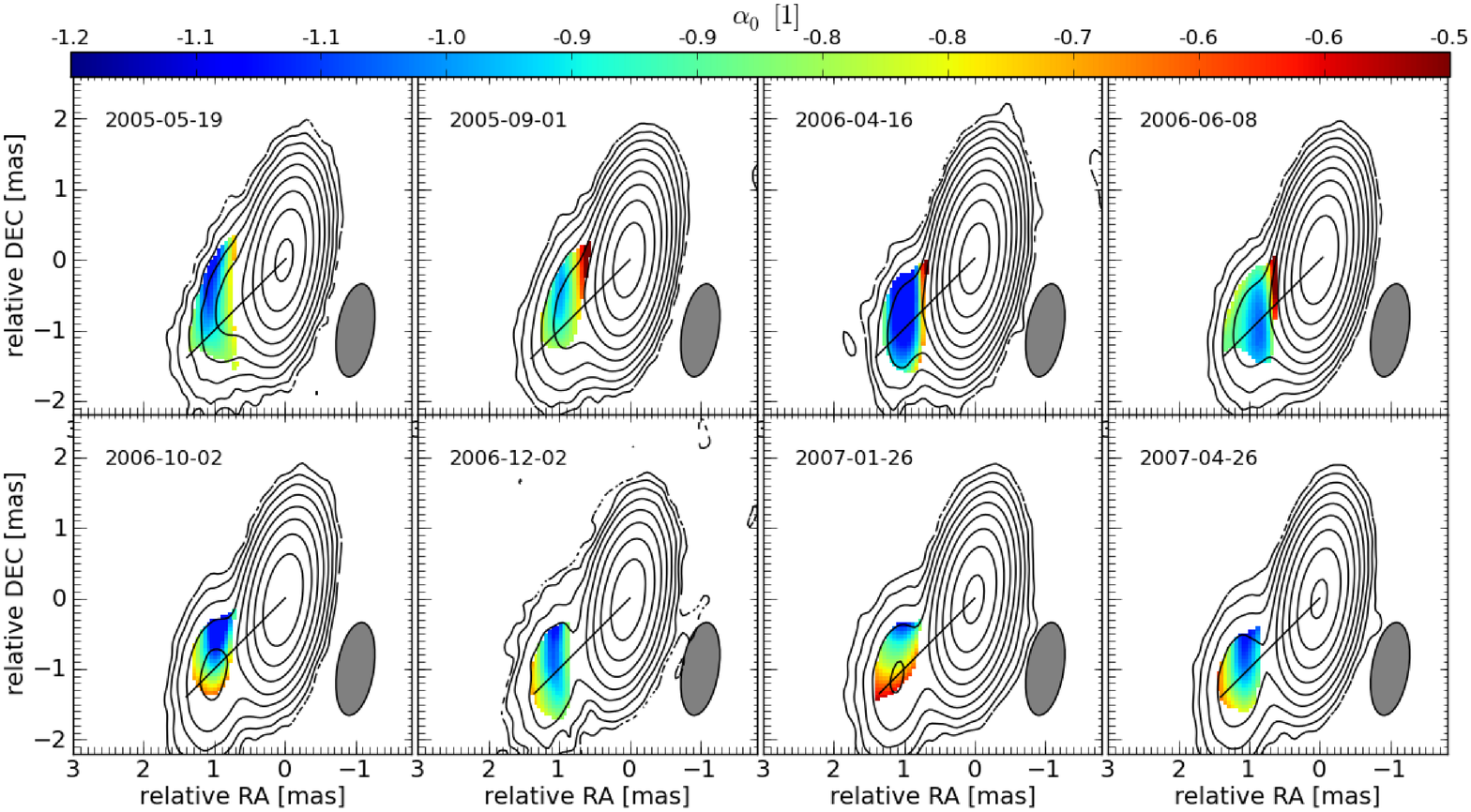} 
\caption{Same as Fig. \ref{Dvm2d} for the optically thin spectral index, $\alpha_0$.} 
\label{Dalpha2d} 
\end{figure*}

\begin{figure*} [h!]
\centering 
\includegraphics[width=17cm]{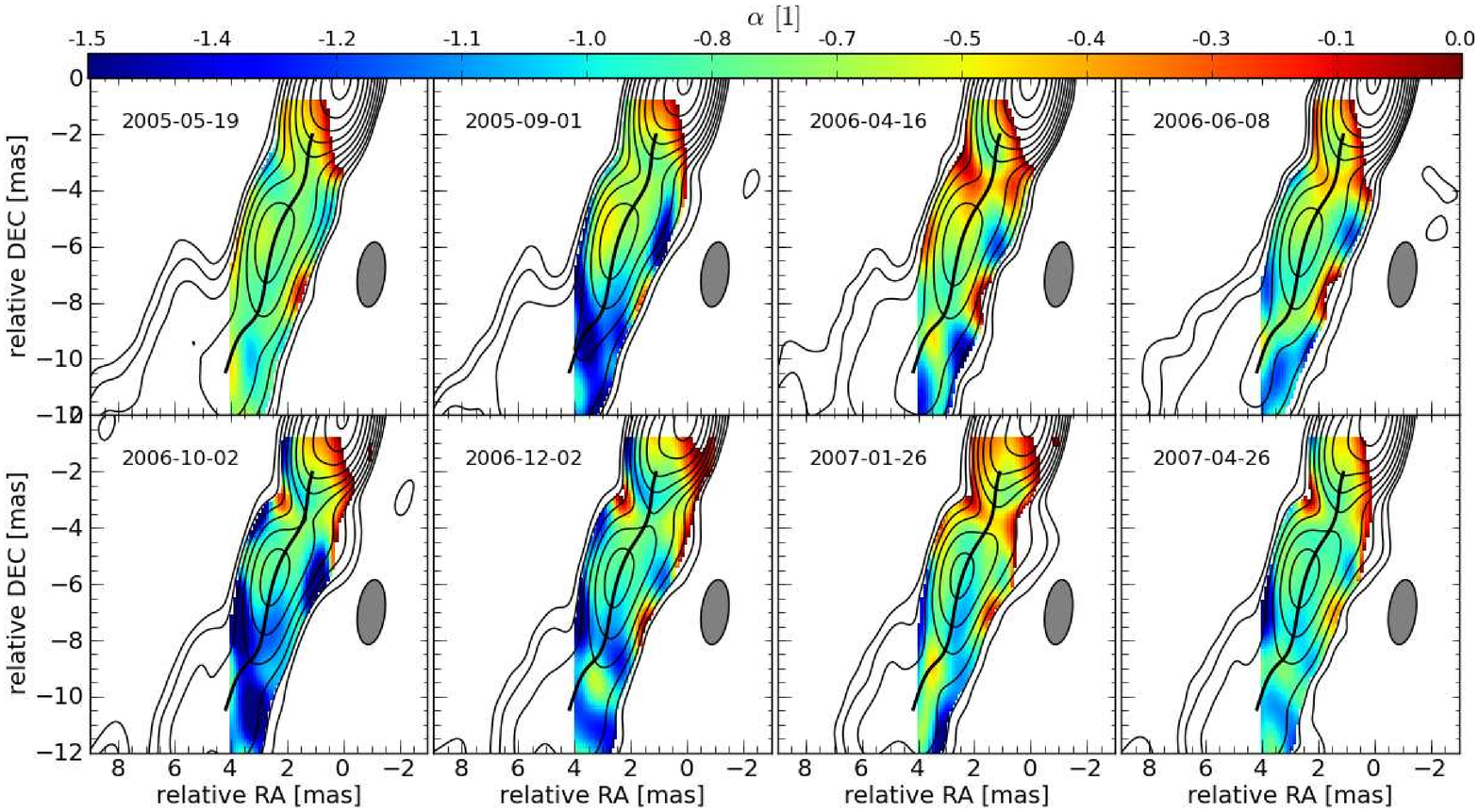} 
\caption{2D distribution of the spectral index, $\alpha$  $(S_\nu\propto \nu^{\alpha})$ for region B $(4\,\mathrm{mas}<r<8\,\mathrm{mas})$ using a beam size of  $2.32\times0.07\,\mathrm{mas}$ with a P.A. of $-7^\circ$ and a pixel size of $0.1\,\mathrm{mas}$. The color map in each panel shows for a given epoch (indicated in the top right corner) the distribution of $\alpha$ and the contours correspond to the $15\,\mathrm{GHz}$ VLBI observations, where the lowest contour is plotted at 5$\times$ the rms value and increase by factors of $2$.} 
\label{Balpha2d} 
\end{figure*}

\clearpage

\begin{figure*} [h!]
\centering 
\includegraphics[width=17cm]{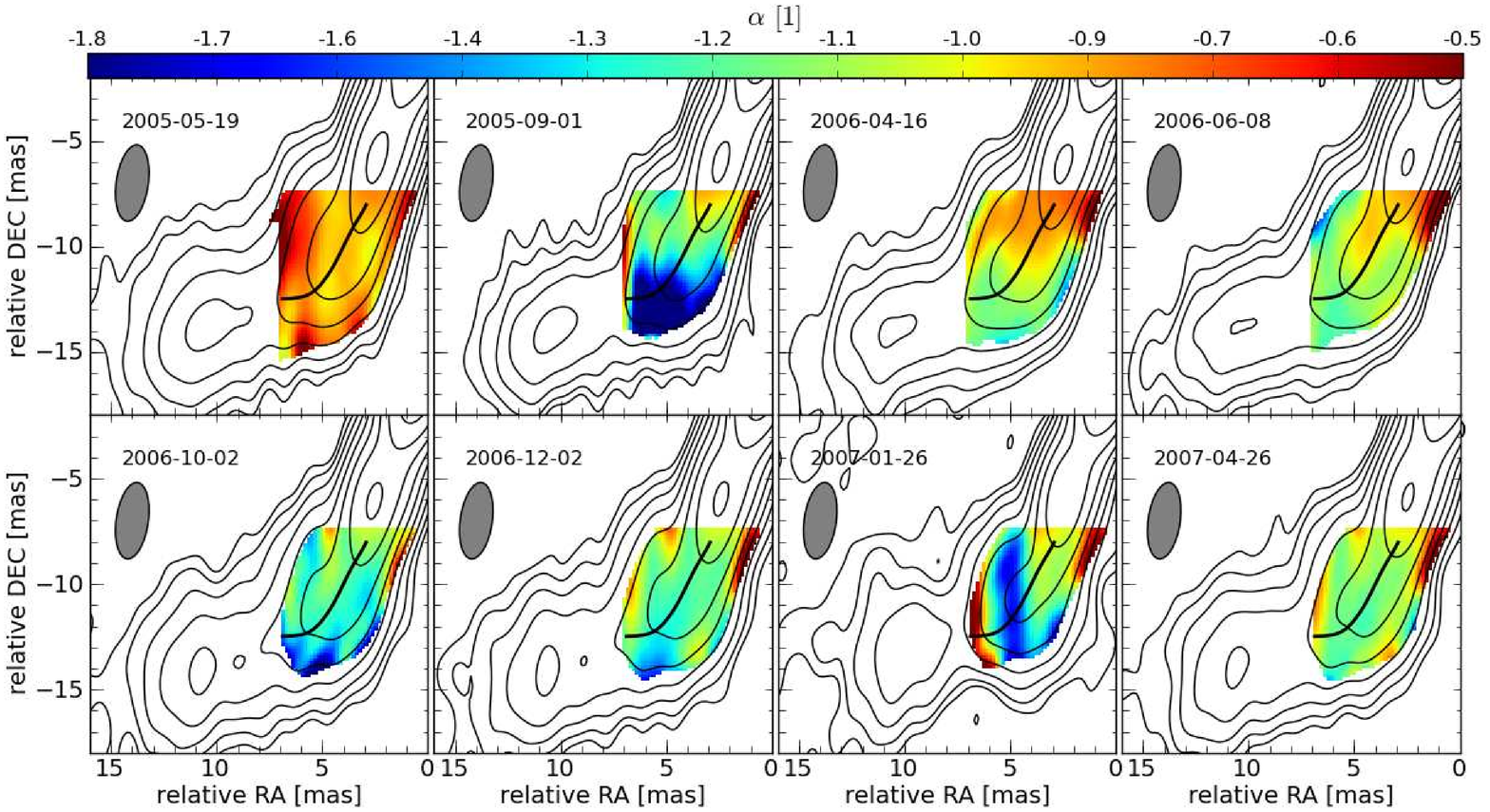} 
\caption{2D distribution of the spectral index, $\alpha$  $(S_\nu\propto \nu^{\alpha})$ for region A $(8\,\mathrm{mas}<r<14\,\mathrm{mas})$ using a beam size of  $3.65\times1.52\,\mathrm{mas}$ with a P.A. of $-8^\circ$ and a pixel size of $0.15\,\mathrm{mas}$. The color map in each panel shows for a given epoch (indicated in the top right corner) the distribution of $\alpha$ and the contours correspond to the $15\,\mathrm{GHz}$ VLBI observations, where the lowest contour is plotted at 10$\times$ the rms value and increase by factors of $2$.} 
\label{A1alpha2d} 
\end{figure*}

\clearpage
\newpage
\section*{Appendix D}
\begin{figure}[h!]
\resizebox{\hsize}{!}{\includegraphics{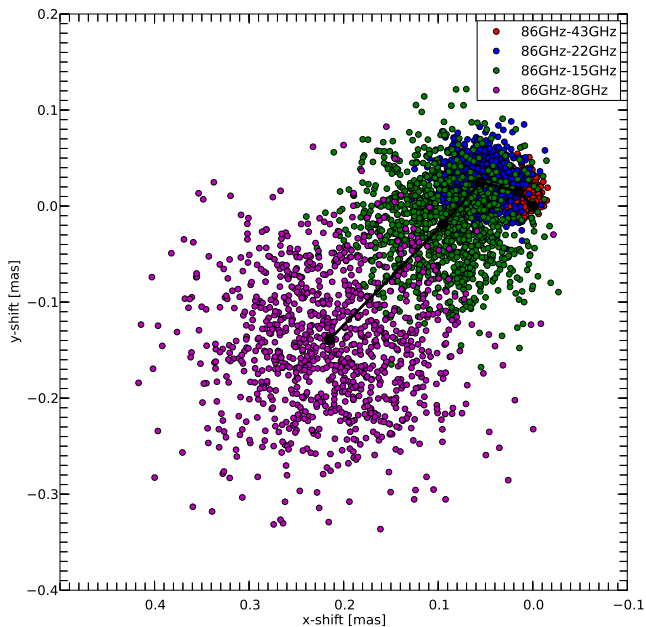}} 
\caption{Calculated random shifts obtained from a normal distribution using the initial shift value as mean and the uncertainty as standard deviation. Different colors correspond to different frequency pairs (see plot legend) and black hexagons indicate the initial shift position. For more details see text.} 
\label{randomshift} 
\end{figure}

An accurate estimate of the uncertainties of the spectral parameters determined in Sect.~\ref{resspecana}, namely, $\alpha_0$, $\alpha$, $S_m$, $\nu_m$, $B$, and $K$ from Eq. \ref{bfield} - \ref{knorm} has to take into account the flux density uncertainties on the individual pixels and the uncertainties caused by the image alignment. We address those uncertainties by using the Monte Carlo technique.\\
We assume that the uncertainties on the obtained image shift are of the order of the used pixel size (see Sect. \ref{align}). Assuming additionally a normal distribution for the scatter of the image shifts, we computed $10^4$ random image shifts and perform for each shift value a spectral analysis (see Sect. \ref{spectralana}). Figure \ref{randomshift} shows the distribution of the image shifts using 1000 random shifts for the May 2005 observations relative to the $86\,\mathrm{GHz}$ image. The different colors correspond to the absolute shifts between the reference VLBI map (here $86\,\mathrm{GHz}$) and the other VLBI maps included in the spectral analysis. The initial shift positions for each frequency are indicated by the hexagon symbol. 

\begin{figure}[h!]
\resizebox{\hsize}{!}{\includegraphics{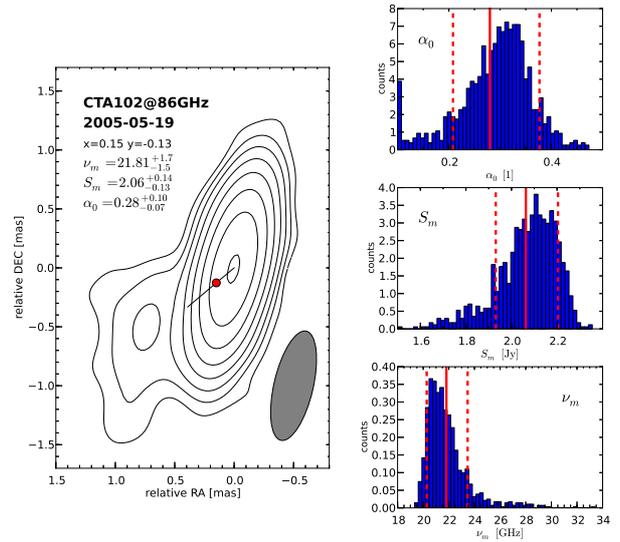}} 
\caption{Result of the Monte Carlo Simulation for the May 2005 observations at the position $x=0.15\,\mathrm{mas}$, $y=-0.13\,\mathrm{mas}$ (relative to the brightness peak). The left panel shows the $86\,\mathrm{GHz}$ contours where the lowest contour is plotted at 5$\times$ the rms value and {increased} by factors of $2$. The solid black line {corresponds} to the jet axis and the red point indicates the selected position. The right panels show from top to bottom the distribution of the optically thin spectral index, $\alpha_0$, the turnover flux density, $S_m$, and the turnover frequency, $\nu_m$. The solid red lines {indicate} the mean of the distribution and the dashed red lines one standard deviation. The values for the spectral values are plotted in the upper left corner of the contour plot. Notice the asymmetric error bars and the tailed distribution of the spectral parameters.} 
\label{vmsmdist} 
\end{figure}

The uncertainties on the spectral parameters for each pixel were calculated from the obtained distribution. Since the equation of the synchrotron spectrum and the spectral slope are highly non-linear (Eq. \ref{snuapprox}), the spectral parameters are log-normal distributed. We computed from those distribution the mean and the standard deviation. The distributions calculated from the random shifts for one selected position are presented in Fig.~\ref{vmsmdist} and show clearly the log-normal distribution of the spectral parameters. 

\begin{figure}[h!]
\resizebox{\hsize}{!}{\includegraphics{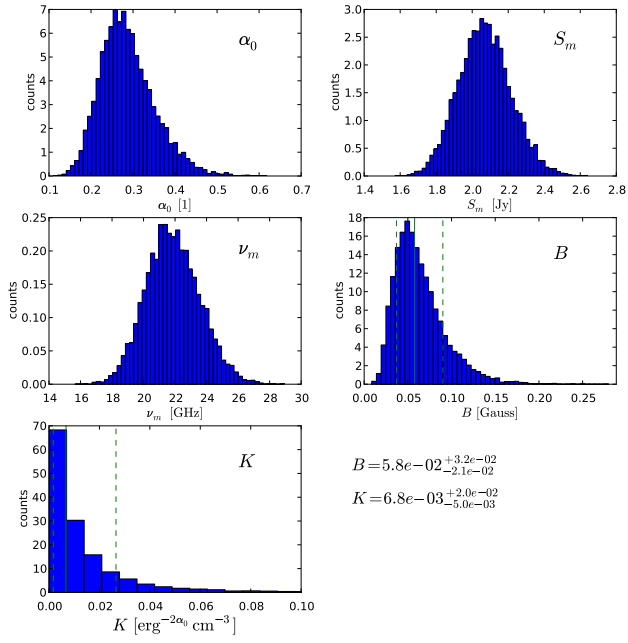}} 
\caption{Monte Carlo simulation results for magnetic field, $B$, and the normalization coefficient, $K$. The other panels show the distribution of the spectral parameters which are used for the calculation of $B$ and $K$. The solid green lines indicates the mean of the distribution and the dashed green lines one standard deviation. The values obtained for $B$ and $K$ are plotted in lower right corner. Notice the asymmetric error bars and the tailed distribution for all parameters.} 
\label{BKdist} 
\end{figure}

Once the variation in the spectral parameters was obtained, we used these results for the calculation of the uncertainties of the magnetic field, $B$, and the normalization coefficient of the relativistic electron distribution, $K$. Again, we used a Monte Carlo approach and select $10^4$ random values from the log-normal distribution of the spectral parameters and computed the scatter in $B$ and $K$ using Eqs.~\ref{bfield} and \ref{knorm} and the estimates of the jet width, $R$, and the Doppler factor, $\delta$ presented in Sect. \ref{physpara}. The dependence of the magnetic field and the normalization coefficient on the spectral parameters is highly non-linear, what results in strongly skewed distributions. In Fig.~\ref{BKdist} we present the distributions of the spectral parameters, the magnetic field and the normalization coefficient. \\

The uncertainties on the spectral index, $\alpha$, ($S_\nu\propto\nu^{\alpha}$) in the different regions are calculated in the same way.

\end{document}